\documentclass[twocolumn]{svjour3}          
\smartqed  

\usepackage{times}
\usepackage{helvet}
\usepackage{courier}

\usepackage{flafter}
\usepackage{graphicx}
\graphicspath{{figs/}}
\usepackage{colortbl} 
\definecolor{grey}{rgb}{0.95,0.95,0.95}
\definecolor{darkgrey}{rgb}{0.6,0.6,0.6}

\usepackage[T1]{fontenc}
\usepackage[utf8]{inputenc}
\fontfamily{ptm}\selectfont
\usepackage[iso]{umlaute}

\usepackage{longtable}
\usepackage{float}
\usepackage{amsmath}
\usepackage{amsfonts}
\usepackage{amssymb}
\usepackage{amscd}
\usepackage{bbm}

\usepackage{algorithm}
\usepackage{algorithmic}
\usepackage{mathrsfs}
\usepackage{wrapfig}
\usepackage{paralist}
\usepackage{url}
\usepackage{xspace}
\usepackage{rotating}
\usepackage{multirow}
\usepackage{cite}

\usepackage[center]{subfigure}

\usepackage{stmaryrd} 
\newcommand{\eg}{e.\,g.,\xspace}
\newcommand{\ie}{i.~e.,\xspace}
\newcommand{\cf}{cf.\xspace}
\newcommand{\del}{del.icio.us\xspace}
\newcommand{\flickr}{Flickr\xspace}
\newcommand{\bibs}{BibSonomy\xspace}
\newcommand{\twitter}{Twitter\xspace}
\newcommand{\wikipedia}{Wikipedia\xspace}
\newcommand{\wordnet}{WordNet\xspace}

\newcommand{\et}{et al.\xspace}
\newcommand{\metis}{\textsf{Metis}}
\newcommand{\mqi}{MQI}
\newcommand{\nmi}{\ensuremath{\text{NMI}}\xspace}

\newcommand{\sbt}{\,\begin{picture}(-1,1)(-1,-3)\circle*{3}\end{picture}\
}
\newcommand{\grey}[1]{\textcolor{darkgrey}{#1}}

\newcommand{\comments}[1]{}
\newcommand{\todo}[1]{}

\renewcommand{\vec}[1]{\overset{\shortrightarrow}{#1}}

\newcommand{\mqc}{\overline{m}_C}

\newcommand{\knn}{\ensuremath{\langle k_{nn}\rangle}}
\newcommand{\webtwo}{so called ``Web 2.0''\xspace}
\newcommand{\qap}{QAP\xspace}

\newcommand{\op}[1]{\ensuremath{\operatorname{#1}}}
\newcommand{\degree}[1]{\ensuremath{\op{d}(#1)}}
\newcommand{\conductance}{\mathit{C}}
\newcommand{\modularity}{\mathit{M}}
\newcommand{\segIndex}{\ensuremath{\mathit{S}}\xspace}
\newcommand{\krackhardt}{KH}
\newcommand{\norm}[1]{\ensuremath{\|#1\|}}
\newcommand{\cov}{\ensuremath{\mathop{cov}}}
\newcommand{\var}{\ensuremath{\mathop{var}}}

\newcommand{\coloneqq}{\mathrel{\mathop{:}}=}       
\newcommand{\walk}[3]{\ensuremath{#1\rightarrow_{#2}#3}}

\newcommand{\N}{\ensuremath{\mathbbm{N}}}           
\newcommand{\R}{\mathbb{R}}

\newcommand{\size}[1]{\ensuremath{|#1|}}

\newcommand{\set}[1]{\ensuremath{\{#1\}}}

\newcommand{\cutset}{\ensuremath{\overline{m}_{C}}\xspace}
\newcommand{\dotcup}{\ensuremath{\mathaccent\cdot\cup}}
\newcommand{\subgraph}[2]{\ensuremath{#1[#2]}}

\newcommand{\Copy}{Copy\xspace}
\newcommand{\Click}{Click\xspace}
\newcommand{\Visit}{Visit\xspace}
\newcommand{\Group}{Group\xspace}
\newcommand{\Friend}{Friend\xspace}

\newcommand{\RT}{ReTweet\xspace}
\newcommand{\Follower}{Follower\xspace}

\newcommand{\Comment}{Comment\xspace}
\newcommand{\Favorite}{Favorite\xspace}
\newcommand{\Contact}{Contact\xspace}

\journalname{}

\begin{document}

\title{User-Relatedness and Community Structure in Social Interaction Networks
}

\author{Folke Mitzlaff,
        Martin Atzmueller,
        Dominik Benz,
        Andreas Hotho and
        Gerd Stumme
}

\institute{
F. Mitzlaff \at University of Kassel, Knowledge and Data Engineering Group, Kassel, Germany\\
\email{mitzlaff@cs.uni-kassel.de}
\and M. Atzmueller \at University of Kassel, Knowledge and Data Engineering Group, Kassel, Germany\\
\email{atzmueller@cs.uni-kassel.de}
\and D. Benz \at University of Kassel, Knowledge and Data Engineering Group, Kassel, Germany\\
\email{benz@cs.uni-kassel.de}
\and A. Hotho \at University of Wuerzburg, Data Mining and Information Retrieval Group, Wuerzburg, Germany\\
\email {hotho@informatik.uni-wuerzburg.de}
\and G. Stumme \at University of Kassel, Knowledge and Data Engineering Group, Kassel, Germany\\
\email{stumme@cs.uni-kassel.de}
}

\date{}

\maketitle
\begin{abstract}
With social media and the according social and ubiquitous applications finding their way into everyday life, there is a rapidly growing amount of user generated content yielding explicit and implicit network structures. We consider social activities and phenomena as proxies for user relatedness. Such activities are represented in so-called \emph{social interaction networks} or \emph{evidence networks}, with different degrees of explicitness.
We focus on evidence networks containing relations on users, which are
represented by connections between individual nodes. Explicit
interaction networks are then created by specific user actions, for
example, when building a friend network. On the other hand, more
implicit networks capture user traces or 'evidences' of user actions
as observed in Web portals, blogs, resource sharing systems, and many
other social services. These implicit networks can be applied for a
broad range of analysis methods instead of using expensive
gold-standard information.

In this paper, we analyze different properties of a set of networks in
social media. We show that there are dependencies and correlations
between the networks. These allow for drawing reciprocal conclusions
concerning pairs of networks, based on the assessment of structural
correlations and ranking interchangeability. Additionally, we show how
these inter-network correlations can be used for assessing the results of structural
analysis techniques, e.g., community mining methods. 

\keywords{social
  networks \and folksonomies \and structural correlation \and
  analysis \and communities}
\end{abstract}

\section{Introduction}\label{sec:introduction}

With the rise of social software, and the increasing availability of mobile internet connections,
social applications are ubiquitously integrated into our daily life.
By interacting with such social systems, the user is leaving traces within the different databases and log files, \eg by copying a post in \bibs, updating the current status via \twitter or putting an image into her favorite list in \flickr.

Ultimately, each type of such traces gives rise to a corresponding network of user relatedness, where users are connected if they interacted either explicitly (\eg by establishing a
``friendship'' link within in an online social network) or implicitly (\eg by visiting a user's profile page). We consider a link within such a network as evidence for user relatedness and call it accordingly
\emph{evidence network}~\cite{HT2010} or \emph{social interaction network}.
This paper analyzes inter-network correlations between such
user-generated networks and uses these correlations for inferring
reciprocal conclusions.\pagebreak

Using information formalized in interaction networks has several advantages:
\begin{compactenum} \item The networks capture explicit and implicit social interactions being collected for a broad range of user actions. \item In every application where users may interact with each other, there are implicit evidence networks, even if no explicit user relationship is being implemented. \item Implicit networks may also be captured anonymously on a client network's proxy server. \item Obtaining the data is rather inexpensive, \eg when automatically being collected in running applications. \item Typically, implicit networks are also significantly larger than explicit networks.
\end{compactenum}

For the evidence networks, we assume that the set of observable social interactions is drawn from a certain ``social population''. Thus, the interactions indicate connections in this distribution, and they manifest themselves with varying degrees in different (proxy) networks. By considering samples of such a ``social population'' we aim to collect evidences for the underlying user relatedness.
%
We analyze data from the social bookmarking system \bibs\footnote{\url{http://www.bibsonomy.org}}\cite{Bibs:VLDB10}, as well as publicly available data from the microblogging service \emph{\twitter}\footnote{\url{http://www.twitter.com}}, and the resource sharing system \emph{\flickr}.\footnote{\url{http://www.flickr.com}}
%
%
\noindent Our contribution can be summarized as follows:
\begin{compactitem}
    \item Considering the notion of evidence networks for inter-network analysis, we analyze them thoroughly with respect to their contained semantic properties and community structure.
    \item We show that there are structural inter-network correlations that allow reciprocal conclusions between the different networks.
    \item We apply standard community evaluation measures on a set of evidence networks.
    We show that there is a strong common community structure across different networks.
    \item We analyze the rankings between a large set of communities mined on the different networks, and show, that the induced rankings are reciprocally consistent.
    \item In summary, we show that the observed correlations and dependencies are strong enough for assessing structure-based analysis techniques, especially community mining methods, for obtaining a relative ranking of community allocations.
\end{compactitem}

The application area of the presented approach is potentially rather broad and
ranges from simple social applications to more advanced ubiquitous applications.
Mobile phones, for example, are equipped with more and more sensors;
%
interactions in mobile web applications then lead to implicit user
relationships which naturally fit into the framework of evidence
networks, for assessing online and offline data at the same time.


The remainder of the paper is structured as follows:
Section~\ref{sec:evidence:networks} summarizes basics of graphs and networks, and introduces the notion of evidence networks.
After that, Section~\ref{sec:analysis} analyzes different networks from our three application systems \bibs, Twitter and \flickr.
Section~\ref{sec:experiments} performs a structure-based analysis for assessing user communities.
Section~\ref{sec:related} discusses related work.
Finally, Section~\ref{sec:conclusion} concludes the paper
with a summary and future work.

\section{Evidence Networks in Social Media}\label{sec:evidence:networks}
\comments{
With the rise of the \webtwo and its applications, information is
both consumed and provided by the same set of users. These interactions happen in environments
where users interact either explicitly (\eg by establishing friendship
relations) or implicitly (\eg by viewing the same video clip). The interactions leave traces which we can consider as evidences for user relatedness.}

This section starts with a brief summary of basic notions of
 graph and network theory. For more details, we refer to standard
literature, \eg~\cite{Diestel2006,newman2003structure,DBLP:conf/dagstuhl/Gaertler04}.
Next, we present all considered data sources and the corresponding graph structures.

\paragraph{Graph \& Network Basics}
An undirected \emph{graph} is an ordered pair $G=(V,E)$, consisting of a finite
set $V$ of \emph{vertices} or \emph{nodes}, and a set $E$ of \emph{edges}, which
are two
element subsets of $V$. In a \emph{directed graph}, $E$
denotes a subset of $V\times V$. For simplicity, we write 
$(u,v)\in E$ in both cases for an edge belonging to $E$ and freely use
the term
\emph{network} as a synonym for a graph. In a \emph{weighted graph},
each edge $l\in E$ is given an edge weight $w(l)$ by some weighting
function $w\colon E\rightarrow\R$.
The \emph{degree} of a node in a network is the number of
connections it has to other nodes. The \emph{adjacency matrix} of a set of nodes $S$ with $n
= |S|$ contained in a (weighted) graph $G=(V,E)$ is a matrix
$A\in\R^{n\times n}$ with $A_{ij}=1$
($A_{ij}=w(i,j)$) iff $(i,j)\in E$ for any nodes $i,j$ in $S$
(assuming some bijective mapping from ${1,\ldots,n}$ to $S$). We
identify a graph with its according adjacency matrix where
appropriate. 
A \emph{path} \walk{v_0}{G}{v_n} of \emph{length} $n$ in a graph $G$
is a
sequence $v_0,\ldots,v_n$ of nodes with $n\ge1$ and $(v_i,v_{i+1})\in
E$ for
$i=0,\ldots,n-1$. A \emph{shortest path} between nodes $u$ and $v$ is
a path
\walk{u}{G}{v} of minimal length. The \emph{transitive closure} of a
graph
$G=(V,E)$ is given by $G^*=(V,E^*)$ with $(u,v)\in E^*$ iff there
exists a path
\walk{u}{G}{v}. A \emph{strongly connected component (scc)} of $G$ is
a subset
$U\subseteq V$, such that \walk{u}{G^*}{v} exists for every $u,v\in
U$. A
\emph{(weakly) connected component (wcc)} is defined accordingly,
ignoring the
direction of edges $(u,v)\in E$.

A binary relation on a set $V$ is a \emph{relation} $R$ as a subset $R\subseteq
V\times
V$. A relation $R$ is naturally mapped to a directed
graph $G_R\coloneqq(V,R)$. We say that a relation $R$ among
individuals $U$ is
\emph{explicit}, if $(u,v)\in R$ only holds, when at least one of
$u,v$
\emph{explicitly} established a connection to the other (\eg user
$u$ added user $v$ \emph{deliberately}
as a friend in an online social network). We call $R$ \emph{implicit},
if
$(u,v)\in R$ can be \emph{derived} from a set of other relations, e.g., it
holds as a side effect of the actions taken by $u$ and $v$ in a social
application.
Explicit relations are thus given by explicit links, e.g., existing
links between users. Implicit relations can be derived or constructed
by analyzing secondary data.

\subsection{Evidence Networks in \twitter}\label{sec:networks:twitter}
As a first case study, we considered the microblogging service
\twitter. Using \twitter, each
user publishes short text messages (called ``\emph{tweets}'') which
may contain freely chosen \emph{hashtags}, \ie distinguished words
which are used for marking keywords or topics. Furthermore, users may
``cite'' each other by ``retweeting'': A user $u$ retweets user $v$'s
content, if $u$ publishes a text message containing ``RT @$v$:''
followed by (an excerpt of) $v$'s corresponding tweet. Users may also
explicitly follow other user's tweets by establishing a corresponding
friendship-like link. For our analysis, we considered the following
networks:
\begin{compactitem}
\item The \emph{Follower graph} is an explicit evidence network given by a directed graph containing an edge $(u,v)$
  iff user $u$ follows the tweets of user $v$.
\item The \emph{ReTweet graph} is an implicit evidence network given by a directed graph; it contains an edge   $(u,v)$ with weight $c\in\N$ iff user $u$ ``retweeted''
  exactly $c$ of user $v$'s tweets.
\end{compactitem}

\paragraph{Data Source.}
We extracted \twitter's ReTweet-network from a publicly available
\twitter data set~\cite{yang2011patterns} which is estimated to cover
$20$-$30\%$ of all public tweets published on \twitter during June 1
2009 to December 31 2009. Additionally, we used the follower network
as made available in \cite{kwak2010twitter} which was crawled during
the time period June 1 2009 until September 24 2009, containing more
than 1.4 billion following relations. For our analysis we only
considered users which were also present in the ReTweet-Network.

\subsection{Evidence Networks in \flickr}\label{sec:networks:flickr}
\flickr focuses
on organizing and sharing photographs collaboratively. Users mainly
upload images and assign arbitrary tags but also interact, \eg by
establishing contacts or commenting images of other users. For our
analysis we extracted the following networks:
\begin{compactitem}
\item The \emph{Contact graph} is an explicit evidence network given by a directed graph; it contains an edge $(u,v)$ iff user $u$ added user $v$ to its personal contact list.
\item The \emph{Favorite graph} is an implicit evidence network given by a directed graph containing an edge
  $(u,v)$ with weight $n\in\N$ iff user $u$ added exactly $n$ of $v$'s
  images to its personal list of favorite images.
\item The \emph{Comment graph} is an implicit evidence network; the directed graph contains an edge $(u,v)$ with a weight $c\in\N$ iff user $u$ posted exactly $c$ comments on
  $v$'s images.
\end{compactitem}

\paragraph{Data Source.}
The \flickr networks were extracted from an own breadth-first crawl
conducted in April until June 2011. The search was regularly reseeded
by randomly selecting a search term from a library catalogue search
term data set\footnote{\url{http://data.gov.au/1277}} which was then
used for querying images using \flickr's
API.\footnote{\url{http://www.flickr.com/services/api/}} In parallel
all incident comments, users, contacts and favorites were crawled.
In total, the considered flickr data set consisted of $588,634$ photos for
$69,104$ users who applied $564,251$ different tags in $5,911,127$ tag
assignments.

Data sets obtained by breadth-first crawl techniques are known to be
biased towards high degree nodes~\cite{gjoka2011practical} and likely
underestimate link symmetry~\cite{becchetti2006comparison}. This work
aims at comparing structural characteristics of different networks
contained within a given social constellation (\eg on the set of users in
\flickr) rather than characterizing the networks. However, the different
networks obtained from \flickr were crawled in parallel. Thus, induced biases
should have a comparable impact on all considered
networks.

\subsection{Evidence Networks in \bibs}\label{sec:networks:bibsonomy}
\comments{ In the context of the BibSonomy system, we distinguish the
  following explicit networks: The follower-graph, the friend-graph,
  and the group graph that are all established using explicit links
  between users. Formally, these graphs can be deﬁned as follows: }

\bibs is a social bookmarking
system where users manage their bookmarks and publication references
via \emph{tag} annotations (\ie freely chosen keywords). Most
bookmarking systems incorporate additional relations on users such as
``\emph{my network}'' in
\del\footnote{\url{http://delicious.com/network/<username>}} and
``\emph{friends}'' in
\bibs\footnote{\url{http://www.bibsonomy.org/friends}}. Each such
network is connected with a certain functionality, \eg for restricting
access to certain resources or for allowing messages to be
sent. Nevertheless, we expect that those networks also have a certain ``social meaning''.
\paragraph{Explicit Evidence Networks}
\begin{compactitem}
\comments{
\item The \emph{Follower graph} is a directed graph containing an edge $(u,v)$
  iff user $u$ follows the posts of user $v$, \ie user $u$ monitors the
  posts and is able to keep track of new posts of user $v$.
}
\item The \emph{Friend graph} is a directed graph containing an edge $(u,v)$
  iff user $u$ has added user $v$ as a friend. In \bibs, the only purpose of the friend graph so far is to restrict
  access to selected posts so that only users classified as
  ``friends'' can observe them.
\item The \emph{Group graph} is an undirected graph containing an edge
  $\{u,v\}$ iff user $u$ and $v$ share a common group, \eg 
  defined by a certain research group or a special interest group.
\end{compactitem}
Due to its limited size we excluded the network obtained from \bibs's follower
feature which enables users to monitor new posts of other users.

Beside those explicit relations among users, different relations
are established implicitly by user interactions within the systems,
\eg when user $u$ looks at user $v$'s resources. Using the \bibs's log
files, a broad range of interaction networks were available.

\paragraph{Implicit Evidence Networks}
\comments{ Concerning the implicit relationships, we propose the
  following networks: The click-graph, the copy graph, and the visit
  graph that are built by analyzing the actions of users, \ie
  clicking on links, copying resources, and browsing in other user's
  resources, respectively. Formally, the graphs are defined as
  follows:
}
\begin{compactitem}
\item The \emph{Click graph} is a directed graph containing an edge $(u,v)$ iff
  user $u$ has clicked on a link on the user page of user $v$.
\item The \emph{Copy graph} is a directed graph containing an edge $(u,v)$ iff
  user $u$ has copied a resource, \ie a publication reference from
  user $v$.
\item The \emph{Visit graph} is a directed graph containing an edge $(u,v)$ iff
  user $u$ has navigated to the user page of user $v$.
\end{compactitem}

Each implicit graph is given a weighting function counting certain events (\eg the number of posts which user $u$ has
copied from $v$ in case of the Copy graph).

\paragraph{Data Source.}
Our primary resource is an anonymized dump of all public bookmark and
publication posts until January 25, 2010.
It consists of 175,521 tags, 5,579 users, 467,291 resources and
2,120,322 tag assignments.  
The dump also contains friendship relations modeled in \bibs
among $700$ users.
Furthermore, we utilized the ``\emph{click log}'' of \bibs, consisting
of entries which are generated whenever a logged-in user clicked on a
link in \bibs. A log entry contains the URL of the currently visited
page together with the corresponding link target, the date and the
user name\footnote{Note: For privacy reasons a user may deactivate
  this feature.}. For our experiments we considered all click log
entries until January 25, 2010. Starting in October 9, 2008, this
dataset consists of 1,788,867 click events in total.
We finally considered the corresponding apache web server log files,
containing around 16 GB compressed log entries.
\section{Comparative Analysis of the Evidence Networks}\label{sec:analysis}
In the following section, we outline general structural properties of the obtained networks and
comparatively discuss major structural characteristics in order to show that there are structural interactions and correlations between the different evidence networks. In particular, we consider general structural properties, the degree distribution and the degree correlation. Furthermore, we analyze topological and semantical distances, the networks' neighborhood, and the inter-network correlations. In the next section, we will see that these are strong enough to draw reciprocal conclusions between the different networks.

\subsection{General Structural Properties}\label{sec:analysis:structure}
Table \ref{tab:statistics:general} summarizes major graph level
statistics for the considered networks which range in size from
hundreds of edges (\eg \bibs's Friend graph) to more than one hundred
million edges (\flickr's Contact graph). All networks obtained from
\bibs are complete and therefore not biased by a previous crawling
process. In exchange, effects induced by limited network sizes have
to be considered.

\begin{figure*}
  \centering
  \includegraphics[scale=1.1]{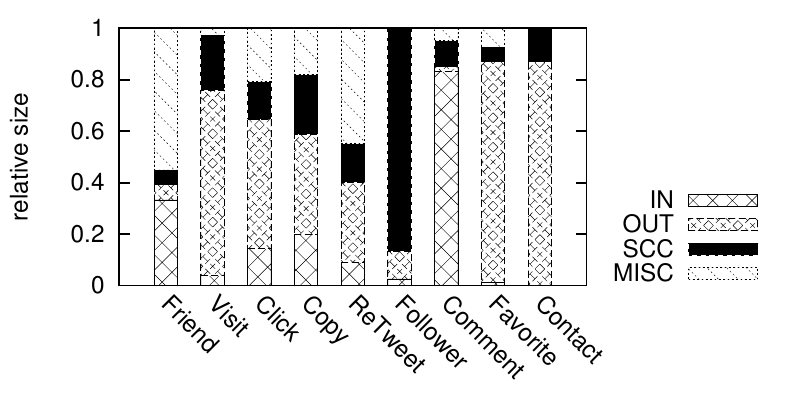}
  \caption{Relative size of the largest strongly connected component
    (SCC) together with the size of its incoming and outgoing node
    sets. }
\label{fig:corestructure}
\end{figure*}

Interestingly, only the Follower graph exhibits a giant strongly
connected component (\ie a large fraction of nodes within a single
strongly connected component) as expected in online social
networks~\cite{mislove2007measurement}. Figure~\ref{fig:corestructure}
shows a more detailed analysis of the graph structure relative to the corresponding
largest strongly connected component (SCC). According to the seminal
work by Broder et al~\cite{broder2000graph} for web graphs, the node set of a graph can be
partitioned into the set of nodes within the largest strongly
connected component, the set of nodes reaching into the SCC (the IN
set) and those reachable from the SCC (the OUT set). All remaining
nodes are comprised in the MISC set. Additionally, all networks contain a giant weakly connected component $\text{WCC*}$, for which $\size{\text{WCC*}} \geq \size{\text{IN}}+\size{\text{SCC}}+\size{\text{OUT}}$.

There is no global common pattern concerning the distribution of the different
node sets.
Only the Click graph, the Copy graph and the ReTweet graph
show a comparable structure. Notably, \flickr's Comment graph is
``inversely'' structured to the FavoriteGraph and the Contact graph. Possible explanations concern the interaction patterns in the different networks, concerning the relation to resources and users. The Follower graph, for example, is densely connected which could be due to the implicitness of the interactions. In contrast, the Comment graph considers reactions to the posts of other users. Favorite and Contact graphs are established explicitly and show similar characteristics. This is also confirmed by the analysis in Section~\ref{sec:analysis:qap}. 
%
\begin{table*}
  \centering
  \begin{tabular}{l r r r r r r}
                &  $|V|$   &    $|E|$    & density           &   \#SCC   & largest SCC      & WCC*  \\\hline
    \Copy       &  $1,427$   &  $4,144$     &  $2\cdot 10^{-3}$  &   $1,108$  &   $309$     & $1,339$      \\
    \Click      &  $1,151$   &  $1,718$     &  $10^{-3}$         &   $963$    &   $150$     & $1,022$      \\
    \Visit      &  $3,381$   &  $8,214$     &  $10^{-3}$         &   $2,599$  &   $717$     & $3,359$      \\
    \Group      &  $550$     &  $6,693$     & $2,2\cdot 10^{-3}$ &   $-$      &   $-$       & $228$        \\
    \Friend     &  $700$     &  $1,012$     & $2\cdot 10^{-3}$   &  $515$     &    $17$     & $238$        \\
    \hline\hline
    \RT         &  $826,104$ &  $2,286,416$ & $3,4\cdot 10^{-6}$ & $699,067$  &  $123,055$  & $702,809$    \\
    \Follower   &$1,486,403$ & $72,590,619$ & $3,3\cdot 10^{-5}$ & $198,883$  & $1,284,201$ & $1,485,356$  \\
    \hline\hline
    \Comment    & $525,902$  &  $3,817,626$ & $1,4\cdot 10^{-5}$ & $472,232$  &  $53,359$   & $522,212$    \\
    \Favorite   &$1,381,812$ & $20,206,779$ & $1,1\cdot 10^{-5}$ & $1,305,350$&  $76,423$   & $1,380,906$  \\
    \Contact    &$5,542,705$ &$119,061,843$ & $3,9\cdot 10^{-6}$ & $4,820,219$&  $722,327$  & $5,542,703$  \\\hline
  \end{tabular}
  \caption{High level statistics for all networks.}
  \label{tab:statistics:general}
\end{table*}
%
The structuring of a network relative to its SCC interplays with its
link symmetry properties (\ie the fraction of links which are
symmetric) and the Krackhardt hierarchy~\cite{Krackhardt:94}, which measures the fraction of
connected pairs of nodes which are reachable only in one
direction. Table~\ref{tab:statistics:paths} reveals a high fraction of
symmetric links in the Follower graph and the Contact graph as typically
observed in online social networks~\cite{mislove2007measurement} but
only the Follower graph shows a low Krackhardt Hierarchy value (\ie
high fraction of connected pairs that are symmetric in the graph's
transitive closure). This deviating behavior can be explained by the
different sizes of the SCCs.

\begin{figure*}
  \centering
  \includegraphics[width=0.32\linewidth]{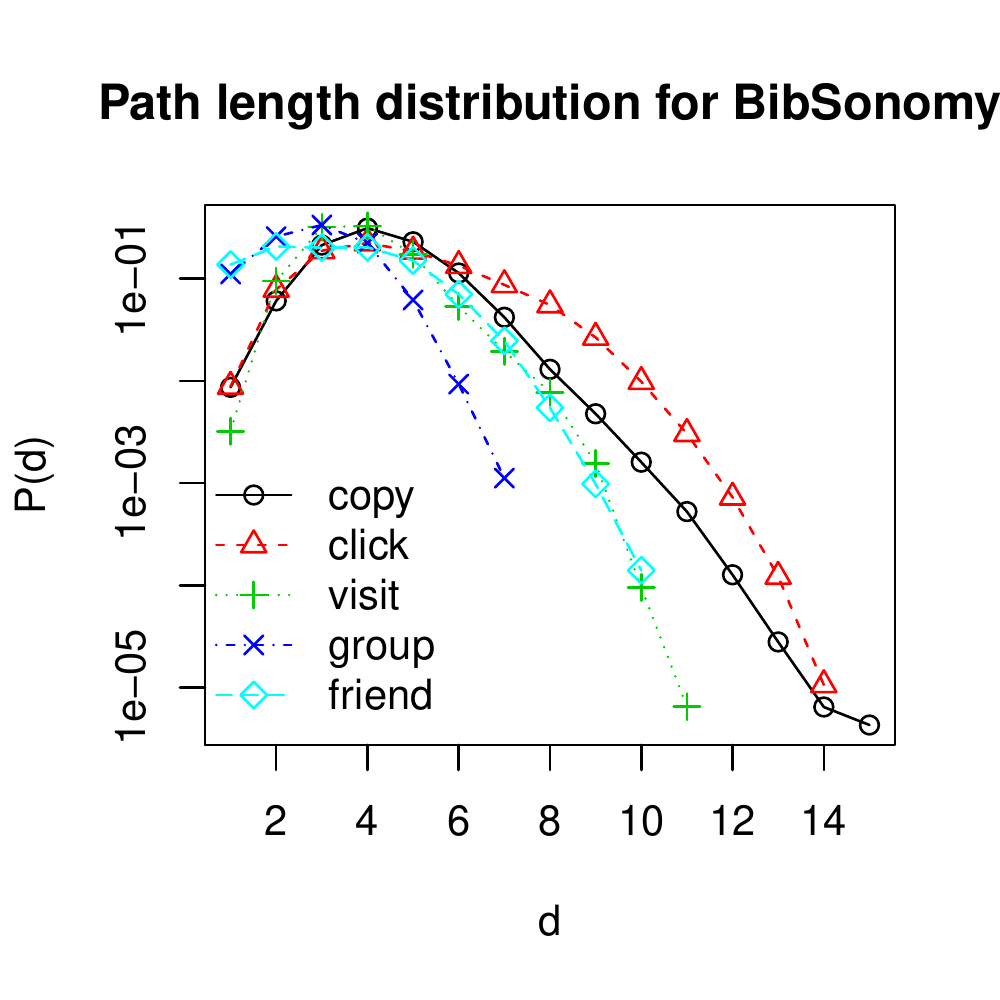}
  \includegraphics[width=0.32\linewidth]{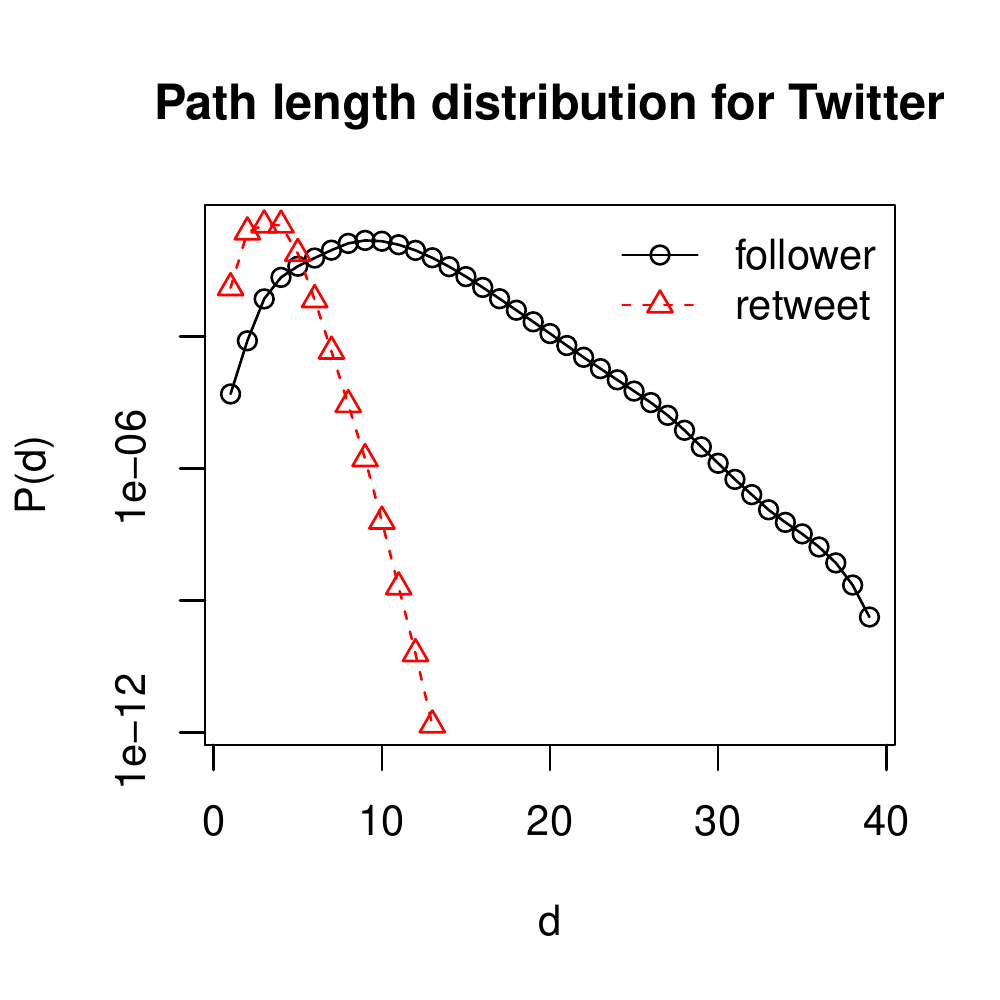}
  \includegraphics[width=0.32\linewidth]{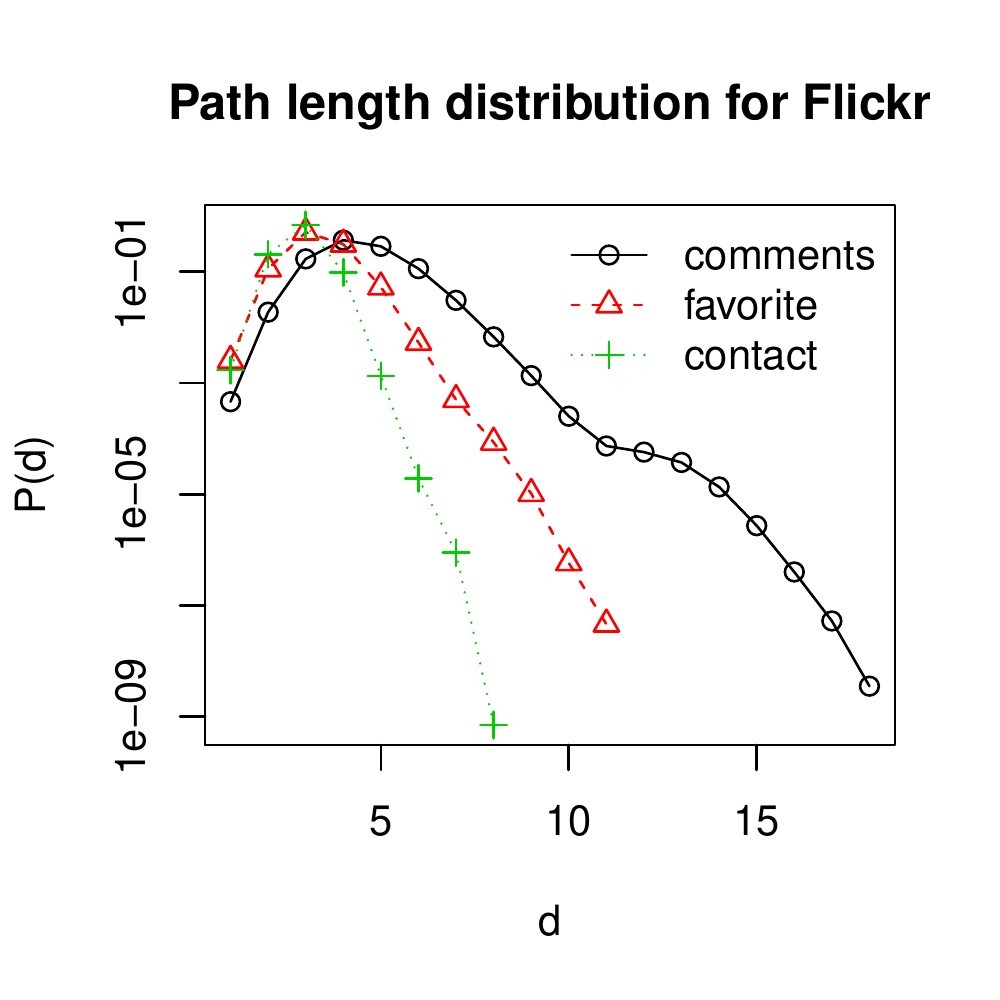}
\comments{
  \includegraphics[width=0.3\linewidth]{figs/similarities/copyGraph-pathlengths}
  \includegraphics[width=0.3\linewidth]{figs/similarities/clickGraph-pathlengths}
  \includegraphics[width=0.3\linewidth]{figs/similarities/visitGraph-pathlengths}
  \includegraphics[width=0.3\linewidth]{figs/similarities/groupGraph-pathlengths}
  \includegraphics[width=0.3\linewidth]{figs/similarities/friendGraph-pathlengths}
  \includegraphics[width=0.3\linewidth]{figs/similarities/tweets2009-RT-pathlengths}
  \includegraphics[width=0.3\linewidth]{figs/similarities/socialgraph-tweets2009-pathlengths}
  \includegraphics[width=0.3\linewidth]{figs/similarities/comments-pathlengths}
  \includegraphics[width=0.3\linewidth]{figs/similarities/favorite-pathlengths}
  \includegraphics[width=0.3\linewidth]{figs/similarities/contact-pathlengths}
}
  \caption{Distribution of the shortest path lengths in the evidence
    networks with logarithmically scaled counts on the $Y$-axis.
    \comments{SCRIPT: plot_path_distribution.R}
  }
\label{fig:pathlength:distribution}
\end{figure*}

Table \ref{tab:statistics:paths} also shows the diameter, average path
length and the transitivity (also called clustering coefficient) for
all considered networks. Except for the Group graph, the Friend graph
and the ReTweet graph, all networks exhibit a comparable magnitude of
these indices. While the Group graph and the Friend graph are
characterized by a large transitivity, the ReTweet graph shows an
unexpected high diameter and average path length.
Figure \ref{fig:pathlength:distribution} breaks down the average to the
distribution of path lengths. The Click graph and the Visit graph, for example, show
a clear common distribution pattern as do the Copy graph, the
Retweet graph, the Follower graph and the Favorite graph where both
groups have a single cluster point around the graph's average path
length.

\begin{table*}\centering
  \begin{tabular}{c|l|r|r|r|r|r}
&               &  diameter  &  APL  & transitivity &  symm. links  & \krackhardt \\\hline\hline
\multirow{5}{*}{{\bibs}} &
     \Copy       &  $15$      &  $4.3$             &   $0.10$     &  $0.09$       &   $0.80$    \\
&    \Click      &  $15$      &  $4.8$             &   $0.02$     &  $0.12$       &   $0.88$    \\
&    \Visit      &  $11$      &  $3.9$             &   $0.01$     &  $0.12$       &   $0.81$    \\
&    \Group      &   $7$      &  $2.9$             &   $0.85$     &  $-$          &   $-$       \\
&    \Friend     &  $10$      &  $3.4$             &   $0.28$     &  $0.12$       &   $0.81$    \\\hline\hline
\multirow{2}{*}{{\twitter}} &
     \RT         &  $39$      &  $9.7$             &   $0.06$     &  $0.12$       &   $0.81^*$  \\
&    \Follower   &  $13$      &  $3.3$             &   $0.01$     &  $0.55$       &   $0.12^*$  \\\hline\hline
\multirow{3}{*}{{\flickr}} &
     \Comment    &  $18$      &  $4.4$             &   $0.03$     &  $0.08$       &   $0.91^*$  \\
&    \Favorite   &  $11$      &  $3.3$             &   $0.02$     &  $0.03$       &   $0.96^*$  \\
&    \Contact    &   $8$      &  $2.9$             &   $0.05$     &  $0.46$       &   $0.87^*$  \\\hline\hline
  \end{tabular}
  \caption{Path statistics with average path length (APL) for all networks where the Krackhardt Hierarchy (\krackhardt) 
    values marked with an asterisk are estimated by repeatedly averaging over random samples 
    of pairs of vertices}
  \label{tab:statistics:paths}
\end{table*}

\subsection{Degree Distribution}\label{sec:analysis:degreedistribution}
\begin{figure*}
  \centering
  \includegraphics[width=0.32\linewidth]{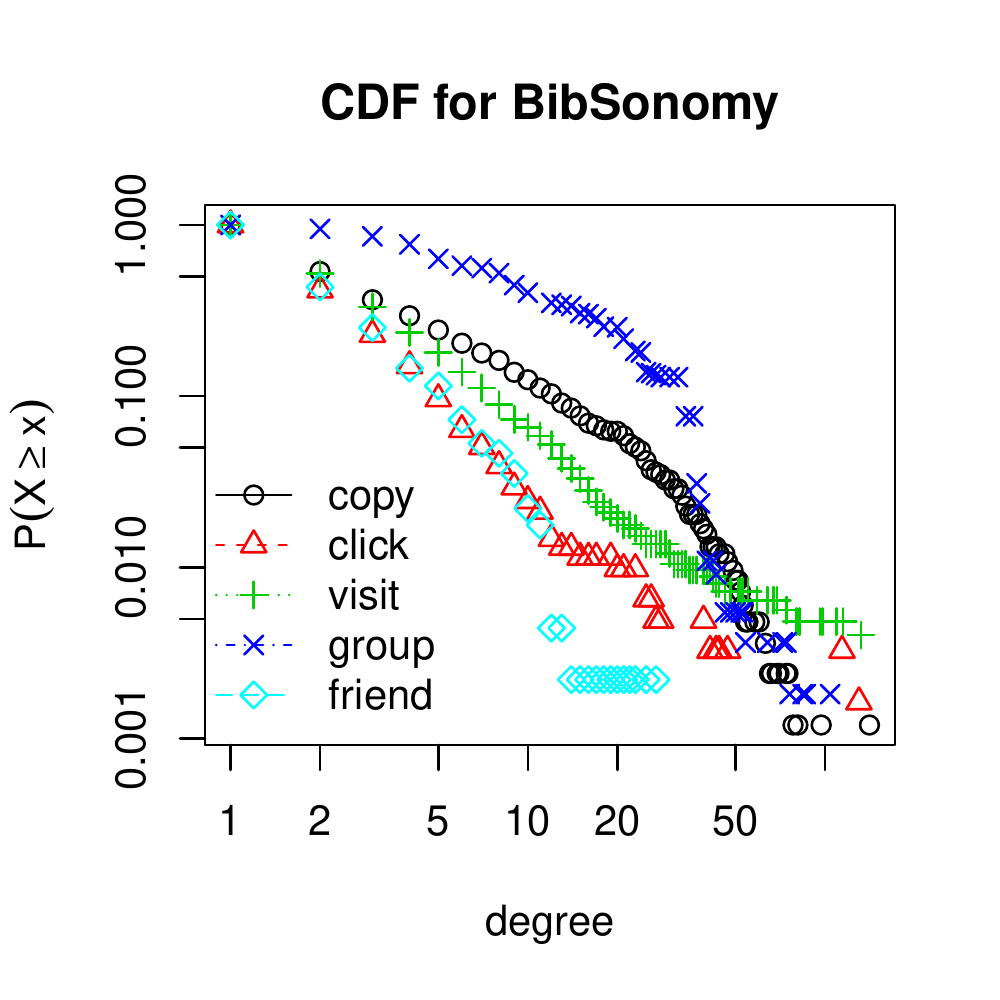}
  \includegraphics[width=0.32\linewidth]{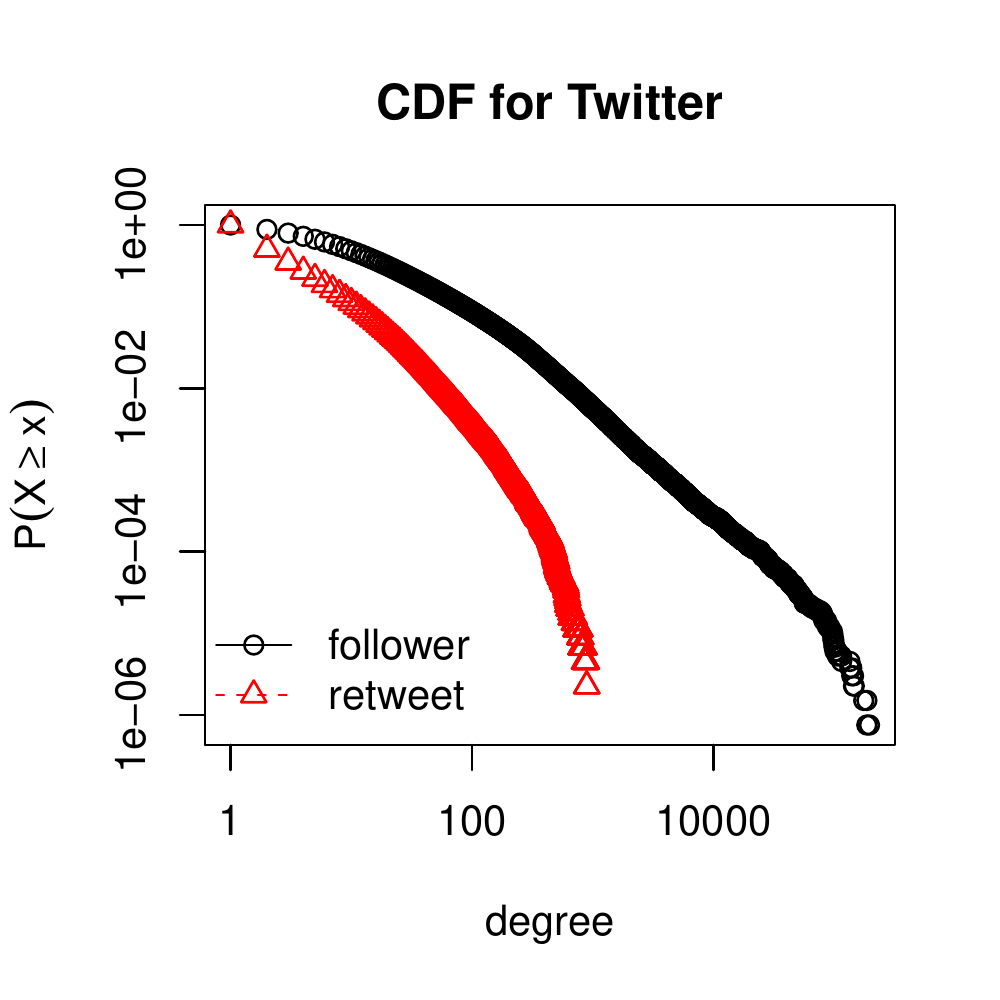}
  \includegraphics[width=0.32\linewidth]{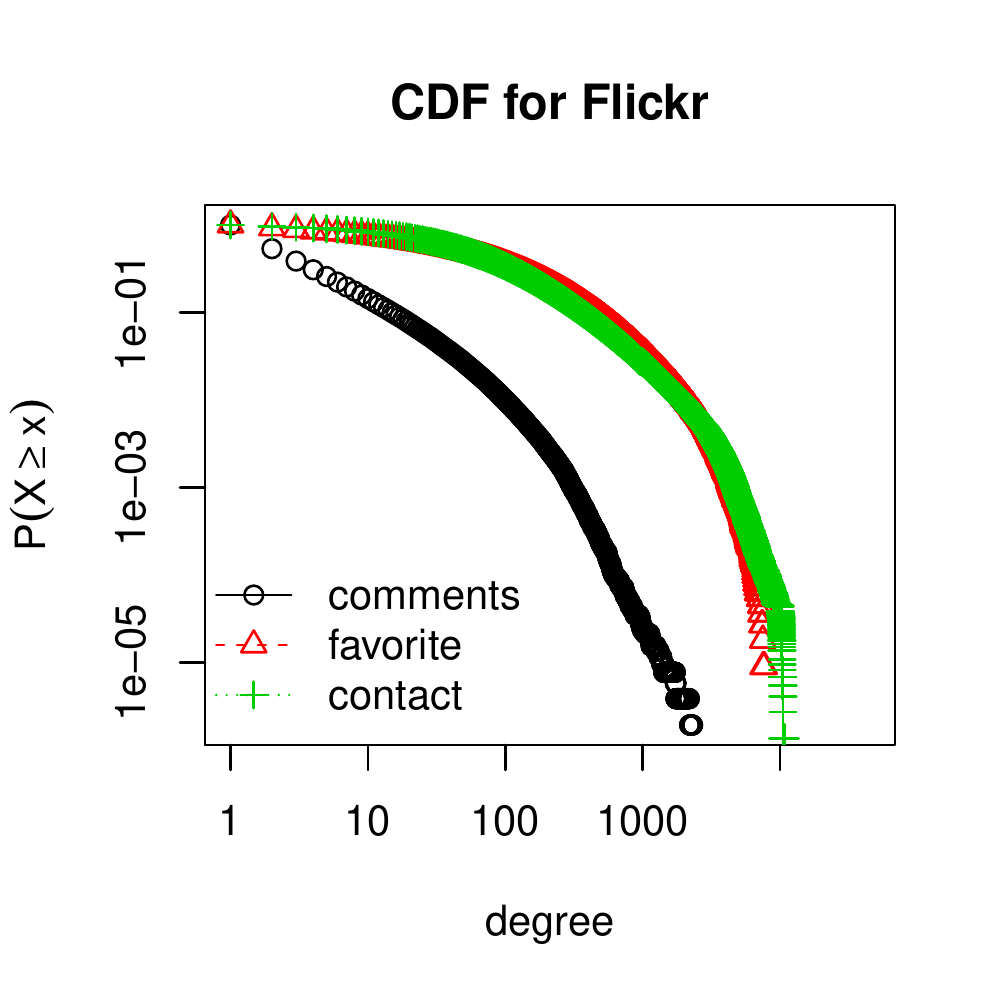}
\comments{
  \includegraphics[width=0.3\linewidth]{figs/statistics/degree/copyGraph-out_degree_cdf}
  \includegraphics[width=0.3\linewidth]{figs/statistics/degree/clickGraph-out_degree_cdf}
  \includegraphics[width=0.3\linewidth]{figs/statistics/degree/visitGraph-out_degree_cdf}
  \includegraphics[width=0.3\linewidth]{figs/statistics/degree/groupGraph-out_degree_cdf}
  \includegraphics[width=0.3\linewidth]{figs/statistics/degree/friendGraph-out_degree_cdf}
  \includegraphics[width=0.3\linewidth]{figs/statistics/degree/tweets2009-RT-out_degree_cdf}
  \includegraphics[width=0.3\linewidth]{figs/statistics/degree/socialgraph_tweets2009-out_degree_cdf}
  \includegraphics[width=0.3\linewidth]{figs/statistics/degree/comments-out_degree_cdf}
  \includegraphics[width=0.3\linewidth]{figs/statistics/degree/favorites-out_degree_cdf}
  \includegraphics[width=0.3\linewidth]{figs/statistics/degree/contacts-out_degree_cdf}
}
  \caption{Logarithmically scaled cumulative degree distribution in
    the evidence networks.
    \comments{SCRIPT: C/gdd,  R/plot_similarities.R}
  }
\label{fig:degree:distribution}
\end{figure*}

The number of adjacent nodes (degree) is one of the most important
features of a node within a given network. Accordingly, the
distribution of all node degrees is widely accepted as one of the key
features for summarizing a given network's
connectivity~\cite{kolaczyk2009statistical,newman2003structure,rapoport1957contribution}. Figure \ref{fig:degree:distribution}
shows the cumulative degree distribution for all networks. With the
exception of the Friend graph (most probably due to its limited size),
all networks exhibit a long tailed degree distribution. Again, there
are strong deviations among the different networks. Especially the favorite and contact networks
obtained from \flickr clearly show a bias towards high degree nodes
induced by the applied breadth-first crawling approach.

\subsection{Degree Correlation}\label{sec:analysis:degreecorrelation}
The graph indices considered so far describe the overall
structure of a network, but they do not provide insights into the
individual connection structure of the network.  A connection
pattern which was observed in many social networks, called
assortativity, homophily or mixing patterns, describes the phenomenon
that similar nodes tend to connect with each
other~\cite{newman2003structure}. This especially applies to the node
degree where this property is also called \emph{degree
  correlation}. In contrast, other types of networks, such as
information networks, technical networks or biological networks,
exhibit a opposite pattern, which is called
disassortativity~\cite{newman2003structure}. Several approaches for
analyzing the degree correlation in networks were proposed (see
\cite{kolaczyk2009statistical} for a discussion). We apply the
approach proposed in \cite{pastor2001dynamical,vazquez2002large}
which compactly visualizes the degree correlation by calculating the
mean degree of all neighbors of a node as a function of the node
degree. Formally, let $p(k'|k)$ denote the conditional probability
that a node of degree $k$ is adjacent to a node of degree $k'$. If a
network exhibits nontrivial correlations among the nodes' degree, this
conditional probability is not independent of $k$ which can be
visualized by plotting the direct nearest neighbors average degree
$\knn\coloneqq\sum_{k'}k'p(k'|k)$. 

\begin{figure*}
  \centering
  \includegraphics[width=0.4\linewidth]{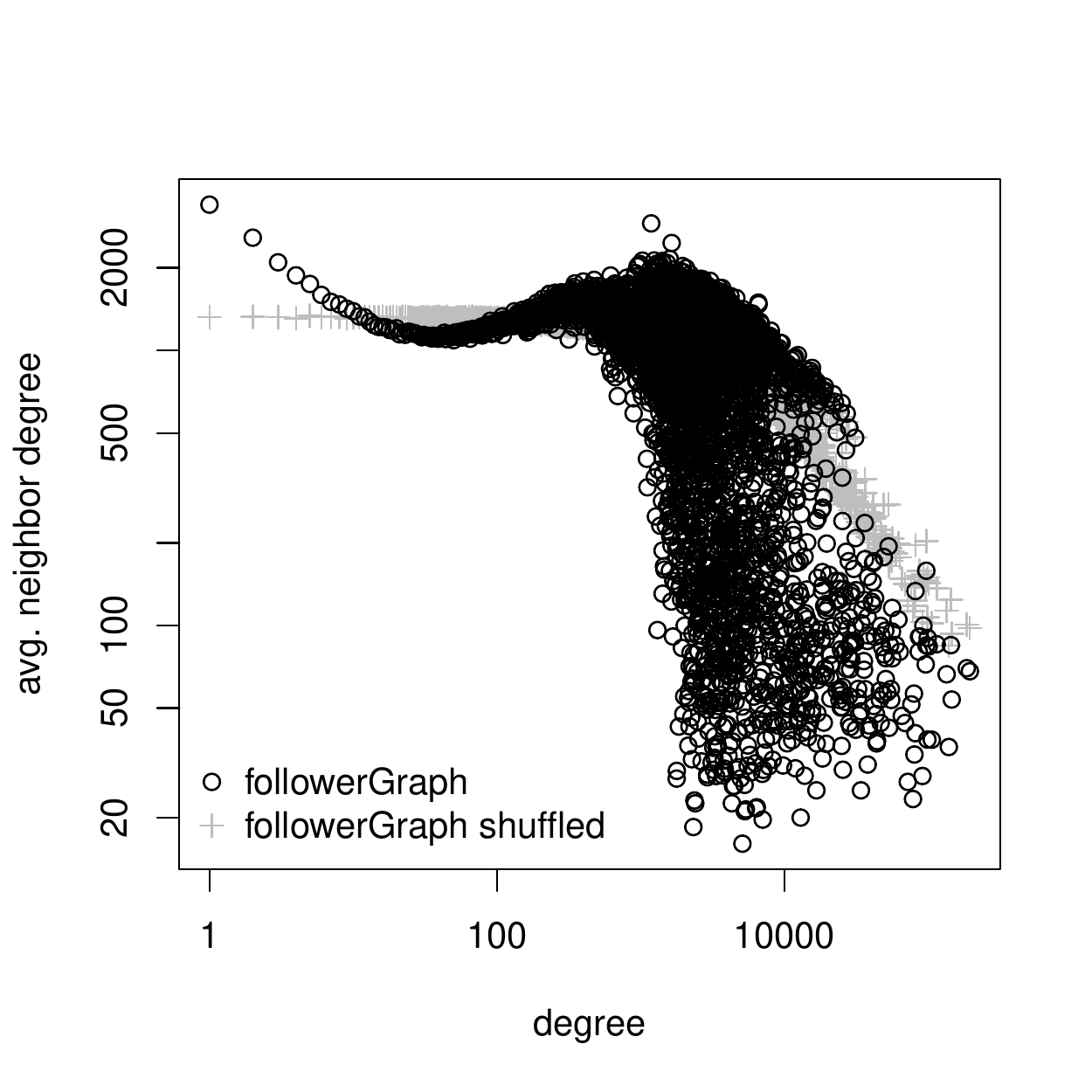}
  \includegraphics[width=0.4\linewidth]{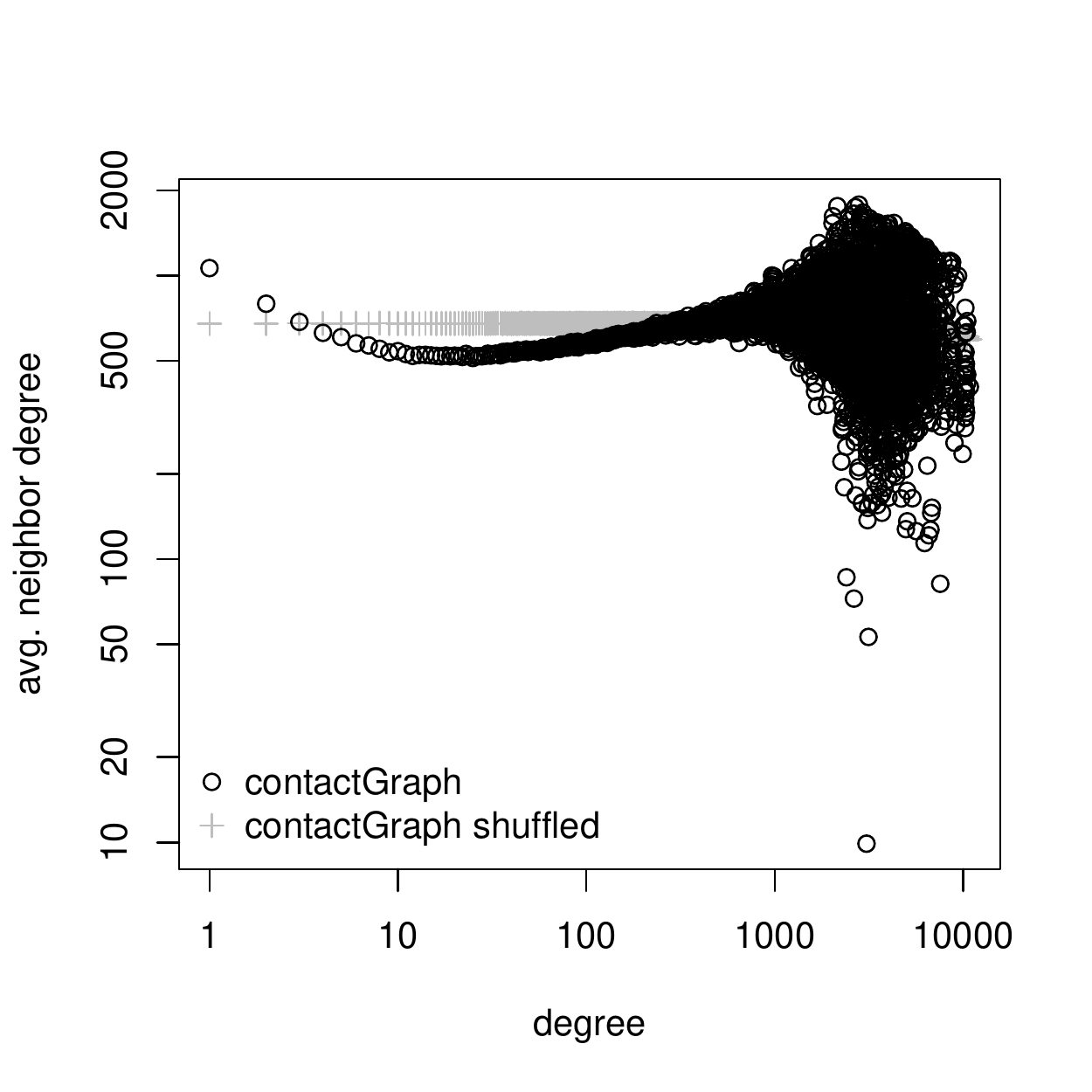}
  \includegraphics[scale=0.54]{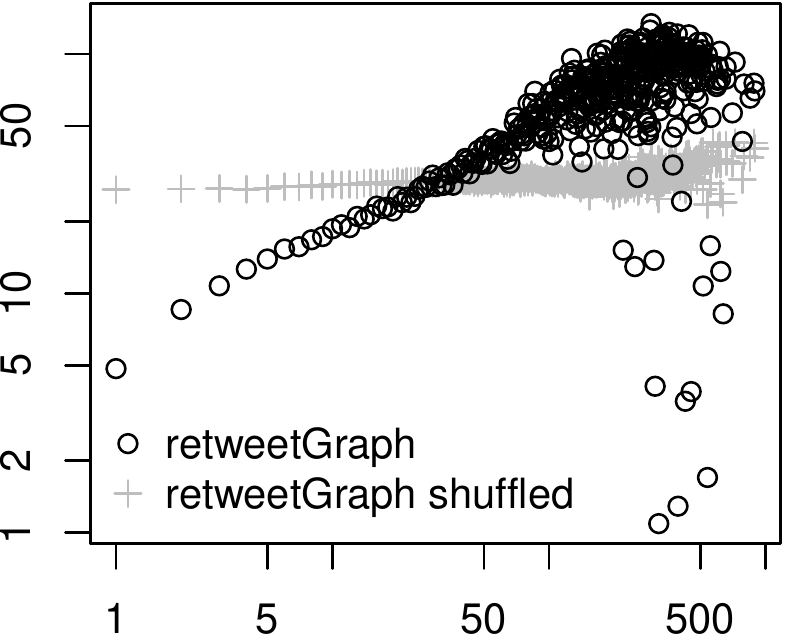}
  \includegraphics[scale=0.54]{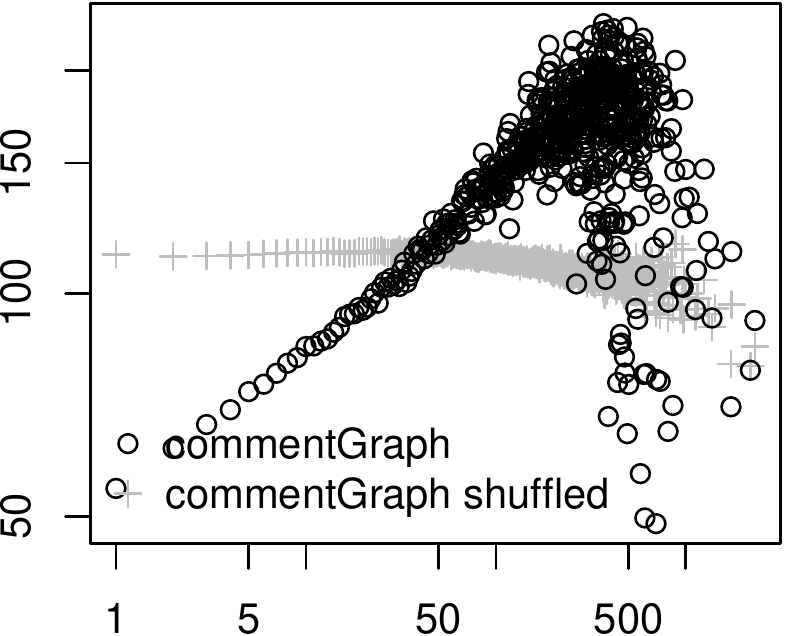}
  \includegraphics[scale=0.54]{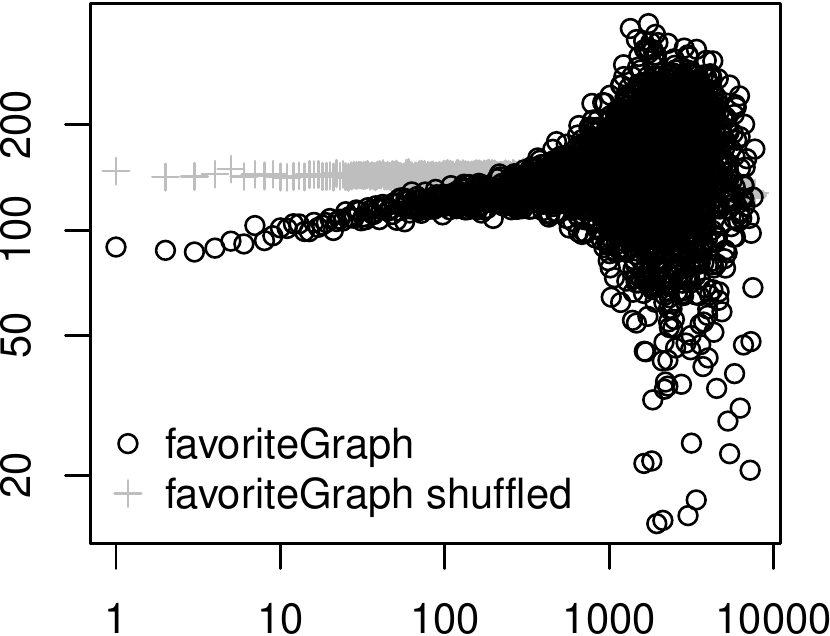}
  \caption{Nearest neighbors average degree \knn{} for evidence
    networks in \twitter and \flickr together with \knn{} averaged
    over five corresponding reshuffled networks.
  }
\label{fig:degree:correlation}
\end{figure*}

Figure \ref{fig:degree:correlation} clearly shows a common pattern for
both \twitter's as well as \flickr's explicit evidence networks. For
lower node degrees, both networks exhibit disassortative
characteristics whereas for higher degrees ($k\ge 100$) assortative
mixing can be observed. Limited network sizes explain the noisy tail
in all plots. In contrast, \twitter's ReTweet graph and \flickr's
Comment graph show consistent strong assortativity. The Favorite graph
shows a slight transition from disassortativity to assortativity but
far less pronounced than in the explicit evidence networks. Please
note that due to the sparsity of the corresponding plots induced by
the limited size of all \bibs networks they were omitted for this
analysis. In the shuffled networks, we shuffled the vertex to vertex assignments.
%
\subsection{Topological and Semantical Distance}
The analysis of the last section has focused on several inherent statistical 
network properties of the evidence network under consideration.
In this section, we will go one step further and take into account
information which is not present in the networks themselves --- namely
background information about the \emph{semantic profile} of each node.
Despite the differences to a typical social network reported above, it
is a natural hypothesis to assume that, \eg two users which are close
in the click network can be expected to share some common interest,
which is reflected in a higher ``semantic similarity'' between these
user nodes. In this way we establish a connection between structural
properties of the respective networks and a \emph{semantic} dimension of user
relatedness.
For measuring the ``true'' semantic similarity between two users we build on our prior work on
semantic analysis of folksonomies~\cite{markines2009evaluating}, where
we discovered that the similarity between tagclouds is a valid proxy
for semantic relatedness.
We compute this similarity in the vector space $\R^T$, in which each user is represented by the vector $v_u := \bigl(w(u, t)\bigr)_{t \in T}\,$, where $w(u,t)$ is the number of times
user $u$ has assigned tag $t$ to one of her resources (in case of
\bibs and \flickr) or the number of times user $u$ has used hash tag
$t$ in one of her tweets (in case of \twitter).

Each vector can be
interpreted as a ``semantic profile'' of the respective user,
represented by the distribution of her tag usage.  
In this vector space, we compute the cosine similarity between two vectors
$\vec{v}_{u1}$ and $\vec{v}_{u2}$ according to
$$\cos\measuredangle(\vec{v}_{u1},\vec{v}_{u2})=
\frac{\vec{v}_{u1}\cdot\vec{v}_{u2}}{||\vec{v}_{u1}||_2\cdot||\vec{v}_{u2}||_2}.$$
This measure is thus independent of the length of the vectors. Its
value ranges from $-1$ (for vectors pointing into opposite directions) over 0 (for orthogonal vectors) to $1$ (for vectors pointing into the same direction). In our case, the similarity
values lie between 0 and 1 because the vectors only contain positive
numbers (refer to~\cite{markines2009evaluating} for details).

\begin{figure*}
  \centering
  \includegraphics[width=0.32\linewidth]{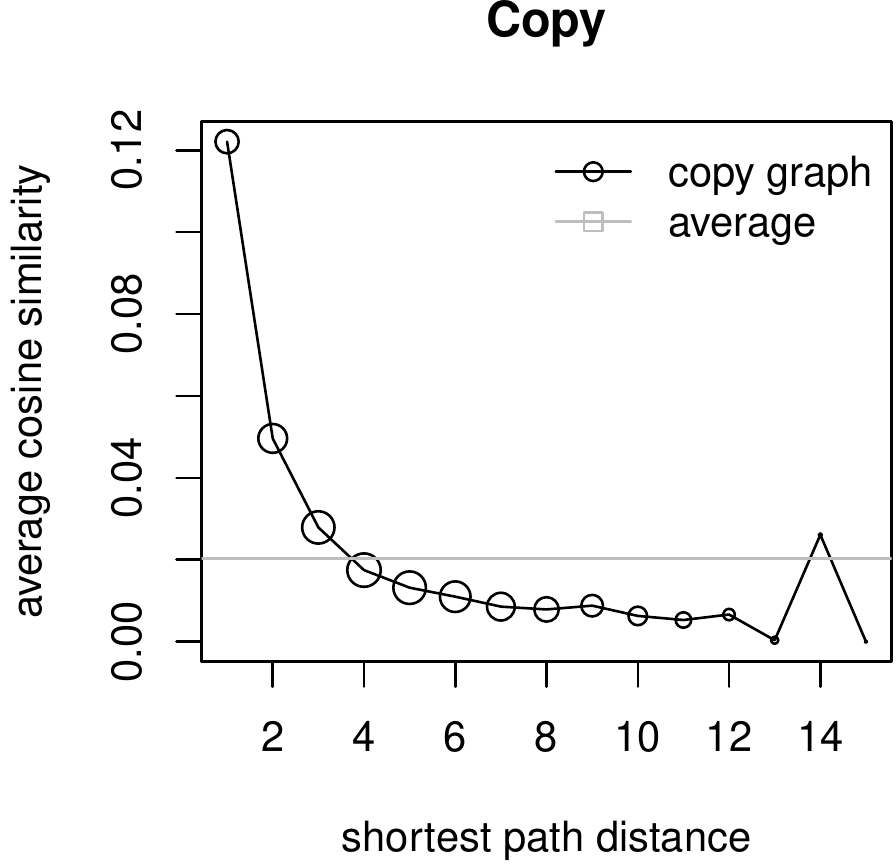}\ \ \ 
  \includegraphics[width=0.28\linewidth]{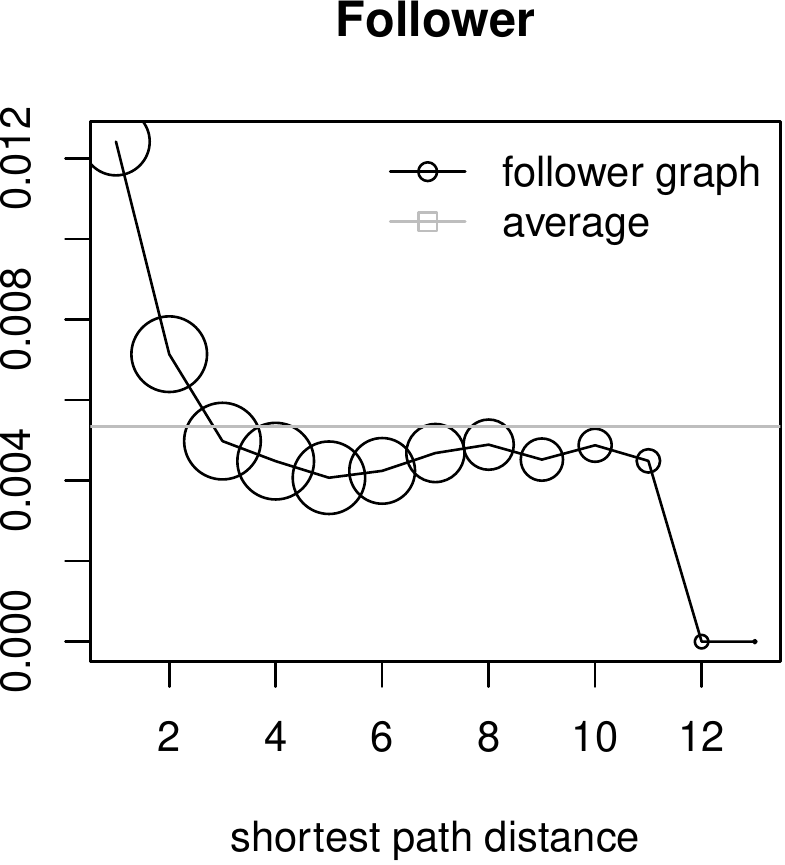}\ \ \ 
  \includegraphics[width=0.28\linewidth]{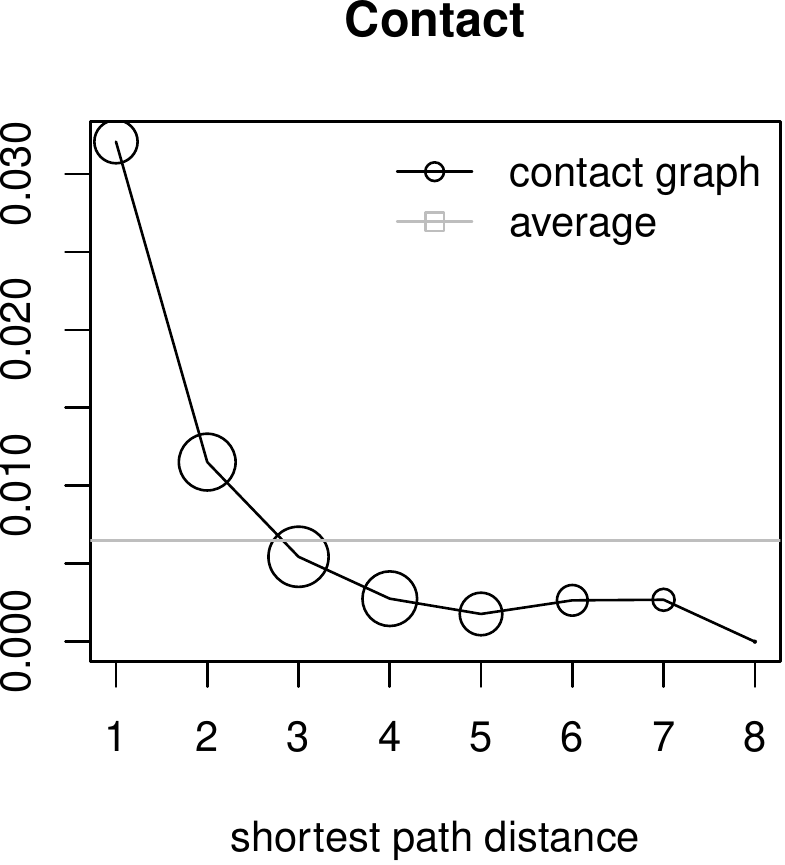}
\comments{
  \includegraphics[width=0.3\linewidth]{figs/similarities/copyGraph}
  \includegraphics[width=0.3\linewidth]{figs/similarities/clickGraph}
  \includegraphics[width=0.3\linewidth]{figs/similarities/visitGraph}
  \includegraphics[width=0.3\linewidth]{figs/similarities/groupGraph}
  \includegraphics[width=0.3\linewidth]{figs/similarities/friendGraph}
  \includegraphics[width=0.3\linewidth]{figs/similarities/tweets2009-RT}
  \includegraphics[width=0.3\linewidth]{figs/similarities/socialgraph-tweets2009}
  \includegraphics[width=0.3\linewidth]{figs/similarities/comments}
  \includegraphics[width=0.3\linewidth]{figs/similarities/favorite}
  \includegraphics[width=0.3\linewidth]{figs/similarities/contact}
}
  \caption{Average cosine similarity (y-axis) between tagclouds of users as a function of
    their distance in each of our networks, exemplarily for the \bibs Copy, the \twitter Follower and the \flickr Contact graphs. The other networks showed similar results.
    \comments{SCRIPT: R/plot_similarity.R}    
  }
  \label{fig:shortpath_vs_sim_tagclouds_all}
\end{figure*}

As in~\cite{schifanella2010folks}, we plot
the average semantic similarity between all pairs $(u, v)$ of users
(as obtained by the cosine similarity in the tag vector space) against
the shortest path between $u$ and $v$ in each of our networks -- as shown in Figure~\ref{fig:shortpath_vs_sim_tagclouds_all}.
The size of each data point scales logarithmically with the corresponding number
of shortest paths existing in the respective network. 
We additionally ruled out effects
induced by shortest path length distribution of a network: We shuffled the feature to vertex assignment, effectively calculating the average pairwise similarity of vertex sets according to the
network's path length distribution.
As expected, the shuffling process had eliminated the correlation observed before almost completely, affirming that the observed correlation of topological and semantical relatedness is not a statistical effect. 

The first obvious observation is that the highest average
\emph{semantic} similarity is found for the smallest
\emph{topological} distance of one. With growing lengths of the
shortest paths (x-axis), the average semantic similarity decreases
quickly. In the Follower graph and all of \flickr's networks, the
similarity even drops below the global average similarity at a
(topological) distance of three to four. 
Some networks show a peak for distant pairs (like the copy and contact graphs in Figure~\ref{fig:shortpath_vs_sim_tagclouds_all} but also the visit graph,
friend graph, and comment graph). Previous work explains these peaks
with the sparsity of the considered data~\cite{HT2010}, specifically for the \bibs networks.
While the larger networks do not exhibit those characteristics, e.g., the Comment graph has several thousand pairs of nodes at the corresponding distances, those peaks show only a marginal increase in the similarity and tend toward the average similarity in the whole network, as shown in the null-model plots.
Furthermore, an observed bow shape in the pathlength distribution (\cf Figure \ref{fig:pathlength:distribution}) correlates
with the corresponding similarity plot.

In summary, these results indicate a correlation of topological
proximity and semantic similarity between the nodes in each of our
observed networks.  This confirms in a more formal way that shared
interests between users are reflected in a higher degree of
interaction between them --- in an implicit or explicit manner.

\subsection{Common Neighborhood}\label{sec:analysis:neighborhood}
\begin{figure*}
  \centering
 \includegraphics[width=0.35\linewidth]{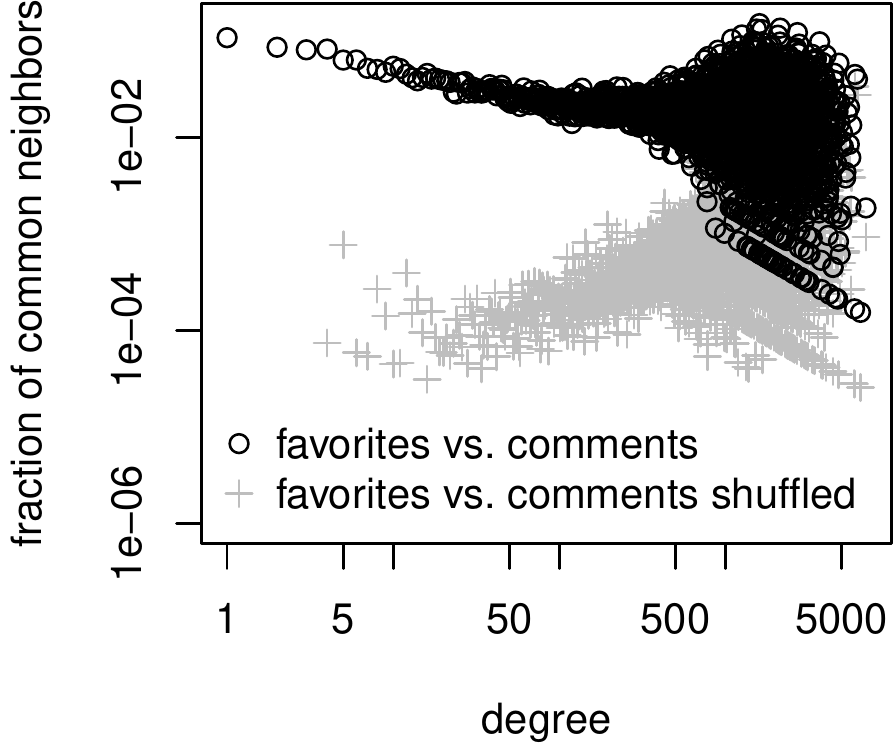}
 \includegraphics[width=0.35\linewidth]{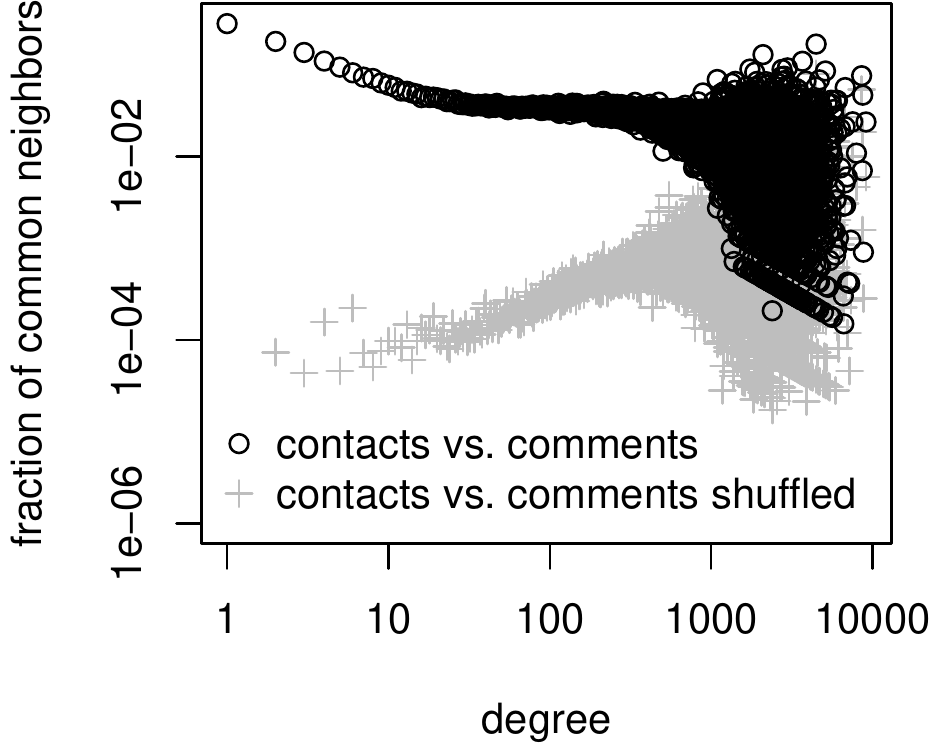}
 \includegraphics[width=0.35\linewidth]{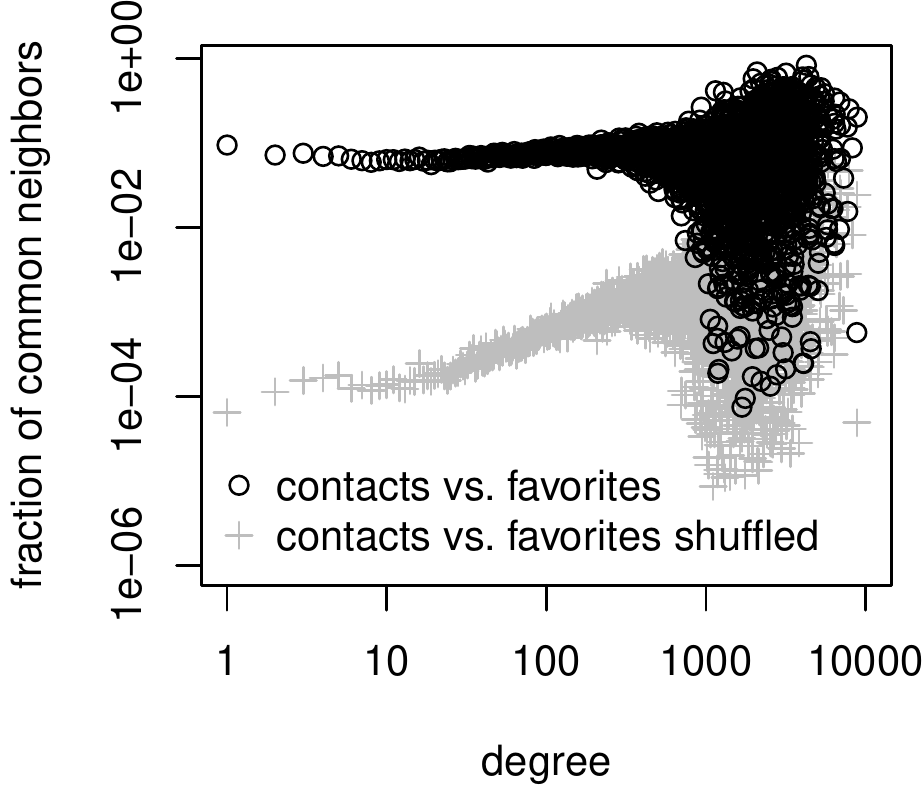}
 \includegraphics[width=0.35\linewidth]{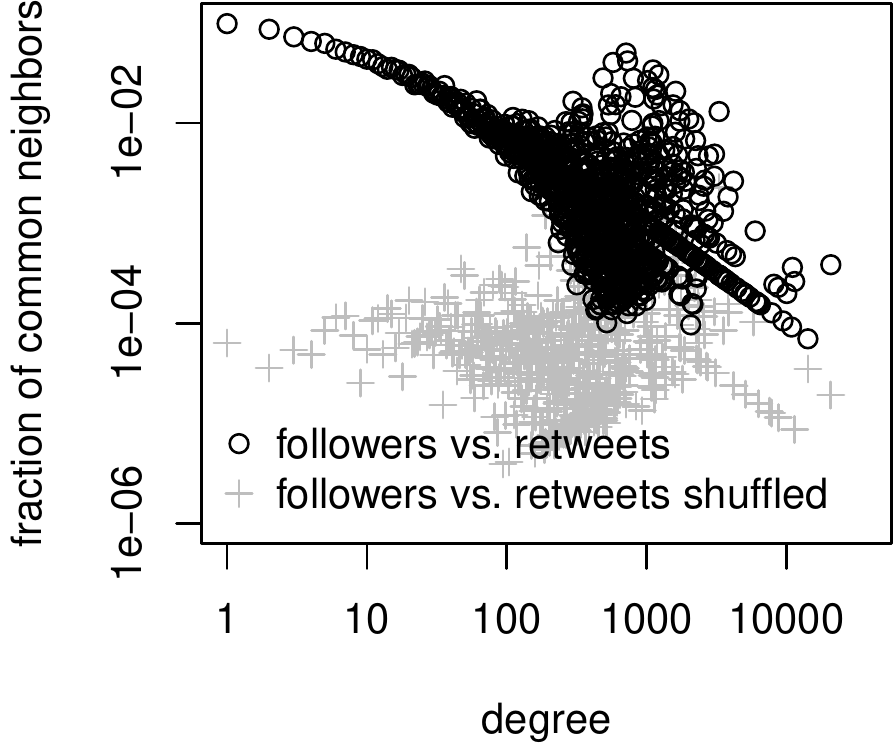}
 \caption{Precision profile for common neighborhood of different
   evidence networks within \twitter and \flickr.}
 \comments{SCRIPT: C/eopp (edgeOverlapPrecision.cpp),  R/plot_edgeOverlaps.R}
\label{fig:neighborhood:precision}
\end{figure*}

So far, we calculated several (descriptive) graph level statistics, and showed deviations
among the global network structures. But also common
patterns across the networks were revealed by analyzing the nearest
neighbors average degree in the previous section. It is
straightforward to consider the properties of similar nodes of different
networks within a system. That is, given two evidence networks
$G_1=(V_1, E_1)$ and $G_2=(V_2, E_2)$ with $V\coloneqq V_1\cap
V_2\ne\emptyset$, we compare, for each node $u\in V$ the local
neighborhoods $N_1(u)\coloneqq\set{(u,v)\mid (u,v)\in E_1}$ in $G_1$
and $N_2(u)\coloneqq\set{(u,v)\mid (u,v)\in E_2}$ in $G_2$. We
considered several similarity scores
\begin{align*}
       J(N_1(u), N_2(u)) \coloneqq{}& \frac{N_1(u)\cap N_2(u)}{N_1(u)\cup N_2(u)} && \text{ \emph{Jaccard}}\\
       P(N_1(u), N_2(u)) \coloneqq{}& \frac{N_1(u)\cap N_2(u)}{N_1(u)} && \text{ \emph{Precision}}\\
  \cos(A_1[u,\sbt\,], A_2[u,\sbt\,]) \coloneqq{}& \frac{A_1[u,\sbt\,]\cdot A_2[u,\sbt\,]}{\norm{A_1}\norm{A_2}} && \text{ \emph{Cosine}}
\end{align*}
where $A_i$ denotes the (weighted) adjacency matrix of ${G_i}_{|V}$, that is, $G_i$
reduced to the common vertex set $V$.
For selected pairs of networks, Figure \ref{fig:neighborhood:precision} shows the average precision score per node degree; the Jaccard index and cosine similarity yielded similar results. For contrasting the obtained result to patterns emerging from random graphs sharing the same degree
distribution, each plot also shows the resulting precision score for
the second network's rewired null model,
averaged over five independently generated null models.

For all considered pairs of networks, the precision profiles shows a clear deviation from the respective null
models -- in shape as well as in magnitude. Whereas the latter show an ascending tendency (less pronounced
for the Follower-/ReTweet graph), the former show a descending
(Favorites vs. Comments and Follower vs. ReTweet) or a descending to
constant (Contacts versus Comments, Contacts vs. Favorites) tendency.
The observed descending tendency for the Follower- and ReTweet graph means that
increasing the number of social contacts does not increase the
number of interactions accordingly. In contrast, a constant pattern as
observed in the Contact- and Favorite graph shows, that people
interact with around ten
percent of their contacts, regardless of the number of their contacts.

\subsection{Inter-Network Correlation Test}\label{sec:analysis:qap}
In the previous section, we compared pairs of networks based on local
similarity per node degree. For a condensed comparison of network
pairs, these similarity scores could simply be averaged. The quadradic
assignment procedure (\qap) test is a standard approach for
inter-network comparison common in literature; it is based on the
correlation of the adjacency matrices of the considered
graphs~\cite{butts2008social,BC:05}. QAP tests a given graph level statistic, for example, the graph covariance against a \qap null hypothesis.

For given graphs $G_1=(V_1, E_1)$ and $G_2=(V_2,E_2)$ with $V\coloneqq
V_1\cap V_2\ne\emptyset$ and (weighted) adjacency matrices $A_i$ corresponding to
${G_i}_{|V}$ ($G_i$ reduced to the common vertex set $V$, \cf~Section~\ref{sec:analysis:neighborhood}, the graph \emph{covariance} is given by
\[
\cov(G_1,G_2) \coloneqq
\frac{1}{n^2-1}\sum_{i=1}^{n}\sum_{j=1}^{n}(A_1[i,j]-\mu_1)(A_2[i,j]-\mu_2)\,,
\]
where $\mu_i$ denotes $A_i$'s mean ($i=1,2$). Then,
$\var(G_i)\coloneqq\cov(G_i, G_i)$ leads to the graph correlation
\[
\rho(G_1,G_2)\coloneqq
\frac{\cov(G_1,G_2)}{\sqrt{\var(G_1)\var(G_2)}}\,.
\]

\begin{figure*}
  \centering
 \includegraphics[width=0.3\linewidth]{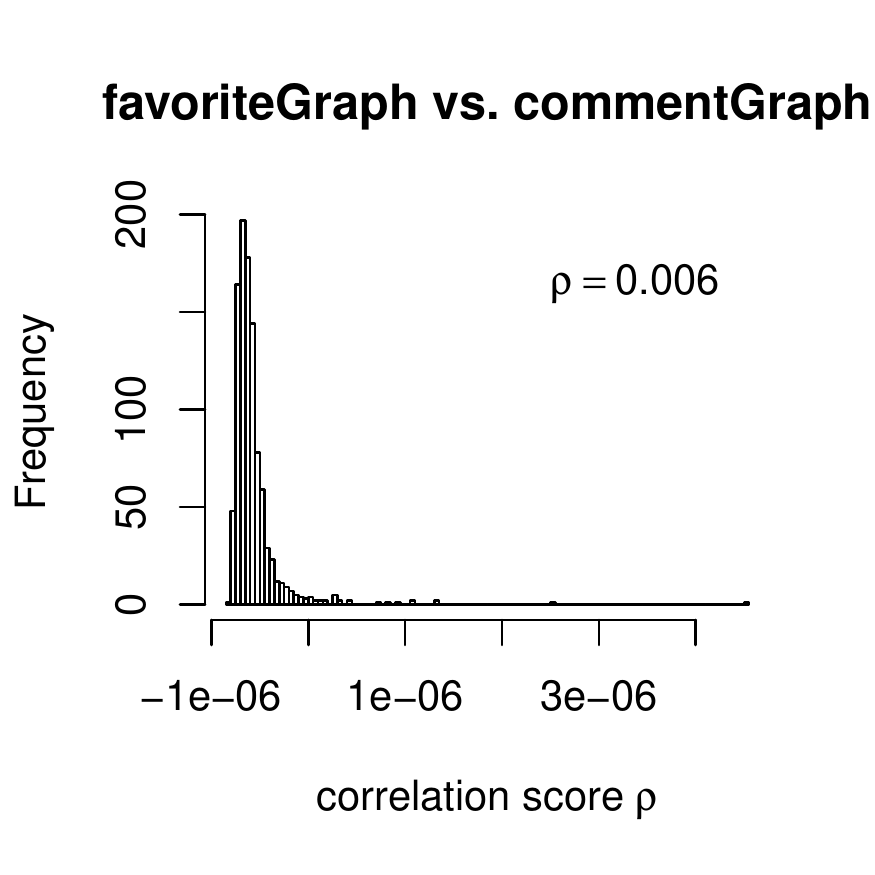}
 \includegraphics[width=0.3\linewidth]{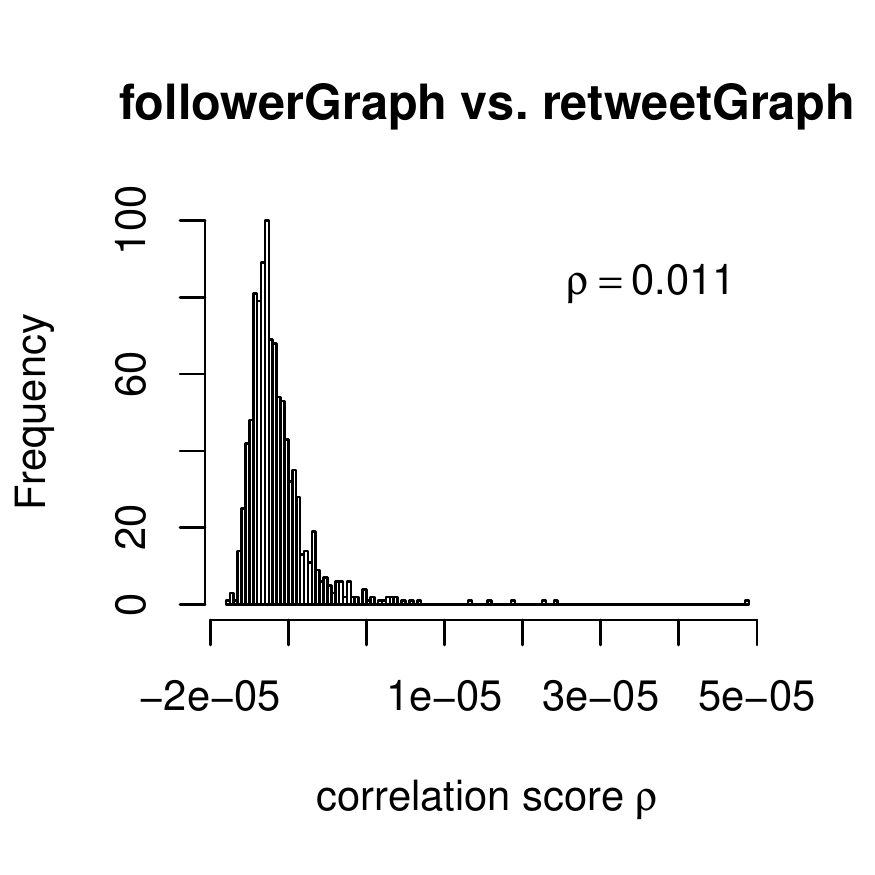}
 \includegraphics[width=0.3\linewidth]{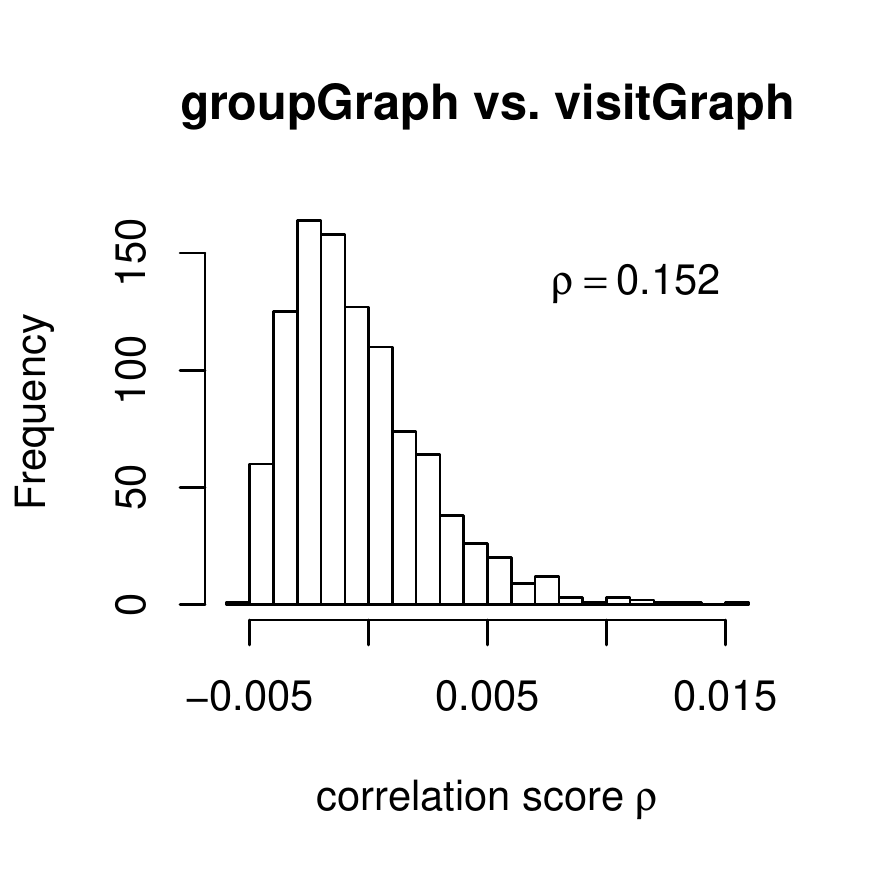}
\comments{
 \includegraphics[width=0.3\linewidth]{figs/qap/favorites-comments-qap}
 \includegraphics[width=0.3\linewidth]{figs/qap/contacts-comments-qap}
 \includegraphics[width=0.3\linewidth]{figs/qap/contacts-favorites-qap}
 \includegraphics[width=0.3\linewidth]{figs/qap/socialgraph_tweets2009-tweets2009_RT-qap}
 \includegraphics[width=0.3\linewidth]{figs/qap/copyGraph-clickGraph-qap}
 \includegraphics[width=0.3\linewidth]{figs/qap/copyGraph-friendGraph-qap}
 \includegraphics[width=0.3\linewidth]{figs/qap/copyGraph-groupGraph-qap}
 \includegraphics[width=0.3\linewidth]{figs/qap/copyGraph-visitGraph-qap}
 \includegraphics[width=0.3\linewidth]{figs/qap/clickGraph-friendGraph-qap}
 \includegraphics[width=0.3\linewidth]{figs/qap/clickGraph-groupGraph-qap}
 \includegraphics[width=0.3\linewidth]{figs/qap/clickGraph-visitGraph-qap}
 \includegraphics[width=0.3\linewidth]{figs/qap/friendGraph-groupGraph-qap}
 \includegraphics[width=0.3\linewidth]{figs/qap/friendGraph-visitGraph-qap}
 \includegraphics[width=0.3\linewidth]{figs/qap/groupGraph-visitGraph-qap}
}
\caption{Distribution of the graph correlation statistic $\rho$ over
  1000 permutations for selected pairs of networks.} 
\label{fig:qap:distribution}
\end{figure*}

Table~\ref{tab:qap} shows the pairwise correlation scores for all
considered networks, while Figure \ref{fig:qap:distribution} relates this by comparison with a set of null-model experiments. All networks within \bibs show a quite similar level
of correlation with a significant peak for the explicit \Friend- and
\Group networks. Considering the results for the networks obtained
from \flickr it is worth noting that, though low in magnitude, the
Favorite graph shows a significant higher correlation with the
Contact graph than the other pairs of networks do. A potential explanation considers the interaction between persons: On the one hand, people in the favorites list might be or become contacts -- on the other hand the photos of contacts might be very interesting for the favorites list. 
But comparing the level of correlations for the different
networks tells just one part of the story.
\begin{table}[ht]\centering\footnotesize

  \begin{tabular}{l|r|r|r|r}
                &  \Copy   &  \Click  & \Visit  & \Group \\\hline\hline
    \Copy       &      $-$ &          &         &          \\\hline
    \Click      &  $0.254$ &      $-$ &         &          \\\hline
    \Visit      &  $0.214$ &  $0.332$ &     $-$ &          \\\hline
    \Group      &  $0.170$ &  $0.148$ & $0.152$ &     $-$  \\\hline
    \Friend     &  $0.248$ &  $0.190$ & $0.193$ & $0.396$  \\\hline
  \end{tabular}\ \\[1em]

  \begin{tabular}{l|r|r|r}
                & \Comment  & \Favorite  & \Contact  \\\hline\hline
    \Comment    &     $-$   &            &           \\\hline
    \Favorite   & $0.006$   &     $-$    &           \\\hline
    \Contact    & $0.005$   & $0.074$    &     $-$   \\\hline
  \end{tabular}\ \\[1em]

  \begin{tabular}{l|r|r}
                & \RT      & \Follower \\\hline\hline
    \RT         &     $-$  &           \\\hline
    \Follower   & $0.011$  &     $-$   \\\hline
  \end{tabular}

  \caption{Pairwise graph correlation observed in \bibs, \flickr and \twitter.}
  \label{tab:qap}
\end{table}
%
The \qap test compares the observed graph correlation to the
distribution of resulting correlation scores obtained on repeated
random row/column permutations of $A_2$. The relative frequency of a permutation
$\pi$ with a correlation $\rho^\pi\ge\rho_o$ is used for assessing the
significance of an observed correlation score $\rho_o$. Intuitively,
the test determines (asymptotically) the fraction of all graphs with
the same structure as $G_{2|V}$ having at least the same level of
correlation with $G_{1|V}$.
Figure \ref{fig:qap:distribution} shows the distribution of correlation
scores obtained for the repeatedly permutated adjacency matrices, \cf~\cite{butts2008social}.
For all pairs of networks, the
correlation is significantly higher than one
would expect by considering random graphs of the same
structure (shown by the QAP test results with a zero score). These results clearly reject the null hypothesis that the degree of correlation can be
explained just by $G_{2|V}$'s graph structure~\cite{butts2008social}. This shows the structural similarity and correlation between the different networks.

\section{Community Rating using Evidence Networks}\label{sec:experiments}
Our comparative analysis in Section \ref{sec:analysis} suggests that
the considered evidence networks exhibit deviating structural
properties on a global network level but also show common local
interaction patterns. 
In the following, we explore how evidence networks can be used to
assess the quality of communities in social applications, where users
collaboratively publish content. A key difficulty in assessing the quality of a given group of
users is the lack of an established ground truth for measuring the
level of coherency. If clusters of tags, for example, within a resource sharing
system like \bibs are to be evaluated, external data sources such as
\wikipedia or \wordnet can be
consulted~\cite{newman2010automatic}. For users there is typically no such
semantic grounding. For the assessment of a community of users, gold-standard data can be applied, if available, or communities can be ranked by humans using introspection techniques for the respective user subgroups, e.g.,~\cite{AP:08}.
Therefore, the approach presented in the following section presents a cost-effective way for assessing the ranking of community allocations using only evidence networks (secondary data) for rating the communities. 
 
We begin with consolidating our notions and vocabulary with respect to
user communities in social media. We then present our approach
for rating the quality of a given community assignment and detail on
the experimental setup. Finally, we present the results obtained from
all considered evidence networks. The results indicate, that the inter-network correlations that we analyzed in the previous section are indeed strong enough to draw reciprocal conclusions between different evidence networks.

\subsection{Community Basics}
The concept of a \emph{community} can be
intuitively defined as a group $C$ of individuals out of a
population $U$ such that members of $C$ are
densely ``related'' to each other but sparsely ``related'' to
individuals in $U\setminus C$. We denote a
\emph{community allocation} of a population $U$ as a
set of communities
$\mathcal{C}=\set{C_1,\ldots,C_n}$ with
$C_i\ne\emptyset$
and $\bigcup_{1\le i\le n}C_i \subseteq U$
for $1\le i \le n$.

In a graph $G=(V, E)$ is in this sense a vertex set $C \subseteq V$,
where nodes in $C$ are densely connected but sparsely connected to
nodes in $V\setminus C$. Though defined in terms of graph theory, the
community concept remains vague, unless the notions of sparse and dense connectedness are specified further.
Several approaches for formalizing communities in graphs exist and
corresponding community structures were observed and analyzed in a
variety of different networks
\cite{newman2003structure,Newman04communityStructure,Leskovec2008,Leskovec2010}.
For a given graph $G=(V,E)$ and a
community $C\subseteq V$, we set $n\coloneqq\size{V}$,
$m\coloneqq\size{E}$, $n_C\coloneqq\size{C}$,
$m_C\coloneqq\size{\set{(u,v)\mid u,v\in C}}$,
$\cutset\coloneqq\size{\set{(u,v)\mid u\in C, v\not\in C}}$. For a
node $u\in V$ its degree is denoted by $d(u)$. Finally, we write
$\subgraph{G}{V'}$ for the induced subgraph of a subset $V'\subseteq
V$.

\subsection{Community Quality Functions}\label{sec:background:qualityfunctions}
Different quality functions $f\colon\mathcal{P}(V)\rightarrow\R$ for
modeling the intuitive concept of a community exist (also called cluster
indices).  \comments{, \eg \todo{just select modularity, conductance
    and segregation index}
\begin{compactitem}
\item \textbf{Conductance:} $f(C) = \frac{\overline{m}_C}{2m_C+\overline{m}_C}$
\item \textbf{Expansion:} $f(C) = \frac{\overline{m}_C}{n_C}$
\item \textbf{Average-ODF:} 
$\frac{1}{n_C}\sum_{u\in C}\frac{\size{\set{(u,v)\in E\colon v\not\in
C}}}{d(u)}$
\item \textbf{Volume:} $\sum_{u\in C}d(u)$
\item \textbf{Edges cut:} $\overline{m}_C$
\end{compactitem}
}We refer to corresponding related work for more details, \eg to
\cite{Leskovec2008} and \cite{DBLP:conf/dagstuhl/Gaertler04}.
This work focuses on
the \emph{modularity}~\cite{Newman04communityStructure}, the \emph{segregation index}~\cite{freeman1978segregation}, and the \emph{conductance}~\cite{kannan2004clusterings}.

\emph{Modularity} is based on comparing the number of edges within a
community with the expected number given a null-model (\ie a
randomized model). Thus, the modularity of a disjoint community clustering is
defined to be the fraction of the edges that fall within the given
clusters minus the expected fraction if edges were distributed at
random. This can be formalized as follows: The modularity
$\modularity$ of a set of nodes and its assigned adjacency matrix
$A\in\N^{n\times n}$ is given by
\[
\modularity(A) = 
   \frac{1}{2m}\sum_{i,j}\bigl(A_{i,j} 
 - \frac{d(i) d(j)}{2m}\bigr)\delta(C_i,C_j)\,,
\] 
where $C_i$ is the cluster to which node $i$ belongs and $C_j$ is the
cluster to which node $j$ belongs; $d(i)$ and $d(j)$ denote $i$ and
$j$'s degrees respectively; $\delta(C_i, C_j)$ is the \emph{Kronecker  
delta} symbol that equals $1$ iff $C_i = C_j$, and $0$ otherwise. For
\emph{directed networks}~\cite{directedModularity} with in- and out-
degree $d(i)^{\text{in}}$ and $d(j)^{\text{out}}$ for $i$ and $j$
respectively the modularity becomes
\[
\modularity(A) =
\frac{1}{m}\sum_{i,j}\bigl(A_{i,j} - \frac{d(i)^{\text{in}}
    d(j)^{\text{out}}}{m}\bigr)\delta(C_i,C_j)\,.
\]
The modularity $\modularity(A)$ obtains its maximum value $\modularity(A) = 1$ for a perfect partitioning, and its minimum value $\modularity(A) = -1$ for the opposite.

While modularity considers the intra-cluster links compared to the
corresponding null-model, the \emph{segregation
  index}~\cite{freeman1978segregation} compares, for a given cluster
$C$, the number of {expected} links accross the cluster boundary to the number of
{observed} inter-cluster links, normalized by the expectation. By
averaging the segregation over all clusters one obtains the
segregation of a community allocation.
\[
\segIndex(C) = 
\begin{cases}
  0                              &{} \text{if } E(\mqc)\le\mqc \\
  \frac{E(\mqc) - \mqc}{E(\mqc)} &{} \text{otherwise}
\end{cases}
\]
The segregation index lies within the interval $[0,1]$ where higher
values indicate more pronounced community structure, following the
intuition that good communities are sparsely connected among each
other.

Finally, the \emph{conductance} captures the intuition of a
``bottleneck'': If a cluster $C$ contains a non-trivial small cut (\ie
$C=C_1\dotcup C_2$ with a small value of \cutset), the cluster is
probably too coarse. Otherwise, if a cluster is strongly
connected to the remainder of the graph (relative to the clusters
internal density), the cluster is probably too fine. The former case
is assessed by a small \emph{intra-cluster conductance} $\alpha(C)$
whereas the latter case corresponds to a small \emph{inter-cluster
  conductance} $\beta(C)$. To avoid trivial cuts, both measures
consider the cut size \cutset relative to the density of the smaller
subset. We define the \emph{conductance} $\phi(C)$ of a
community $C$ in $G=(V,E)$ and the \emph{conductance}
$\phi(\subgraph{G}{V'})$ of a subgraph \subgraph{G}{V'} as follows:
\begin{align*}
  \phi(C)\coloneqq&{}
  \begin{cases}
    1, & C\in\set{\emptyset, V}\\
    0, & C\not\in\set{\emptyset, V}\\
       &  \text{ and } \cutset=0 \\
    \frac{\cutset}{\min\left(\sum_{u\in C}\degree{u}, \sum_{v\in
          V\setminus C}\degree{v}\right)} & \text{otherwise}
  \end{cases}\\
  \phi(\subgraph{G}{V'})\coloneqq&{} \min_{C\subseteq V'}\phi(C)
\end{align*}
For a given clustering $C\coloneqq\set{C_1,\ldots,C_n}$, the
\emph{intra-cluster conductance} $\alpha(C)$ and the
\emph{inter-cluster conductance}
$\beta(C)$~\cite{brandes2003experiments} are defined accordingly.
\[
\alpha(C)\coloneqq\min_{i\in 1,\ldots, n}\phi(\subgraph{G}{C_i})
\quad\text{and}\quad
\beta(C)\coloneqq 1-\max_{i\in 1,\ldots, n}\phi(C_i)
\]
Intra-cluster conductance was pioneered in
\cite{kannan2004clusterings} and found many applications in graph
clustering~\cite{brandes2003experiments,spielman2004nearlylinear,andersen2006local} and community mining
literature~\cite{leskovec2008statistical,Leskovec2010}. Calculating
$\alpha(C)$ inherently comprises the NP-hard problem of minimizing the
conductance for all cuts of a graph and thus in practice approximate
algorithms must be applied. We applied the
\metis\footnote{\url{http://glaros.dtc.umn.edu/gkhome/views/metis}}\cite{karypis1999fast}
graph clustering algorithm with Max-flow Quotient-cut Improvement
(\mqi)~\cite{lang2004flowbased}.

For comparing different community allocations, we consider
the applied quality function as a ranking on the community
allocations and apply the Kendall rank correlation
coefficient~\cite{kendall1938measure}, which is commonly used in
information retrieval for comparing different rankings. In the
following we refer to it with ``Kendall's $\tau$'' for short.
The range is in $[-1,1]$: While $1$ indicates perfect (positive) correlation, a value of $-1$ indicates perfect negative correlation. A value of $0$ is expected for independent rankings.
We discuss its application in more detail below.

\subsection{Evaluation Paradigm}\label{sec:experiments:paradigm}
For the evaluation of user recommendations, Siersdorfer proposed to
consult existing social structures~\cite{siersdorfer2009social}. We
apply this paradigm to the evaluation of user communities and assess
the quality of a given community allocation relative to the community
structure within evidence networks. Specifically, we assume a set of
users $U$ within a social application and an evidence network
$G=(V,E)$ with $V\subseteq U$.
\begin{compactitem}
\item We first mine for communities in an appropriate feature space,
  resulting in a community allocation \[\mathcal{C}=C_1\dotcup C_2\dotcup \ldots \dotcup C_k\]
  with $\bigcup_{C_i}^{i = 1 \ldots k}\subseteq U$.
\item We then calculate some community quality function, such as
  modularity, within $G$ relative to $C_1\cap V,\ldots, C_k\cap
  V$, for an evidence network $G= (V, E)$.
\end{compactitem}
Since we do not utilize a gold-standard network, the resulting score is not considered as an absolute value measuring the quality of the respective community directly. However, it induces a ranking on
different community allocations on $U$ which can be used to select the
top ranked allocations obtained from different clustering
algorithms. In this way, it indirectly represents the quality of the respective community using the respective rankings.

Of course the question arises, whether different evidence networks
and quality functions induce consistent rankings on the set of all
possible community allocations. Ideally we would list all possible
community allocations and calculate each quality function in every
evidence network. This would result in corresponding rankings of the
community allocations which then could be compared. In the best case,
all rankings would coincide, whereas in the worst case all rankings
would be statistically independent.
Unfortunately it is not feasible to list all possible community
allocations due to its combinatorial explosion. On the other hand,
Monte Carlo methods would tend to compare rankings on ``bad''
community allocations, since the fraction of sensible allocations is very
small and it is thus more likely to randomly choose an allocation
which just does not make sense. We therefore chose a different
practical approach. Using a broad class of different clustering
methods with comprehensive parameter sets we produce as many community
allocations as possible: We try to get a representative sample of all
``reasonable'' community allocations which is ranked accordingly for
comparing the quality measures induced by the different evidence
networks.

\subsection{Experimentation}\label{sec:experiments:description}
For mining communities, we represent every user $u$ by a vector
$\vec{u}$ in a corresponding vector space. For the considered
applications, the $i$'th component of user $\vec{u}$ respectively
denotes, how often user $u$
\begin{compactitem}
\item applied tag $i$ to resources (\bibs),
\item used hash tag $i$ within tweets (\twitter) and
\item applied hash tag $i$ to photographs (\flickr).
\end{compactitem}

We used the freely available clustering tool
Cluto.\footnote{\url{http://glaros.dtc.umn.edu/gkhome/views/cluto/}}
Cluto incorporates three different classes of clustering algorithms
(agglomerative, partitional and graph based) which can be broadly
parameterized with respect to similarity functions, normalization,
number of clusters etc.  We used a brute-force approach for the
parametrization, resulting in 16,128 parameter combinations, out of
which only a subset succeeded, due to resource limitations and due to
incompatible parameter combinations. In total, we were able to collect
3,665 community allocations for \bibs, 993 for \twitter and 2,985 for
the \flickr dataset.  We calculated for each community allocation and
each evidence network all of the community quality functions described
in Section \ref{sec:background:qualityfunctions}, namely
intra-conductance, inter-conductance, modularity and segregation
index. For comparison, we further constructed to each evidence network
a corresponding null model by randomly rewiring pairs of
edges~\cite{maslov2002specificity}. In these null model networks,
clustering and community structure is destroyed while the 
 degree distribution of the networks is fixed. Rankings induced by these networks are
thus independent of the respective community structure. We averaged
corresponding quality functions over five independently constructed
null model graphs.

For summarizing the consistency between a pair of induced rankings, we
calculate Kendall's $\tau$ as described above. The drawback of such a global
measure is that all positions are considered equally important. Yet,
in our case we are mainly interested in the top ranked positions, since they are expected to be the ``most reasonable'' cluster assignments. For
assessing the consistency between two rankings in more detail, we
calculate the \emph{size of the overlap} of the two rankings,
considering only the \emph{top $k$ positions}~\cite{FKS:03}; for example, we consider how many
community allocations are placed in both rankings among the top $10$
positions. The absolute figures alone are hard to interpret, as we
expect a certain overlap merely by statistical effects (consider, \eg
the case were $n=k$).

We therefore compare our results with the expected overlap
assuming two randomly ordered rankings. Let $\pi_1$ and $\pi_2$ be two random permutations on \{1, \ldots, n\},
and let $m$ denote the number of common community allocations among the top
$k$ positions, \ie
$m\coloneqq\size{\set{C_{\pi_1(1)},\ldots,C_{\pi_1(k)}}\cap\set{C_{\pi_2(1)},\ldots,C_{\pi_2(k)}}}$. This
corresponds to a random sampling process without replacement. There are $\binom{k}{m}$ possible
combinations of $m$ common elements among the top $k$ positions. The
remaining $k-m$ top elements can be chosen out of $n-k$ community allocations
which are not contained in \set{C_{\pi_1(1)},\ldots,C_{\pi_1(k)}},
that is $\binom{k}{m}\cdot\binom{n-k}{k-m}$ options exist. Normalizing
by the number of possible rankings of length $k$ we obtain
\[
P(m)=\frac{\binom{k}{m}\cdot\binom{n-k}{k-m}}{\binom{n}{k}}
\]
which is a special case ($n_1 = n_2 = k$) of the hypergeometric
distribution $h(m; n, n_1, n_2)$ with mean $n_1\frac{n_2}{n}$. The
expected size of the overlap is therefore given by $\frac{k^2}{n}$.

For all considered systems and evidence networks, nearly all community
allocations obtained a inter-conductance score of (almost) 0. This is due to
the sparsity of evidence networks: As communities are mined
independently of the evidence networks, most clusters contain more
than one component when mapped into the evidence network which leads
to a corresponding minimal conductance value of 0. We therefore
exclude inter-conductance from presentation below.

\subsubsection{\bibs}\label{sec:experiments:results}
Using the \bibs tagging data we clustered the set of users in the feature space where each
user is represented by a vector corresponding to the user's tag cloud. This
resulted in $3,665$ community allocations. We calculated for each
community allocation intra-conductance, modularity and segregation
index in every evidence network
(Friend graph, Group graph, Copy graph, Click graph, Visit graph). We
thus obtained 15 different rankings on all community allocations.

\begin{figure*}
  \centering
 \includegraphics[width=0.4\linewidth]{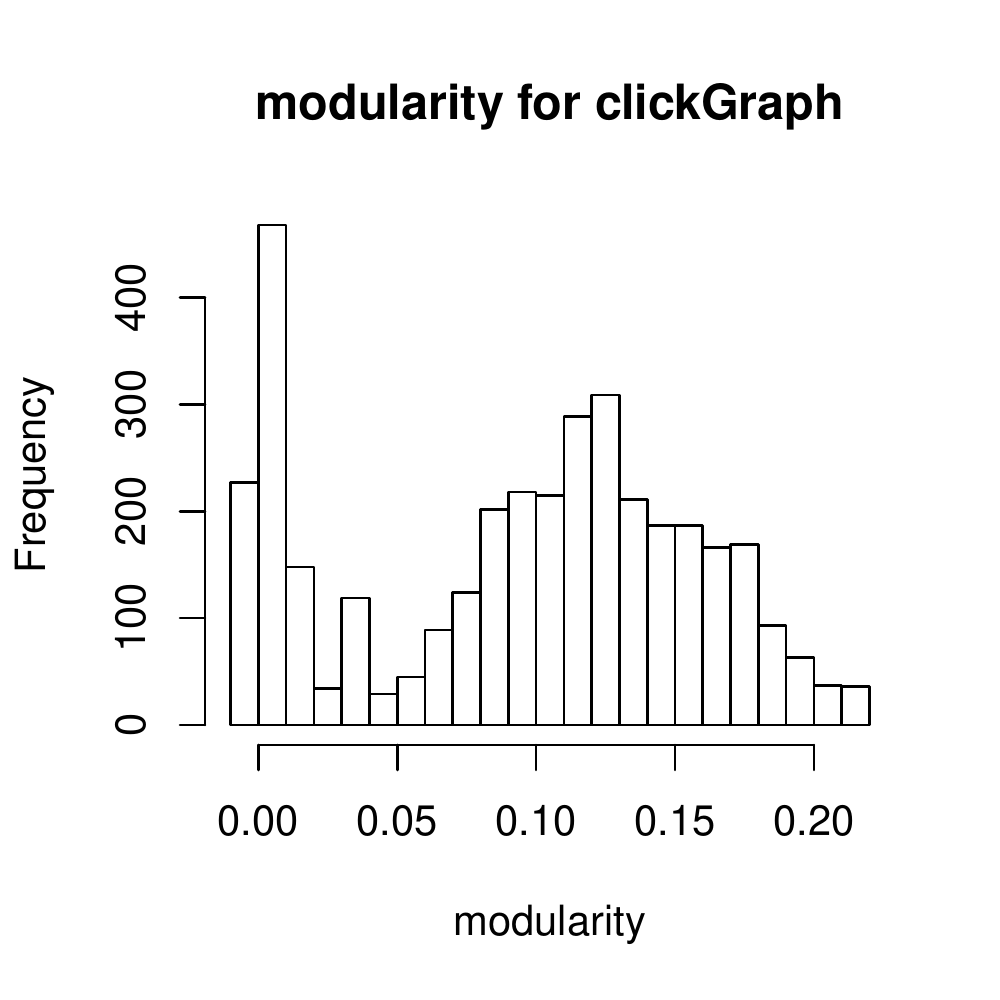}
 \includegraphics[width=0.4\linewidth]{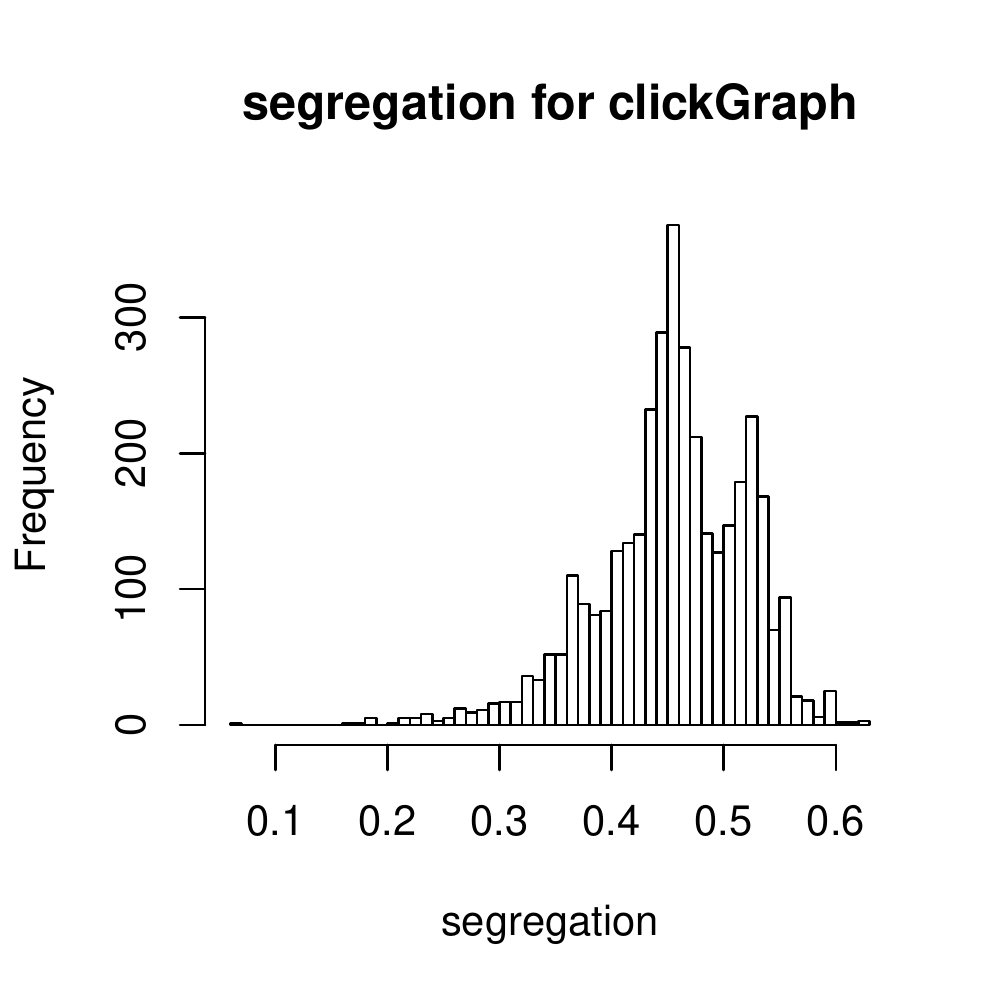}
  \caption{Distribution of the quality functions exemplarily shown for the
    \bibs Click graph.
    \comments{SCRIPT: R/calculateRankCorrelation2.R}
  }
\label{fig:community:bibsonomy:distribution}
\end{figure*}

For intra-conductance, all networks but the
Group graph induced trivial rankings as nearly all community
assignments obtained a zero score. Therefore, we exclude the
intra-con\-ductance rankings in \bibs from further
considerations.
Figure~\ref{fig:community:bibsonomy:distribution} shows the
distribution of obtained quality function scores, exemplarily using the Click graph.
Modularity and segregation index both induce a large number of high quality communities in all networks. In the distributions in the Click graph example, the overall majority focuses on high quality communities indicating the existing community structure.
%

Table~\ref{tab:bibsonomy:modularity:correlation:full:tau} shows
Kendall's $\tau$ calculated on all pairs of all rankings induced by the
different quality functions. The lower left triangle shows the correlation coefficient (Kendall's Tau)
for pairs were one network's community structure was
destroyed by randomly rewiring links (averaged over five
repetitions). Considering modularity, the resulting correlation
coefficients indicate significantly correlated rankings among all
pairs of networks which is absent when one network is shuffled. This
indicates that different networks assess community structure
consistently (to a certain extent) in terms of modularity and that the
ranking is in fact dependent on the community structure of the network.
Correlation coefficients for rankings induced by segregation on the
other hand do not indicate similar correlations and the obtained
correlation coefficients for rewired networks do not support the
required dependence on the network's community structure. 

\begin{table}\footnotesize
  Modularity:
  \begin{center}
  \begin{tabular}{l|c|c|c|c|c}
          &\Click &\Copy&\Friend&\Group&\Visit\\ \hline

  \Click  &\cellcolor{grey}-0.430
          &~0.806
          &~0.820
          &~0.798
          &~0.822
          \\\hline
  \Copy   &\cellcolor{grey}-0.453
          &\cellcolor{grey}-0.194
          &~0.769
          &~0.747
          &~0.775
          \\\hline
  \Friend &\cellcolor{grey}-0.429
          &\cellcolor{grey}-0.199
          &\cellcolor{grey}~0.250
          &~0.784
          &~0.767
          \\\hline
  \Group  &\cellcolor{grey}-0.423
          &\cellcolor{grey}-0.206
          &\cellcolor{grey}~0.242
          &\cellcolor{grey}~0.450
          &~0.741
          \\\hline
  \Visit  &\cellcolor{grey}-0.370
          &\cellcolor{grey}-0.175
          &\cellcolor{grey}~0.222
          &\cellcolor{grey}~0.413
          &\cellcolor{grey}-0.407
          \\
  \end{tabular}
\end{center}

  Segregation:
  \begin{center}
  \begin{tabular}{l|c|c|c|c|c}
          &\Click &\Copy&\Friend&\Group&\Visit\\ \hline

  \Click  &\cellcolor{grey}-0.013
          &~0.011
          &~0.036
          &~0.069
          &~0.016
          \\\hline
  \Copy   &\cellcolor{grey}~0.020
          &\cellcolor{grey}-0.013
          &-0.080
          &~0.048
          &-0.036
          \\\hline
  \Friend &\cellcolor{grey}-0.019
          &\cellcolor{grey}-0.014
          &\cellcolor{grey}-0.075
          &~0.030
          &~0.067
          \\\hline
  \Group  &\cellcolor{grey}-0.074
          &\cellcolor{grey}-0.022
          &\cellcolor{grey}~0.026
          &\cellcolor{grey}~0.028
          &-0.011
          \\\hline
  \Visit  &\cellcolor{grey}-0.020
          &\cellcolor{grey}~0.027
          &\cellcolor{grey}-0.095
          &\cellcolor{grey}~0.010
          &\cellcolor{grey}~0.021
          \\
  \end{tabular}
\end{center}

  \caption{
    Kendall's $\tau$ correlation coefficient for modularity and segregation
    based rankings relative to evidence networks in \bibs. For $i \leq j$
    the lower triangle shaded in gray shows the correlation for rankings
    induced by network $j$ and network $i$'s null models.
  } 
  \label{tab:bibsonomy:modularity:correlation:full:tau}
\end{table}

\begin{figure*}
\includegraphics[scale=0.35]{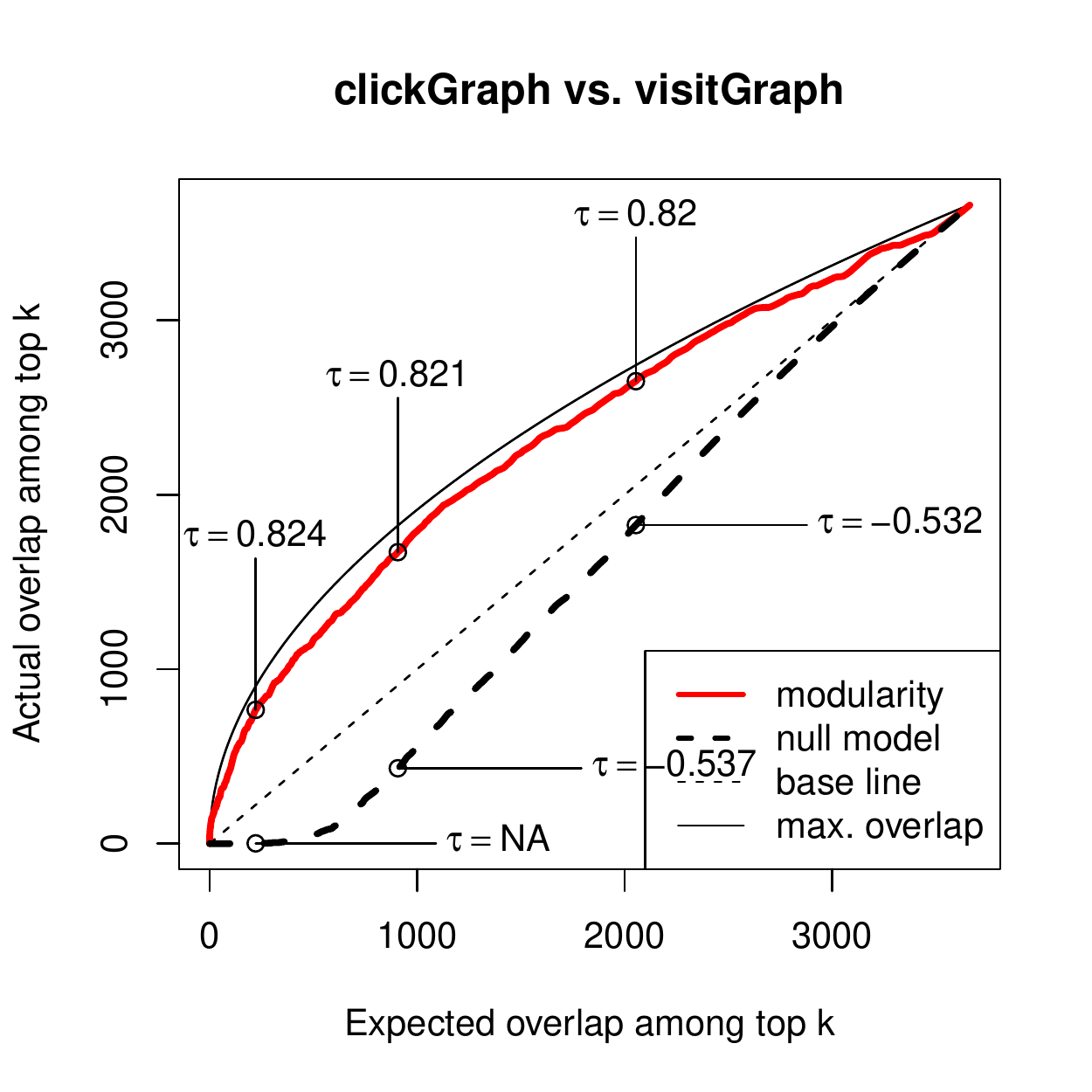}
\includegraphics[scale=0.35]{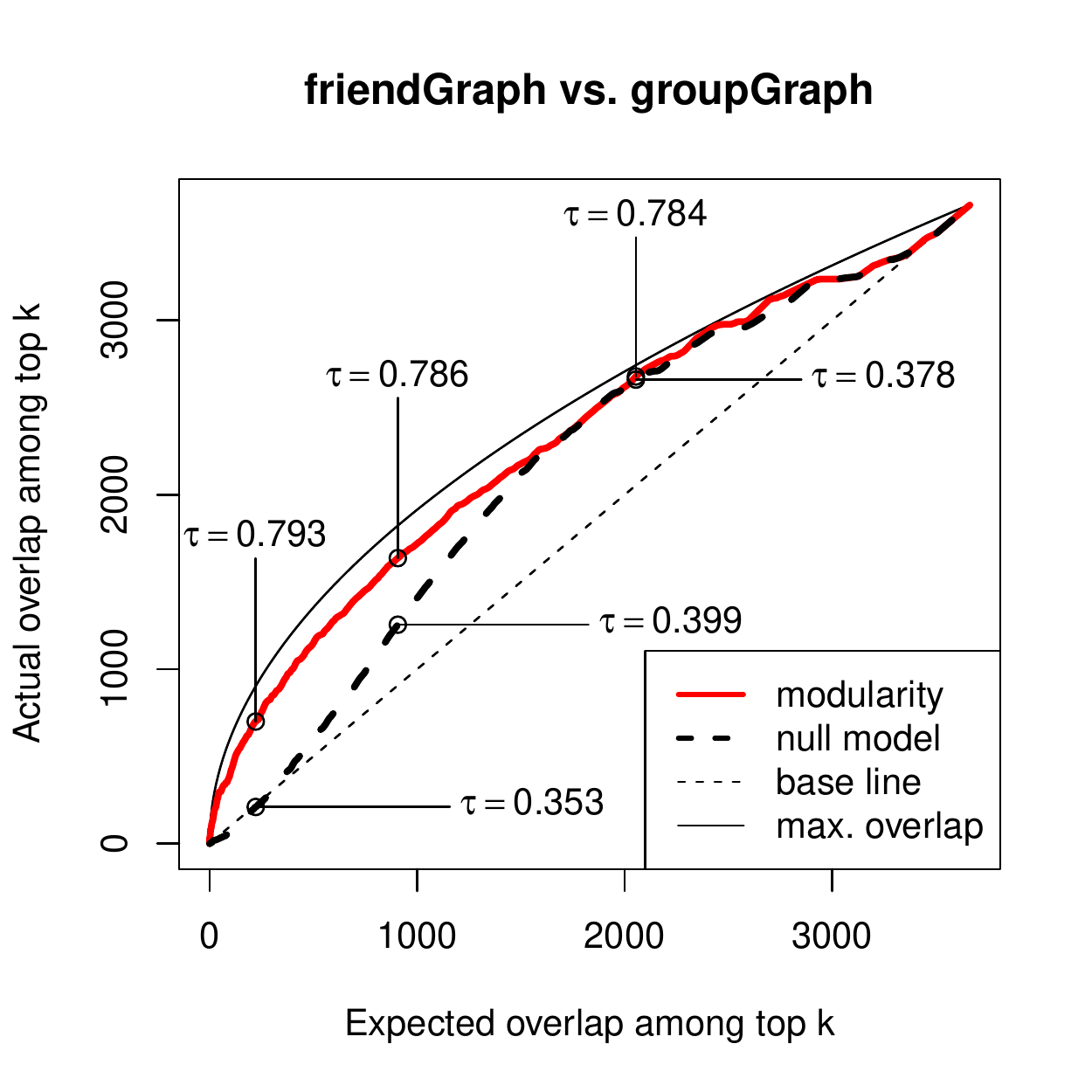}
\includegraphics[scale=0.35]{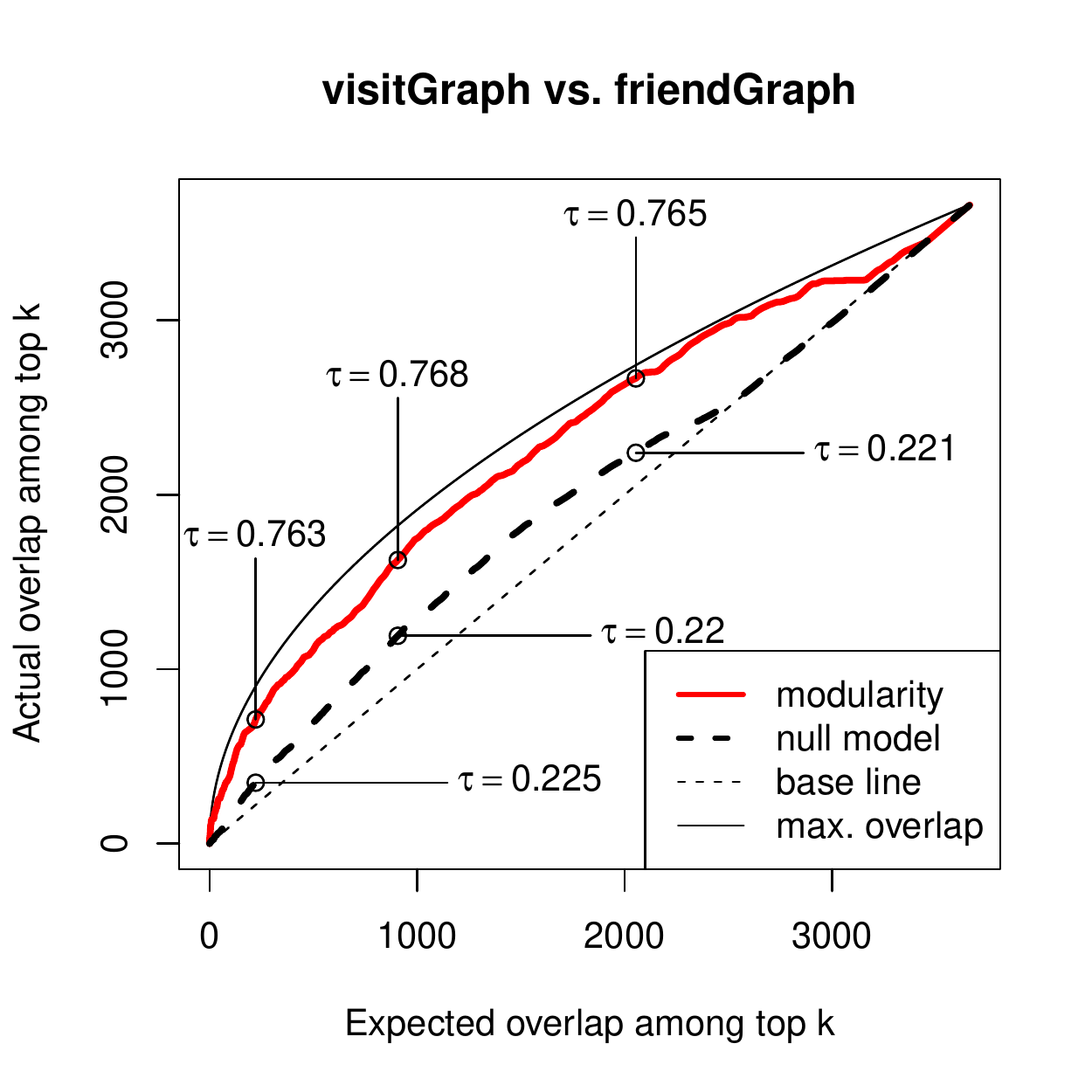}
\includegraphics[scale=0.35]{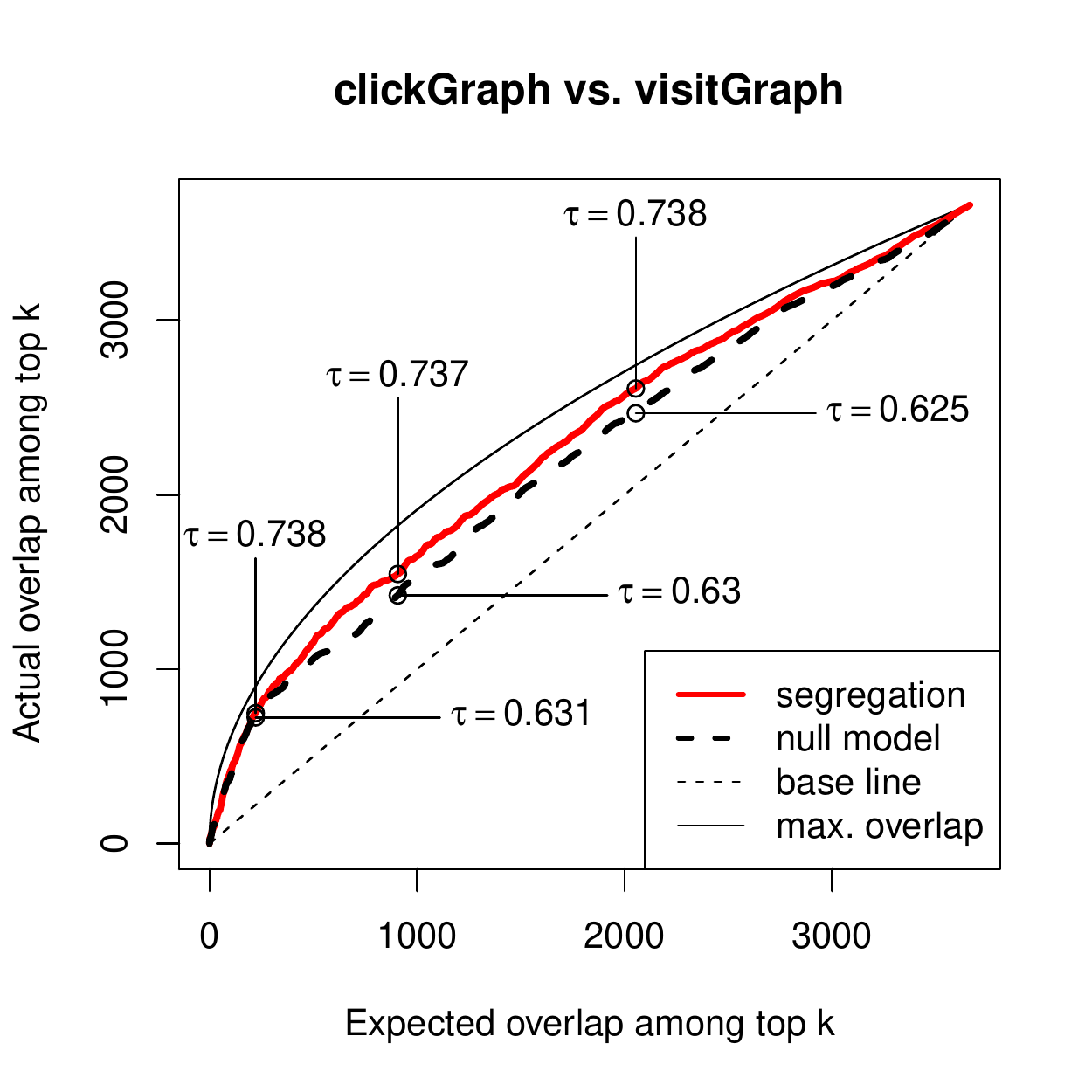}
\includegraphics[scale=0.35]{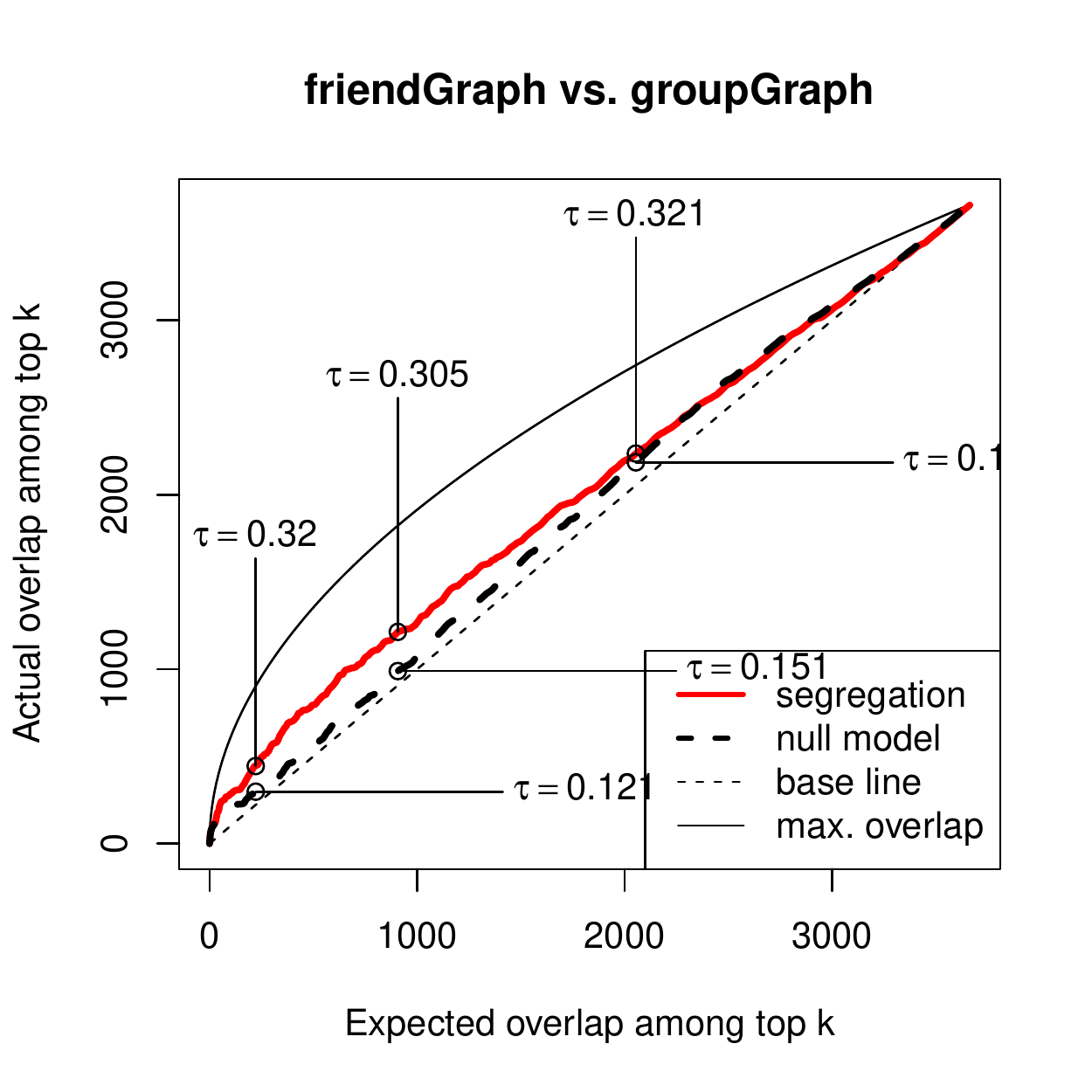}
\includegraphics[scale=0.35]{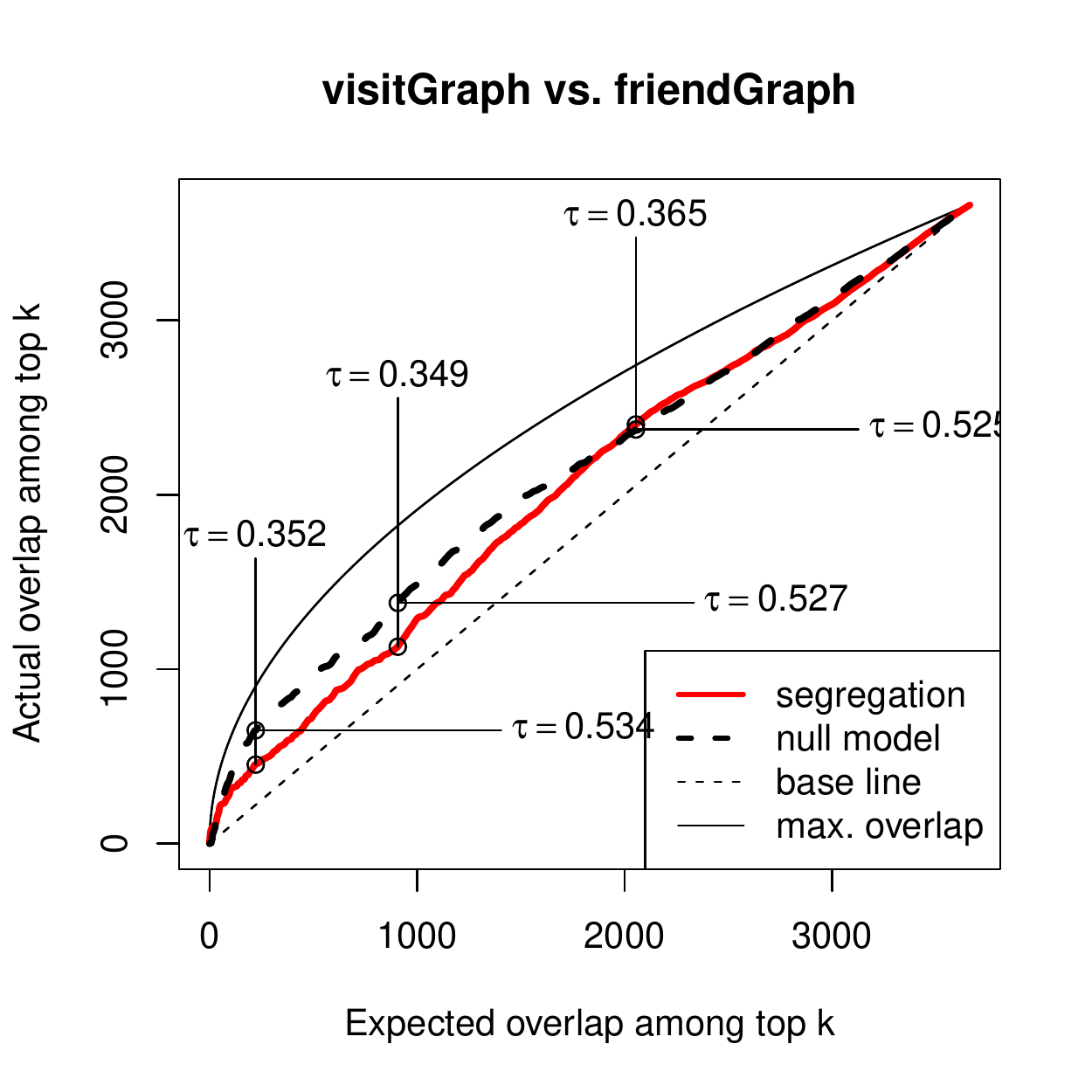}
  \centering
  \caption{Comparison between two rankings by considering the size of the
    intersection between the corresponding top $k$ entry set. The plot
    shows, how much the actual overlap deviates from the expected
    overlap if the rankings were randomly ordered.}
  \label{fig:correlations:bibsonomy:ovlerlap}
\end{figure*}

Finally, Figure~\ref{fig:correlations:bibsonomy:ovlerlap} details on
the dependence of observed correlation on the relative position within
the ranking, exemplarily for selected pairs of networks. Firstly, we note that for \emph{modularity
induced rankings} the overlap between the top $k$ positions is
consistently very high and better than the expected overlap (cf. Section~\ref{sec:experiments:description} above), as shown by the the bold lines above the main diagonal close to the ``max overlap'' line. 
%
If the induced rankings are meaningful with respect to their
inherent community structure, the rankings should depend on the
community structure and not on statistical properties of the network
which merely result from the network's degree distribution. For that, we
calculated for each network the induced rankings on corresponding
randomly rewired networks (in which the community structure is
destroyed) \cite{maslov2002specificity}. The respective overlap curve
is given in Fig. \ref{fig:correlations:bibsonomy:ovlerlap} as the bold dashed
line (``null model'').
Furthermore, the correlation coefficients consistently
indicate significant correlations with increasing magnitude for larger
$k$.

Secondly, the overlap with rankings induced by rewired networks
show varying behavior for the different networks. Comparing the
rankings induced by the Click graph and the Visit graph, the
null-model overlap consistently lies below the expected overlap for independent rankings whereas it coincides with the expected null-model
overlap only in the beginning for both other pairs of networks.

Alltogether, the correlation coefficients indicate independence in rankings for all null-model rankings. These findings indicate a consistent relative ranking behavior in the \bibs networks, which is also confirmed by the null-model experiments.

\subsubsection{\twitter}\label{sec:experiments:twitter}

\begin{figure*}
  \centering
  \includegraphics[width=0.32\linewidth]{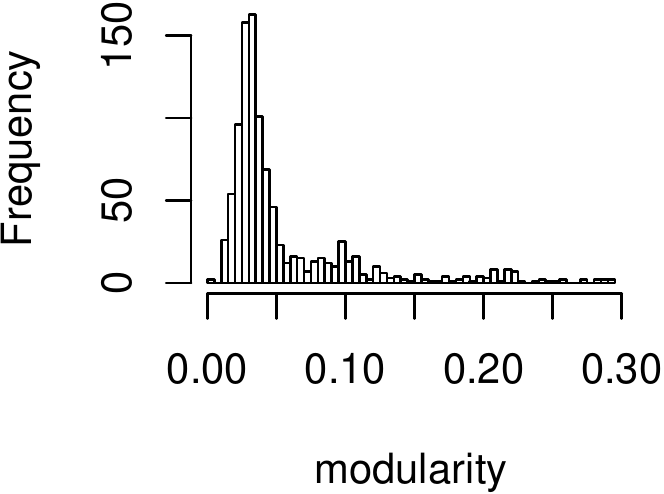}
  \includegraphics[width=0.32\linewidth]{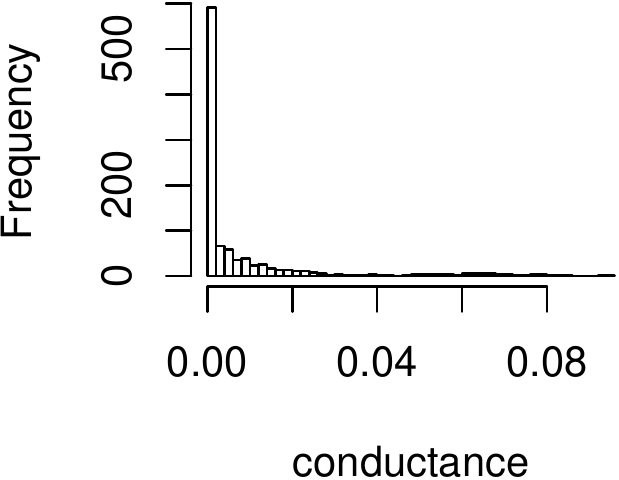}
  \includegraphics[width=0.32\linewidth]{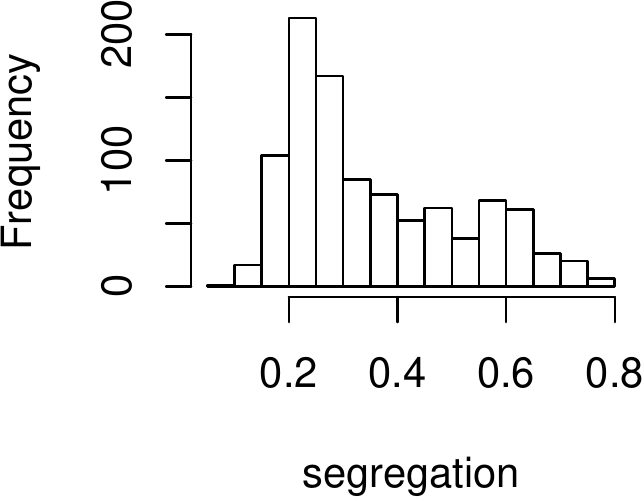}
  \caption{Distribution of the quality functions for the
    follower graph in \twitter.}
\label{fig:community:twitter:distribution}
\end{figure*}

\begin{figure*}
\includegraphics[width=0.32\linewidth]{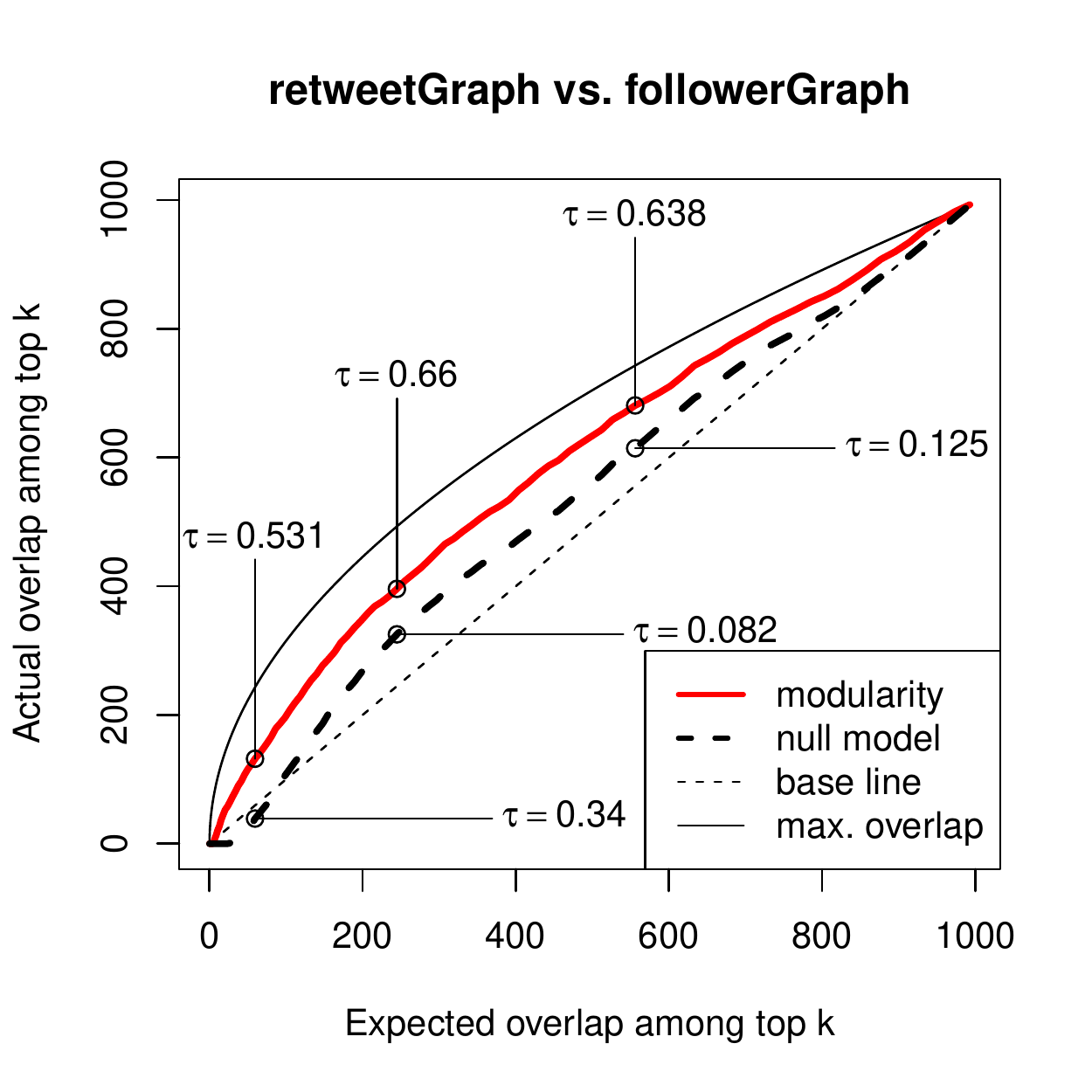}
\includegraphics[width=0.32\linewidth]{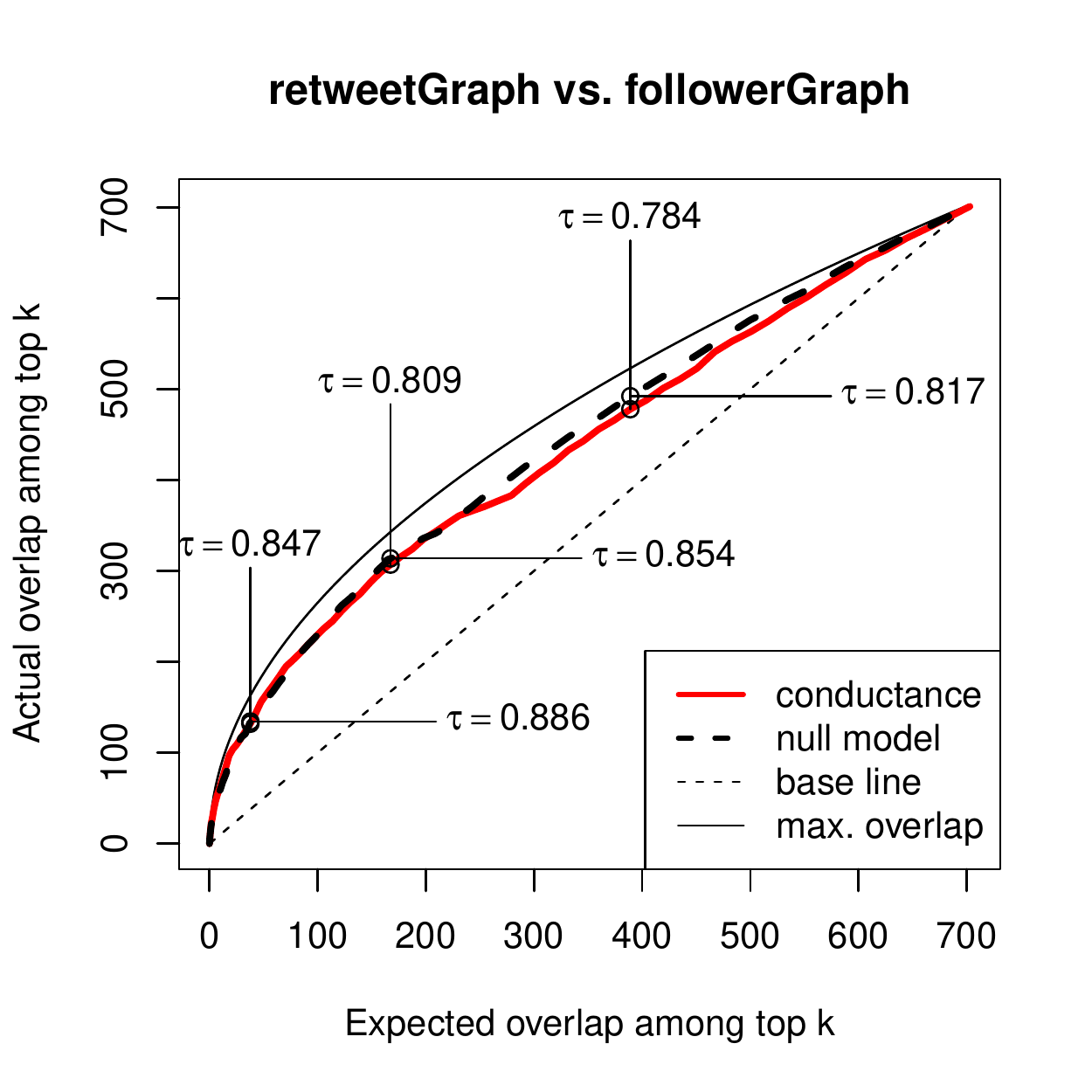}
\includegraphics[width=0.32\linewidth]{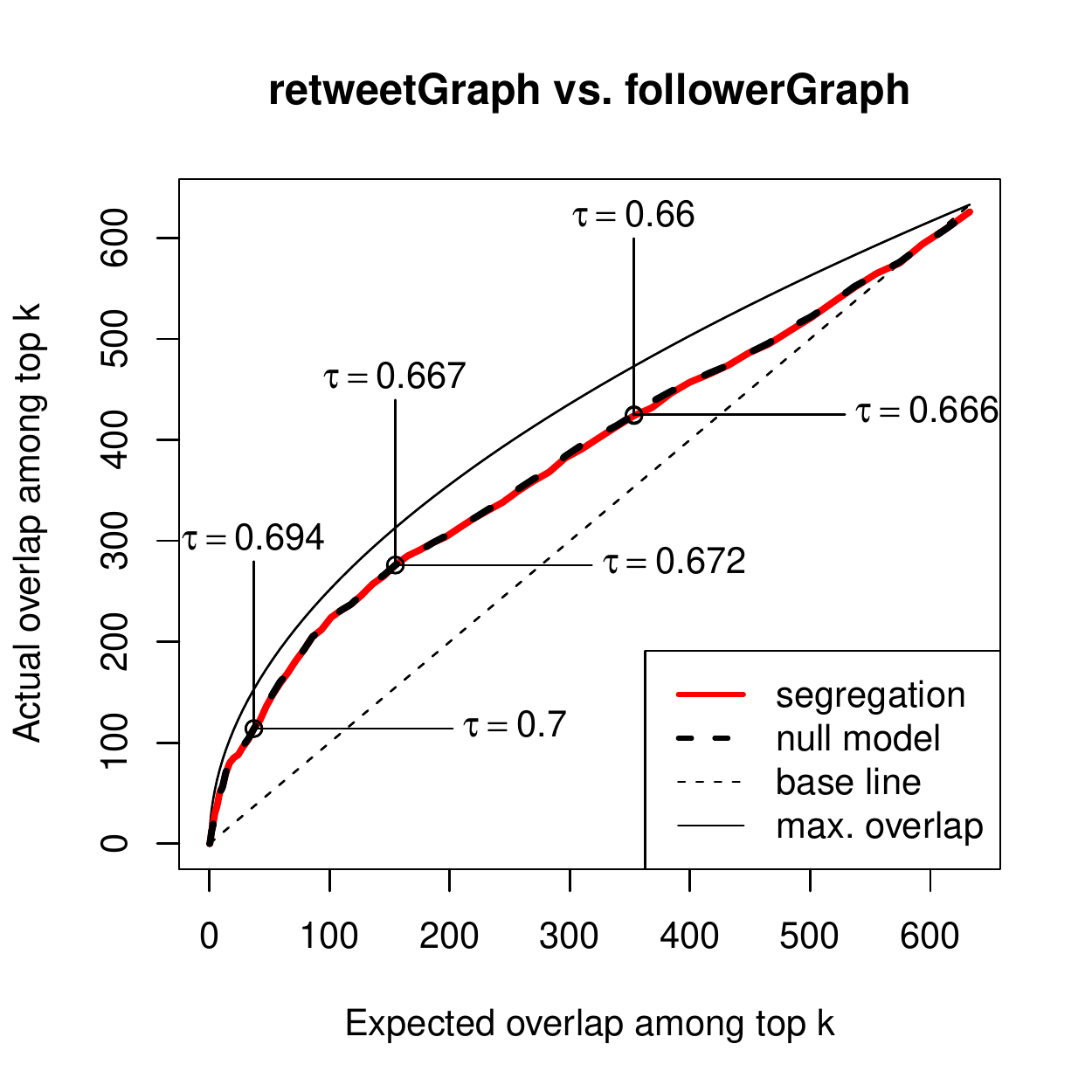}
  \centering
  \caption{ Comparison between rankings by considering the size of the
    intersection between the corresponding top $k$ entry set in
    \twitter: Modularity (left), intra-conductance (middle) and segregation index (right).}
  \label{fig:correlations:twitter:ovlerlap}
\end{figure*}

Applying the same set of clustering algorithms as for \bibs resulted in $993$ clustering
models for \twitter; as described above, the smaller number of community allocations -- compared to the \bibs dataset -- is explained by the fact that during application of the algorithms and parametrizations of the Cluto toolkit many algorithms did not terminate correctly due to resource limitations.
Figure~\ref{fig:community:twitter:distribution} shows the
distribution of all applied quality functions in all considered
networks. Again, modularity and segregation index both lead to a broad
range of quality function scores. It is worth noting that the modularity
distribution does not exhibit a sharply pronounced peak at level 0 as
in the case of \bibs.
Table~\ref{tab:twitter:modularity:correlation:full:tau} shows as in
the case for \bibs Kendall's $\tau$ rank correlation coefficient for
all pairs of networks as well as corresponding null models. For rankings
induced by intra-conductance and segregation index, Kendall's $\tau$
suggests independence, whereas for rankings induced by modularity,
significantly higher correlation is observed than in corresponding
null model experiments. These observations are in line with those for \bibs.
\begin{table}\scriptsize
    Modularity:\vspace{-\baselineskip}
    \begin{center}
  \begin{tabular}{l|c|c}
          & \RT                     & \Follower               \\ \hline

  \RT       &\cellcolor{grey}~0.057
            &~0.658
            \\\hline
  \Follower &\cellcolor{grey}~0.443
            &\cellcolor{grey}~0.097
            \\
  \end{tabular}
\end{center}

    Intra-Conductance:\vspace{-\baselineskip}
    \begin{center}
  \begin{tabular}{l|c|c}
          & \RT                     & \Follower               \\ \hline

  \RT       &\cellcolor{grey}-0.012
            &-0.027
            \\\hline
  \Follower &\cellcolor{grey}~0.024
            &\cellcolor{grey}-0.018
            \\
  \end{tabular}
\end{center}

    Segregation:\vspace{-\baselineskip}
    \begin{center}
  \begin{tabular}{l|c|c}
          & \RT                     & \Follower               \\ \hline

  \RT       &\cellcolor{grey}-0.005
            &~0.049
            \\\hline
  \Follower &\cellcolor{grey}~0.000
            &\cellcolor{grey}~0.004
            \\
  \end{tabular}
\end{center}

  \caption{
    Kendall's $\tau$ correlation coefficient for modularity and segregation
    based rankings relative to evidence networks in \twitter. For $i \leq j$
    the lower triangle shaded in gray shows the correlation for rankings
    induced by network $j$ and network $i$'s null models.
  } 
  \label{tab:twitter:modularity:correlation:full:tau}
\end{table}
Figure~\ref{fig:correlations:twitter:ovlerlap} visualizes these
results. Only the modularity induced rankings show a clear distinction
between the rankings induced by the original graphs and the
corresponding null models, indicating that the obtained
rankings are indeed dependent on the network's community structure.

Fig. \ref{fig:correlations:twitter:ovlerlap} plots the actual overlap between
the \emph{top $k$} positions of the rankings induced by \twitter's
Follower graph and the ReTweet graph versus the size of the expected
overlap if the rankings were independent (cf. Section~\ref{sec:experiments:description} above). 
As for the \bibs networks, we also calculated for each network the induced rankings on corresponding
randomly rewired networks (in which the community structure is
destroyed) \cite{maslov2002specificity}. The respective overlap curve
is given in Fig. \ref{fig:correlations:twitter:ovlerlap} as the bold dashed
line (``null model'').
Exemplary Fig. \ref{fig:correlations:twitter:ovlerlap} additionally shows
Kendall's $\tau$ for the two sequences of common community allocations
between the top $234$ elements (first quarter in the plots) of the rankings induced by twitter's
follower and retweet network (with a $p$ value $p<10^{-6}$) showing a
very strong correspondence between the two rankings. Again, these results confirm the consistent relative ranking of the community allocations as we have observed for the \bibs networks above.

\subsubsection{\flickr}\label{sec:experiments:flickr}
\begin{figure*}
  \centering
  \includegraphics[width=0.245\linewidth]{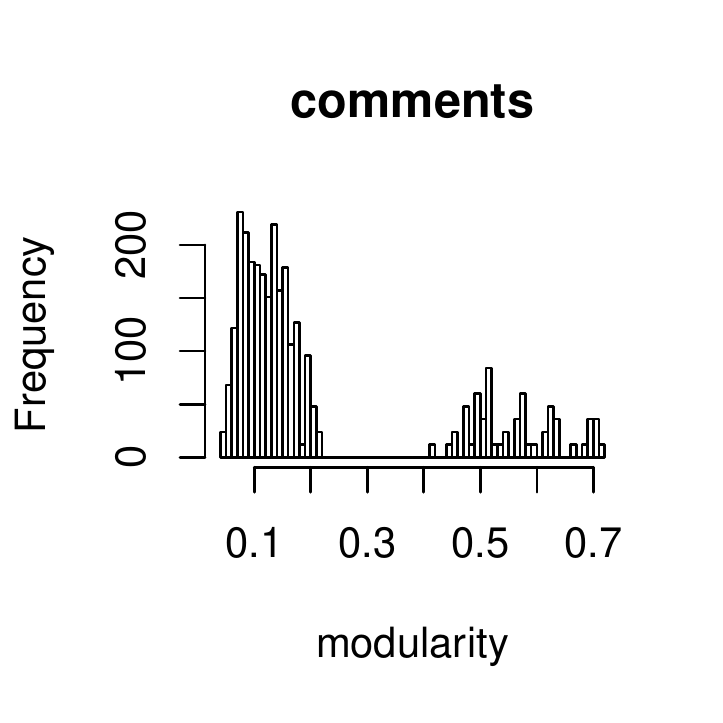}
  \includegraphics[width=0.245\linewidth]{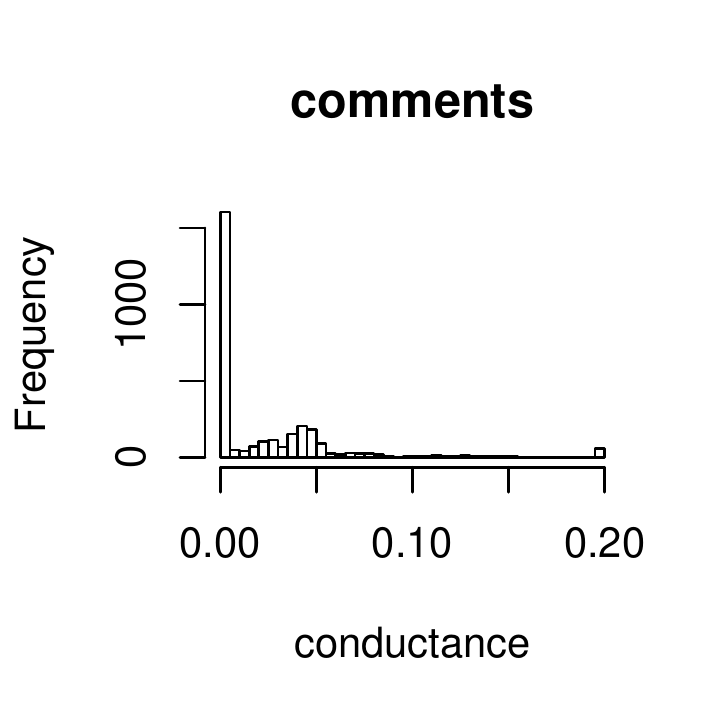}
  \includegraphics[width=0.245\linewidth]{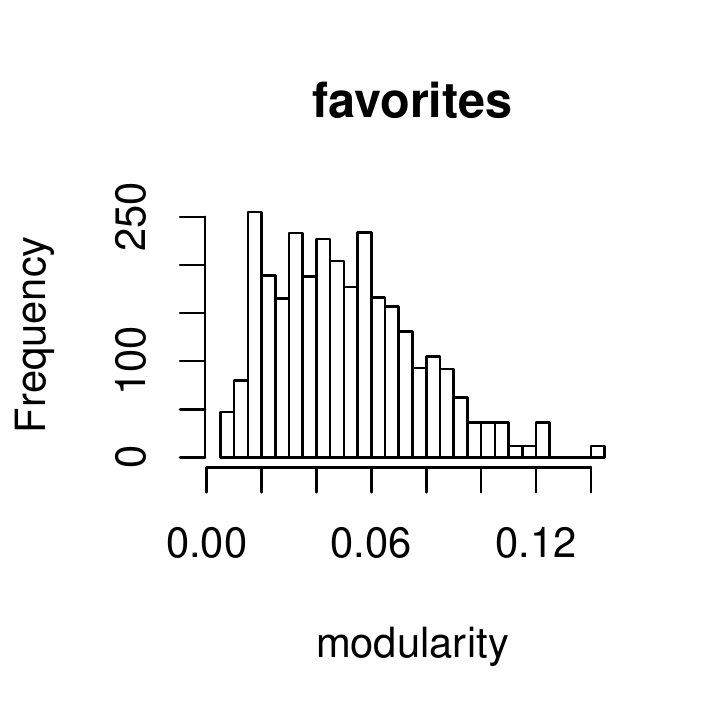}
  \includegraphics[width=0.245\linewidth]{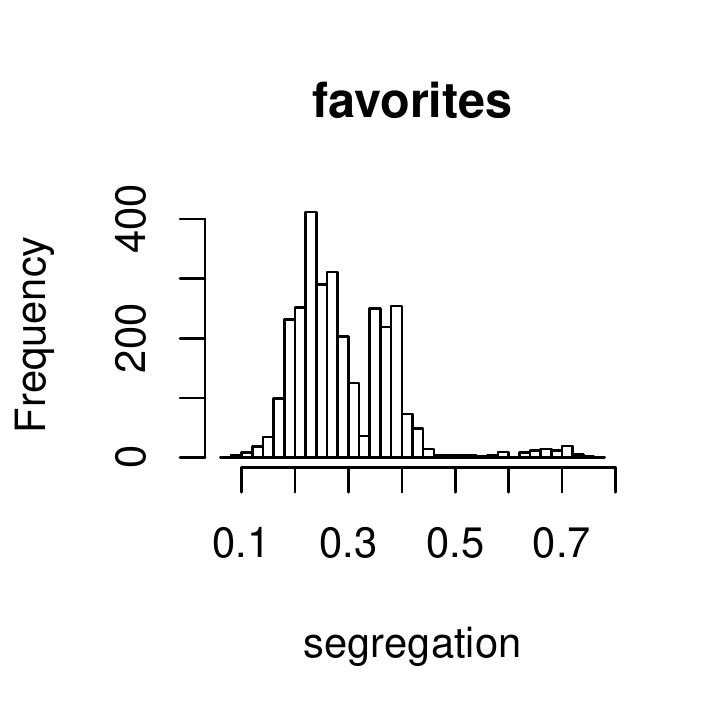}
  \caption{Distribution of the quality functions in all networks in \flickr.}
\label{fig:community:flickr:distribution}
\end{figure*}

For the \flickr dataset, the applied clustering algorithms resulted in $2,985$
community allocations. Figure~\ref{fig:community:flickr:distribution}
shows the distribution of quality function scores for all considered
networks. Again, modularity and segregation index show a broad range
of quality scores whereas (due to sparsity) intra-conductance only
assessed low quality communities. Most notably, the distribution of
modularity show a clear separation between lower and higher quality
ranked community allocations in the Comment graph whereas the
Favorites graph and the Contacts graph does not induce such a
pronounced bimodal distribution. Rankings induced by the segregation
index show a bimodal distribution for all considered networks.

\begin{figure*}
\includegraphics[width=0.32\linewidth]{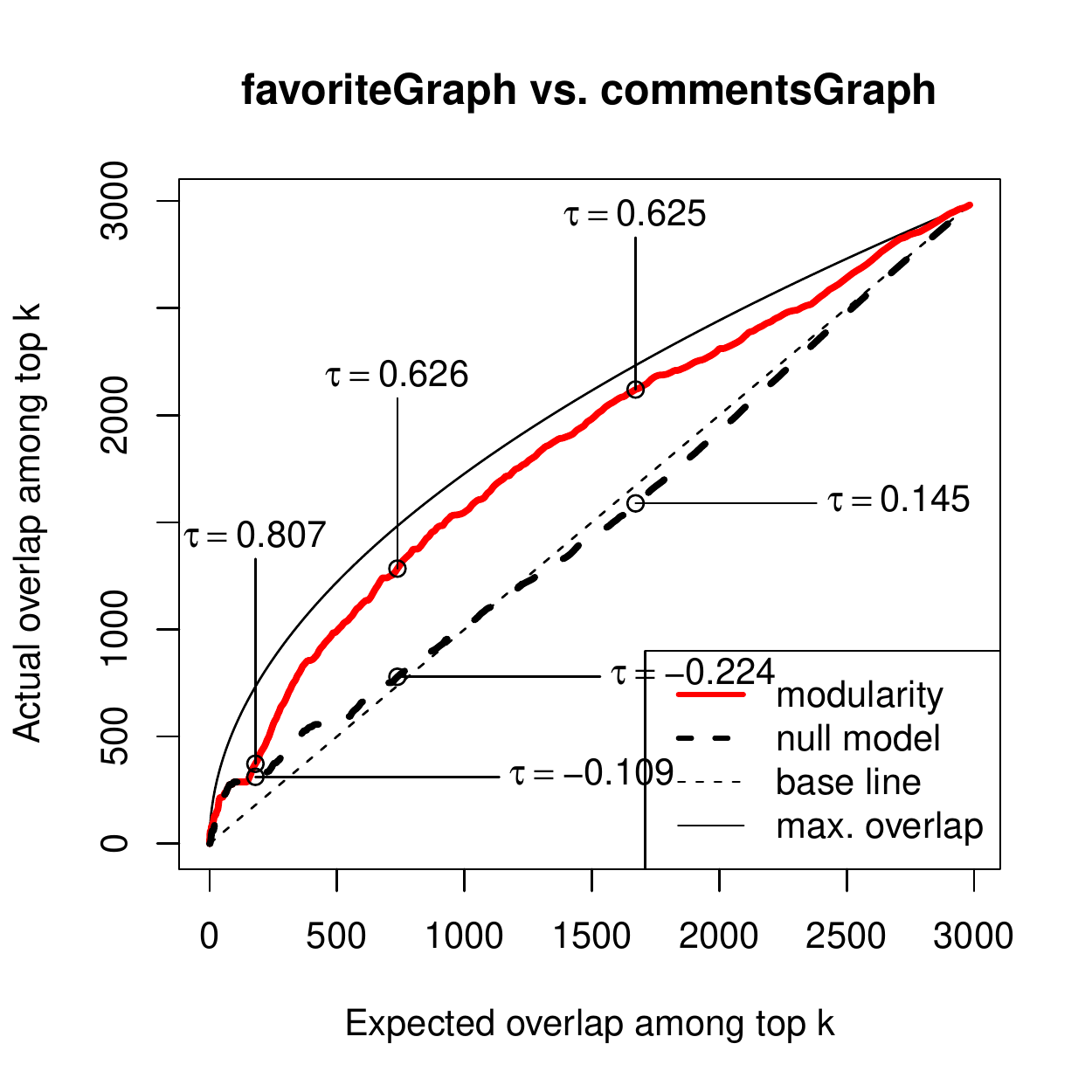}
\includegraphics[width=0.32\linewidth]{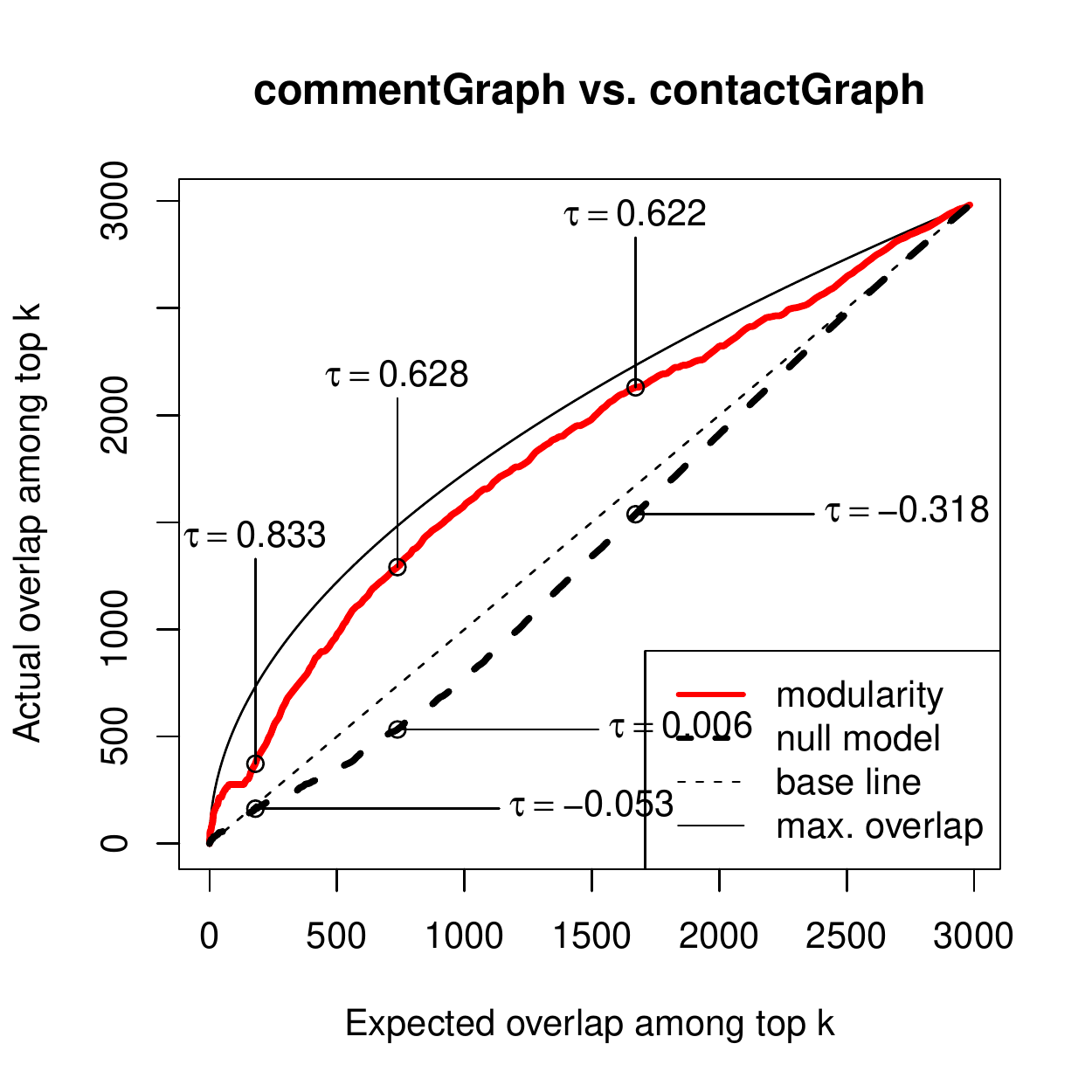}
\includegraphics[width=0.3\linewidth]{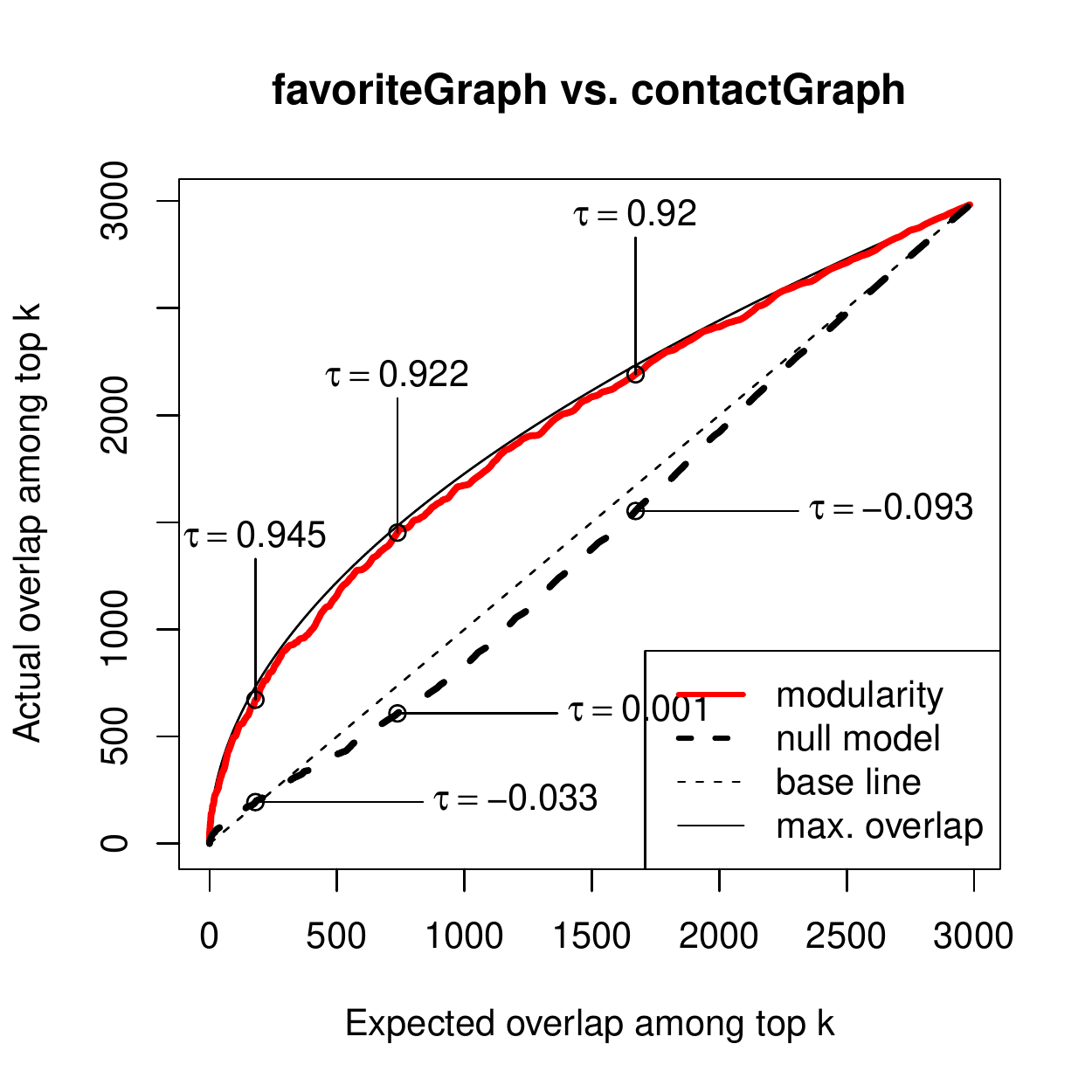}
\includegraphics[width=0.32\linewidth]{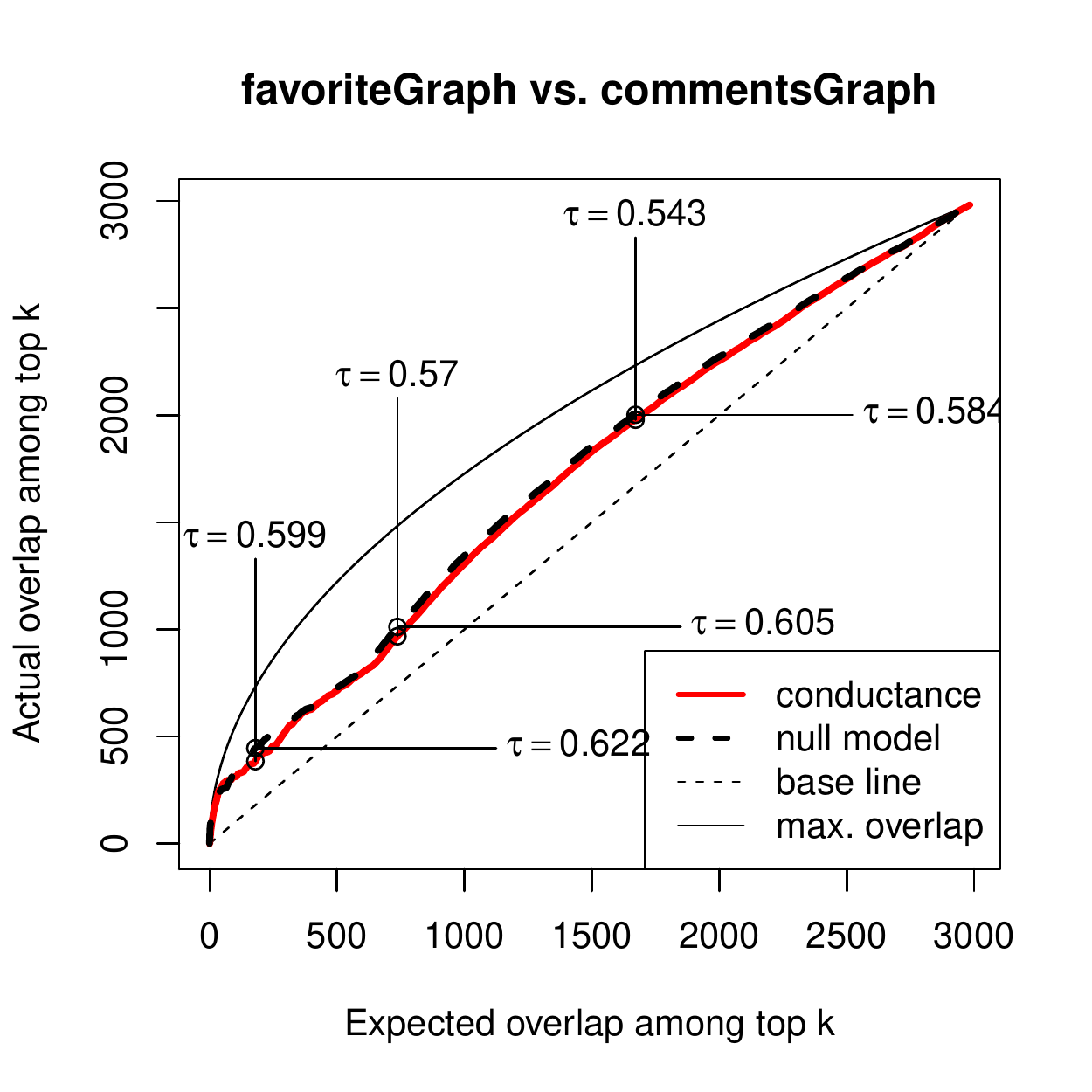}
\includegraphics[width=0.32\linewidth]{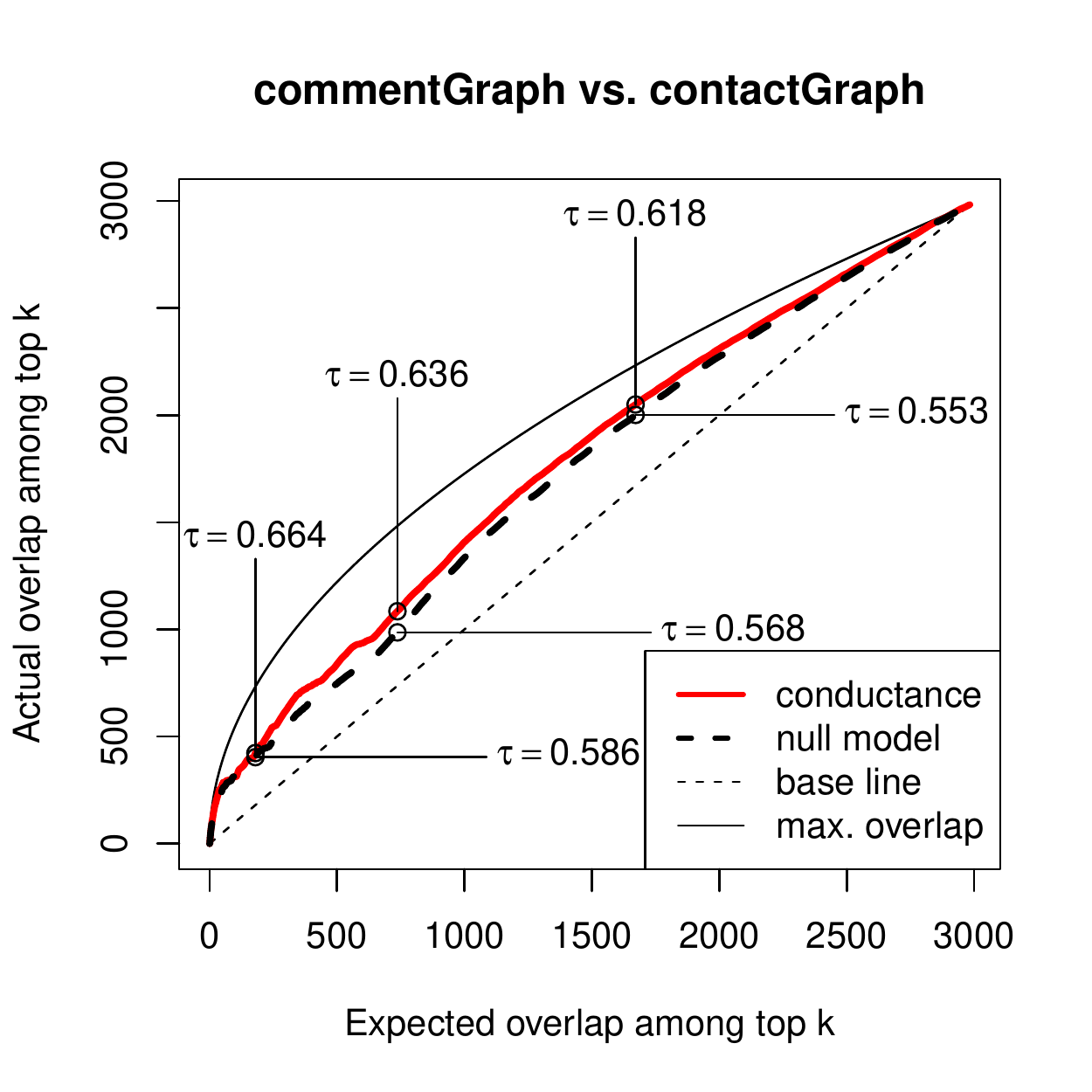}
\includegraphics[width=0.32\linewidth]{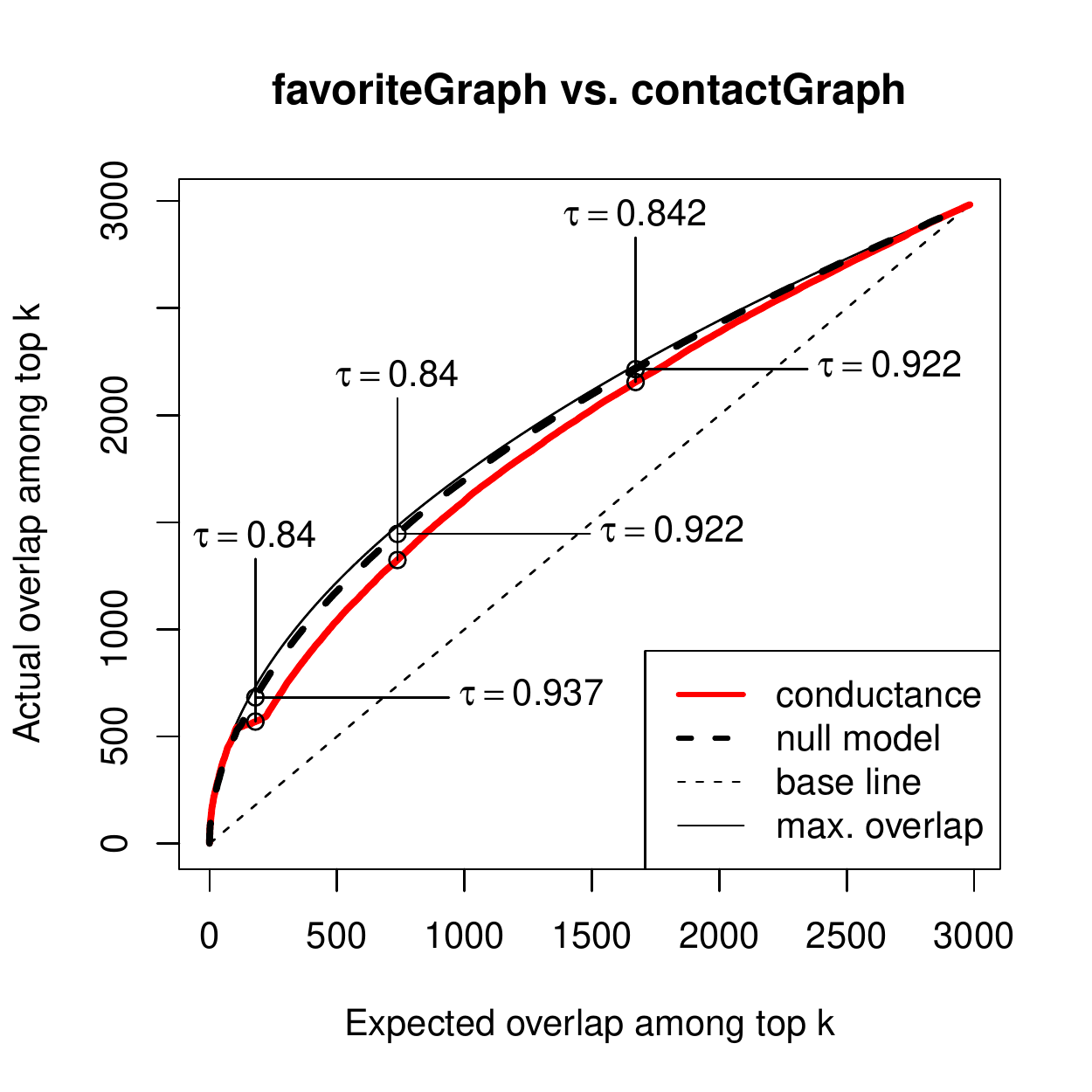}
\includegraphics[width=0.32\linewidth]{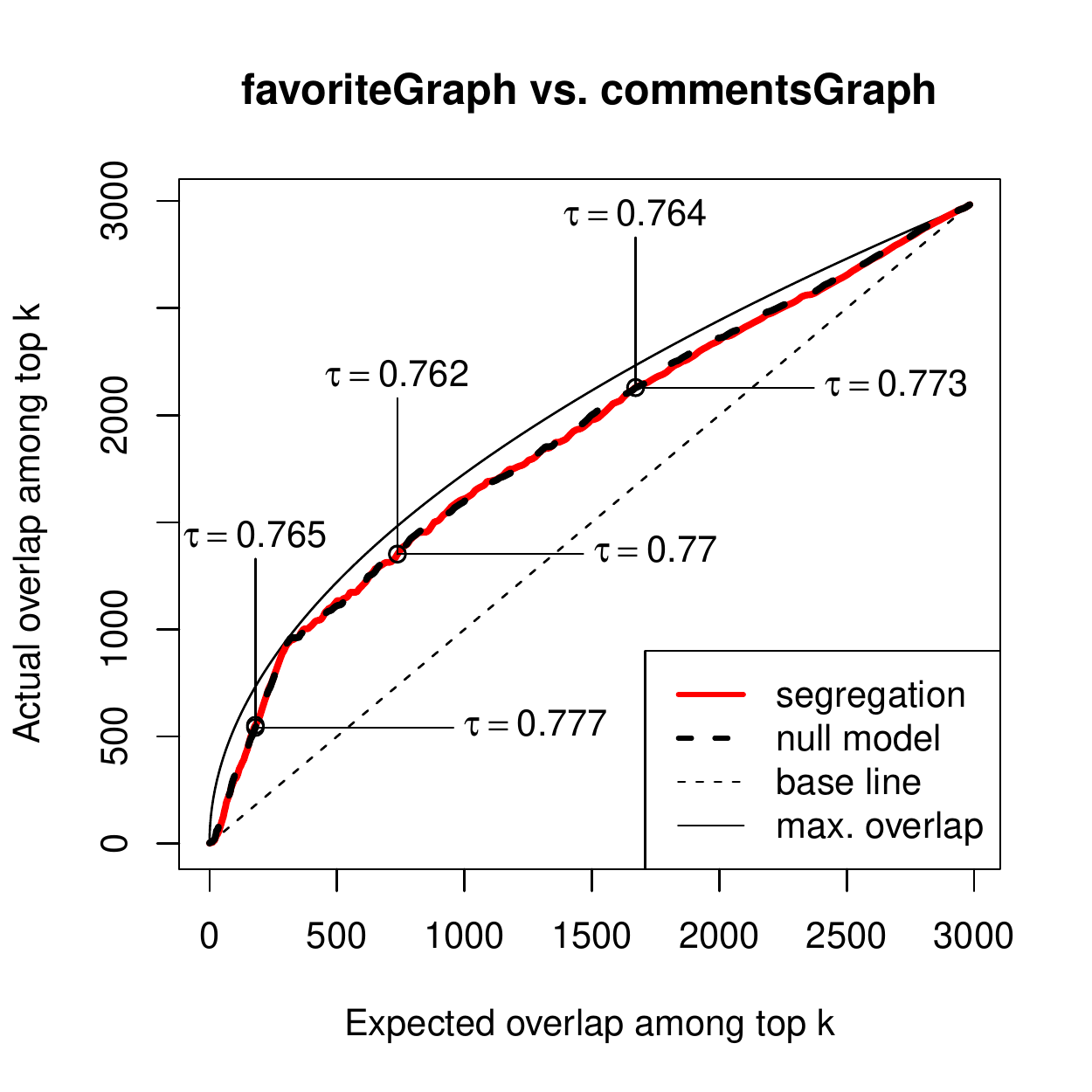}
\includegraphics[width=0.32\linewidth]{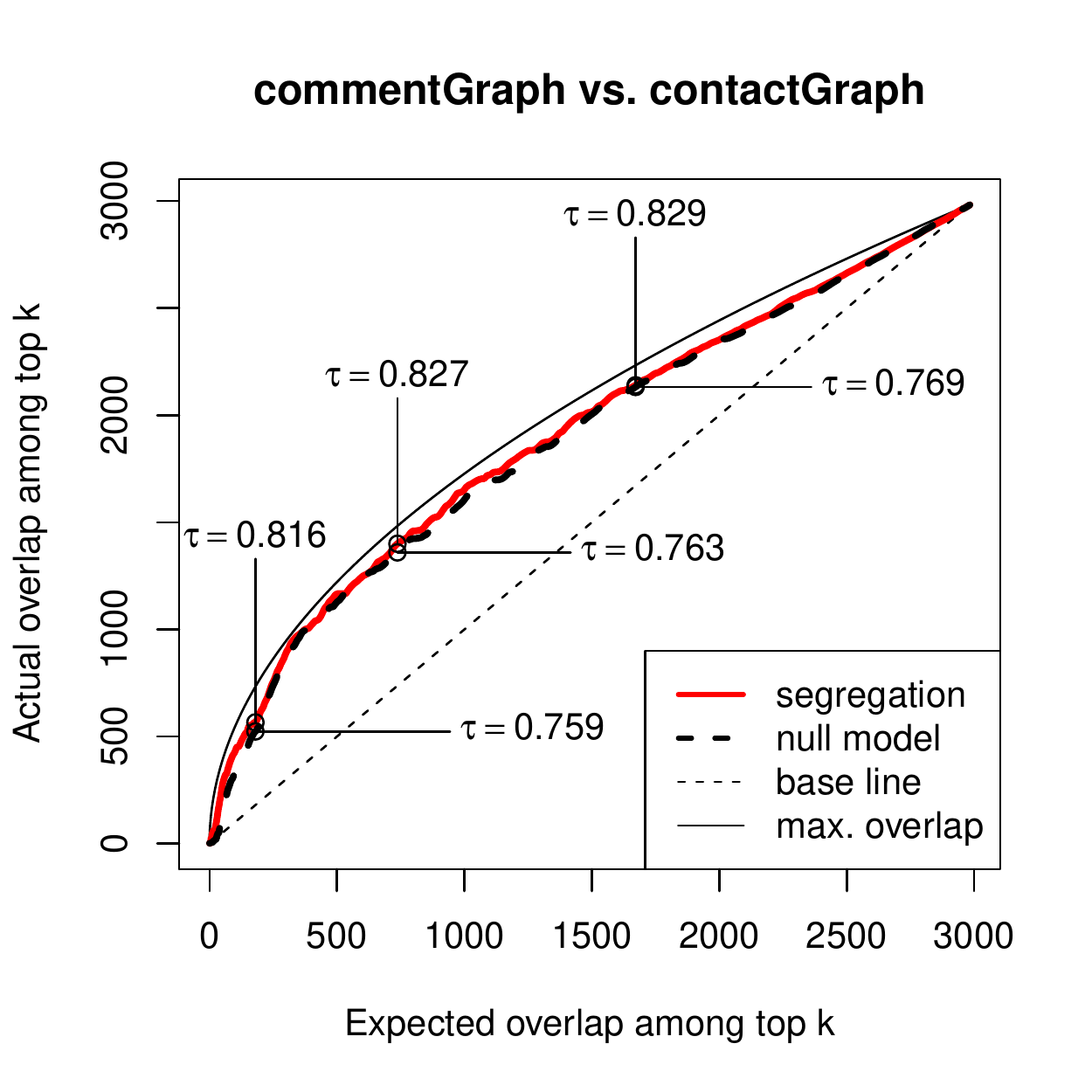}
\includegraphics[width=0.32\linewidth]{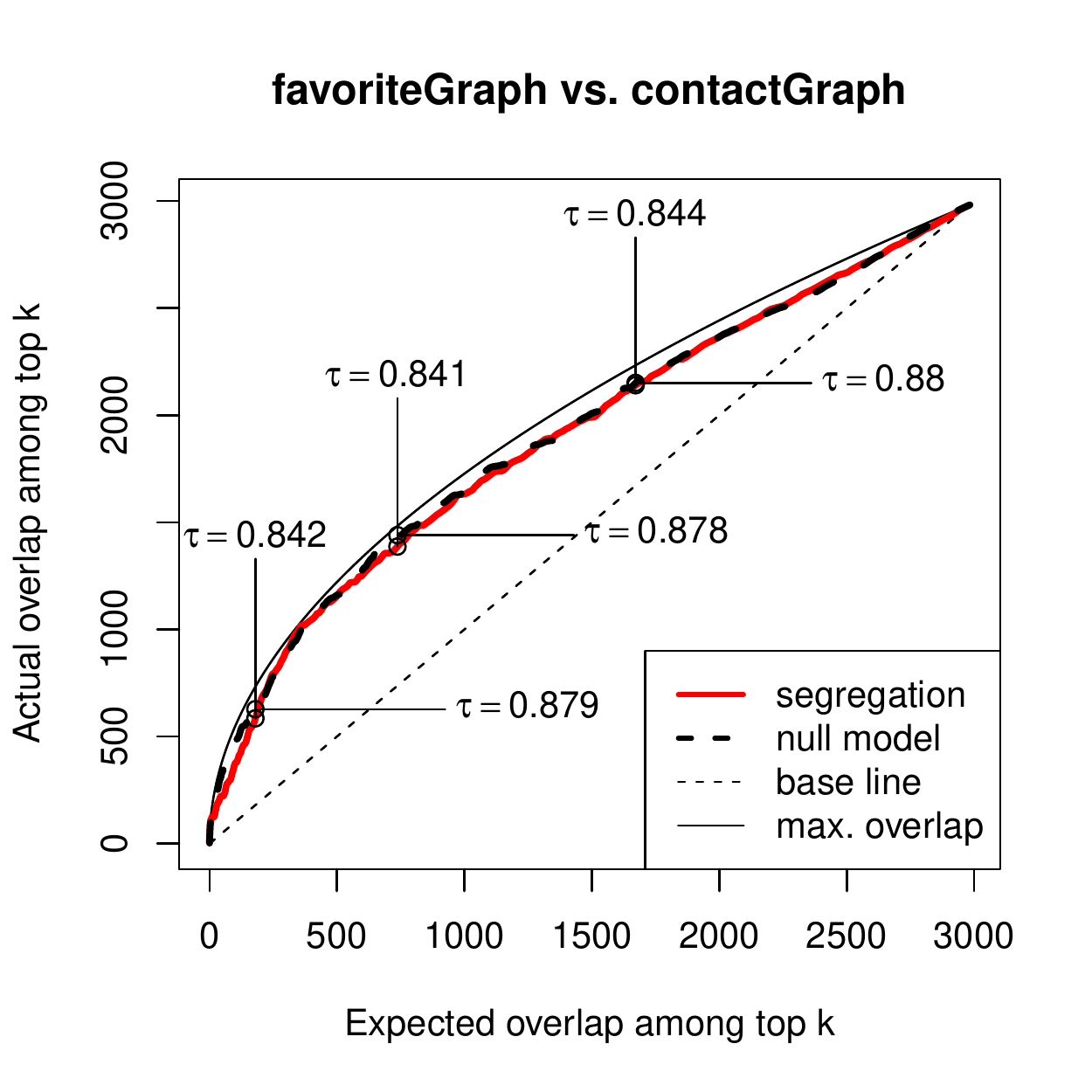}
  \centering
  \caption{ Comparison between two rankings by considering the size of the
    intersection between the corresponding top $k$ entry set in
    \flickr: Modularity (top), intra-conductance (middle) and segregation index (bottom).}
  \label{fig:correlations:flickr:ovlerlap}
\end{figure*}

Table~\ref{tab:flickr:modularity:correlation:full:tau} shows the averages of
Kendall's $\tau$ over every network and all pairs of networks together
with corresponding null model graphs for modularity and segregation
index. Again only for modularity, the correlation coefficient suggests
strong correlations among rankings obtained in different networks
which is missing when community structure is destroyed in corresponding null models.
\begin{table}\footnotesize
    Modularity:\vspace{-1.5\baselineskip}
    \begin{center}
  \begin{tabular}{l|c|c|c}
            & \Comment & \Favorite & \Contact \\ \hline
  \Comment  & \cellcolor{grey}-0.029
            & ~0.789
            & ~0.811
            \\\hline
  \Favorite & \cellcolor{grey}-0.151
            & \cellcolor{grey}-0.220
            & ~0.944
            \\\hline
  \Contact  & \cellcolor{grey}-0.152
            & \cellcolor{grey}-0.007
            & \cellcolor{grey}-0.016
            \\
  \end{tabular}
\end{center}

    Segregation:\vspace{-1.5\baselineskip}
    \begin{center}
  \begin{tabular}{l|c|c|c}
            & \Comment & \Favorite & \Contact \\ \hline
  \Comment  & \cellcolor{grey}~0.027
            & ~0.034
            & ~0.034
            \\\hline
  \Favorite & \cellcolor{grey}~0.014
            & \cellcolor{grey}-0.004
            & ~0.028
            \\\hline
  \Contact  & \cellcolor{grey}~0.016
            & \cellcolor{grey}-0.023
            & \cellcolor{grey}-0.013
            \\
  \end{tabular}
\end{center}

  \caption{
    Kendall's $\tau$ correlation coefficient for modularity and segregation
    based rankings relative to evidence networks in \flickr. The upper triangle shows the correlations between the different networks. For $i<j$
    the shaded lower triangle shows the correlation for rankings
    induced by network $j$ and network $i$'s null models, \ie comparing the shuffled networks.    
  } 
  \label{tab:flickr:modularity:correlation:full:tau}
\end{table}
These results are again visualized in Figure~\ref{fig:correlations:flickr:ovlerlap}. Again, these results confirm the findings obtained by the \bibs and Twitter analysis. 

\subsection{Discussion}

The applied paradigm of evaluating against existing social structures has the significant advantage, that the assessment of community allocations are relative to explicitly established social links (\eg friendship links in \bibs); their intention is multifaceted as it is the case for the underlying social constellation. On the other hand, evidence networks as introduced in this work could directly be used for finding and evaluating communities. The main drawback would be, that evidence networks typically only cover a certain amount of the users (in case of the friendship network in \bibs only $12$\%). Additionally, most community detection algorithms try to optimize a given objective function which assesses the quality of the community allocations and there is a risk that the method thus optimizes towards community allocations biased by the applied objective function (\eg communities which contain half of the users
  in case of modularity \cite{Leskovec2010}). Thus, the proposed method separates the objective function used for optimizing the community allocation from the quality function used for assessing it.

The experimental results presented in the previous section indicate that implicit evidence networks used for assessing the quality of a community structure are surprisingly consistent with the expected behavior as formalized
by the existing explicit social structures, in particular concerning the Friend graph.

In our experiments, we observed a high correlation between the quality measures calculated on the implicit and explicit networks supporting this hypothesis. Furthermore, in our experiments, modularity provided always the best results in contrast to the conductance or the segregation index measures. Therefore, based on our findings, we would always recommend modularity as a quality measure for the ranking based assessment method.

\section{Related Work}\label{sec:related}

In the following, we discuss related approaches concerning the analysis of network structure, communities, and implicit link structures in different systems and networks.

Analyzing Web 2.0 data by applying complex network theory goes back to
the analysis of (samples from) the web graph \cite{broder2000graph}. 
Mislove \et~\cite{mislove2007measurement} applied methods from social
network analysis as well as complex network theory and analyzed large scale crawls from
prominent social networking sites. Some properties common to all considered
social networks are worked out and contrasted to properties of the web graph. 
Newman analyzed many real life networks, summing up characteristics of
social networks
\cite{newman2003social}. 
The analysis of online social media, the interrelations of the involved actors, and the involved geospatial extents have attracted a lot of attention during the last decades, especially for the microblogging system \twitter.
A thorough analysis of fundamental network properties and interaction patterns in \twitter
can be found in~\cite{kwak2010twitter}.

Interdependencies of social links and geospatial proximity are
investigated in \cite{scellato2011sociospatial,mcgee2011geographic,kaltenbrunner2012eyes}, especially concerning the correlation of the probability of friendship links and
the geographic distance of the corresponding users.
%
Silva et al.~\cite{SWZ:10} mine structural correlation patterns in network partitions, \ie correlations between vertex attributes and dense components in undirected graphs. While their approach results in individual patterns, our analysis captures both patterns and the networks/graphs as a whole and provides comprehensive analysis on their combined structure.

Schifanella \et~\cite{schifanella2010folks} investigated the
relationship of topological closeness (in terms of the length of shortest paths) with
respect to the semantic similarity between the users, while similarity measures in the context of semantic analysis of folksonomies are evaluated in~\cite{markines2009evaluating}. We adapt these approaches for our setting. 

Leroy et al.~\cite{LCB:10} discusses a feature-based approach using implicit information for inferring interaction networks in the context of link prediction. 
Eagle et al.~\cite{EPL:09} describe an approach for reconstructing friendship relations from secondary (mobile phone) data. They show, that friendship links can be inferred with a high probability but do not present a comprehensive analysis of different evidence networks and their impact on the predictability. Barrat et al.~\cite{barrat2010lss} discuss the relation between online and offline networks.
Similarly, Chin et al.~\cite{DBLP:conf/socialcom/ChinWXZWZ11} consider ephemeral networks of encounters for inferring contact networks, however, no relations to other evidence networks are discussed.

Another aspect of our work is the analysis of implicit link structures which can be obtained in a running Web 2.0 system and how they relate to other existing link structures, \ie evidence networks.
Butts and Carley~\cite{BC:05} describe simple algorithms for structural comparisons between different kinds of structured objects. 
Furthermore, Baeza-Yates \et~\cite{baeza2007extracting} propose to present
query-logs as an implicit folksonomy where queries can be seen as tags
associated to documents clicked by people making those queries. 
The authors extracted semantic relations between queries from a query-click bipartite graph; nodes are queries and an edge between nodes exists when at least
one equal URL has been clicked after submitting the query. 
Krause \et~\cite{krause2008logsonomy} analyzed term-co-occurrence-networks in
the logfiles of internet search systems.

Fortunato and Castellano~\cite{Fortunato2007} discuss various aspects connected to the concept of community structure in graphs. Basic definitions as well as existing and new methods for community detection are presented.
In \cite{Lancichinetti2009}, Lancichinetti and Fortunato present a thorough comparison of many different state of the art community detection algorithms for graphs. The performance of algorithms are compared relative to a class of adequately generated artificial benchmark graphs.
Karamolegkos \et~\cite{Karamolegkos20091498} introduced metrics for
assessing user relatedness and community structure by means of the normalized
size of user profile overlaps. They evaluate their metrics in a live setting,
focussing on the optimization of the given metrics.
%
%
Using a metric which is purely based on the structure of graphs,
Newman presents algorithms for finding communities and assessing community structure
in graphs~\cite{Newman:2004}.  A thorough empirical analysis of the impact of
different community mining algorithms and their corresponding objective function
on the resulting community structures is presented in \cite{Leskovec2010}, which is based on the size resolved analysis of community structure in graphs as presented in~\cite{Leskovec2008}. 
In the context of this paper, we do not focus on the identification of communities. In contrast, we propose a method for the relative ranking and assessment of communities based on our findings on structural inter-network correlations.
\section{Conclusions}\label{sec:conclusion}

The contribution of this paper can be summarized as follows: We introduce evidence networks as a new tool for structural assessment in social applications and show that there are structural inter-network correlations that allow reciprocal conclusions between the different networks. Furthermore, our conducted experiments suggest that the different networks are not contradictory but complementary. The evidence networks are thoroughly analyzed with respect to the contained community structure (\cf Section \ref{sec:analysis}). It is shown that there is a strong common community structure across different networks.

Furthermore, using standard community evaluation measures, we showed that there is a strong common community structure across different networks; the induced rankings are reciprocally consistent. Therefore, we proposed a method for the relative ranking of communities for their assessment. In general, the task of automatically finding and recognizing meaningful community allocations on a set $U$ of users is still an open problem. Due to the multifaceted characteristics of user relatedness it is impossible to define ``the best'' community allocation. Therefore, the assessment of the quality of a given community is thus always application dependent and \emph{relative} to certain aspects of user relatedness, \eg race of individuals in \cite{newman2003structure}, shared topical interests in social bookmarking systems, or social traces manifested in the evidence networks.

Specifically, the presented analysis is thus not only relevant for the evaluation of community mining techniques, but also for implementing new community detection or user recommendation algorithms, among others.  
The context of the presented analysis is given by social media applications such
as social networking, social bookmarking, and social resource sharing
systems, considering the \twitter microblogging service, our own system
\bibs~\cite{Bibs:VLDB10}, and the \flickr resource sharing system as examples.

\comments{
This paper's contribution is three-fold. \emph{Firstly}, a new method for assessing the quality of user communities in social resource sharing systems is introduced (\cf Section \ref{sec:method}), applied to the real-world data from the social resource sharing system \bibs and evidences for its soundness are collected (\cf Section \ref{sec:experiments}). This method gives \emph{secondly} raise to a work flow for obtaining meaningful community allocations, allowing for exhaustive search in several algorithm's parameter space (\cf Section \ref{sec:method:paradigm}). \emph{Thirdly}, by considering different networks as samples from the same underlying social constellation (\cf Section \ref{sec:method}), the explanatory power of various commonly used cluster indices are explored in a new way, showing significant differences (\cf Section \ref{sec:experiments:results}).
}

For future work, we aim to investigate, how the single evidence networks can be suitably combined into a weighted network. For this, we need to further analyze the individual structure of the networks, and the possible interactions. As another direction of research, we plan to incorporate evidence networks in the community detection process (\eg in terms of constraints).



\bibliographystyle{spmpsci}      
\bibliography{bibliography,sg}







\end{document}